\documentclass{amsart}
\usepackage{amsmath,amssymb}
\usepackage[poorman]{cleveref}
\usepackage{physics}
\usepackage{mathrsfs}
\usepackage{graphicx}
\usepackage{color}
\usepackage{algorithm}
\usepackage{algpseudocode}
\usepackage{subcaption}

\newtheorem{theorem}{Theorem}[section]
\newtheorem{lemma}[theorem]{Lemma}

\theoremstyle{definition}

\theoremstyle{remark}
\newtheorem{remark}[theorem]{Remark}

\newcommand{\dt}{\Delta t}
\newcommand{\e}{\mathrm{e}} 
\newcommand{\id}{\mathrm{id}} 
\newcommand{\ii}{\mathrm{i}} 

\newcommand{\Bcal}[2]{\mathcal{B}_{#1,#2}}
\newcommand{\bk}{\boldsymbol{k}}

\newcommand{\Pcal}{\mathcal{P}}

\newcommand{\yx}[1]{\textcolor{red}{[yx: #1]}}
\newcommand\dgm[2][.5]{\,\includegraphics[scale=#1]{figures/#2}\,}

\numberwithin{equation}{section}

\begin{document}

\title[Second-order discretization of Dyson series]{Second-order discretization of Dyson series: iterative method, numerical analysis and applications in open quantum systems}

\author{Zhenning Cai}
\address{Department of Mathematics, National University of Singapore, 10 Lower Kent Ridge Road, Singapore 119706}
\email{matcz@nus.edu.sg}
\thanks{The work of Zhenning Cai was supported by the Academic Research Fund of
the Ministry of Education of Singapore under grant A-8002392-00-00.}

\author{Yixiao Sun}
\address{Department of Mathematics, National University of Singapore, 10 Lower Kent Ridge Road, Singapore 119706}
\email{s.yixiao@u.nus.edu}

\author{Geshuo Wang}
\address{Department of Applied Mathematics, University of Washington, Seattle, WA 98195, USA (corresponding author)}
\email{geshuo@uw.edu}


\begin{abstract}
We propose a general strategy to discretize the Dyson series without applying direct numerical quadrature to high-dimensional integrals, and extend this framework to open quantum systems. The resulting discretization can also be interpreted as a Strang splitting combined with a Taylor expansion. Based on this formulation, we develop a numerically exact iterative method for simulation system-bath dynamics. We propose two numerical schemes, which are first-order and second-order in time step $\Delta t$ respectively. We perform a rigorous numerical analysis to establish the convergence orders of both schemes, proving that the global error decreases as $\mathcal{O}(\Delta t)$ and $\mathcal{O}(\Delta t^2)$ for the first- and second-order methods, respectively. In the second-order scheme, we can safely omitted most terms arising from the Strang splitting and Taylor expansion while maintaining second-order accuracy, leading to a substantial reduction in computational complexity. For the second-order method, we achieves a time complexity of $\mathcal{O}(M^3 2^{2K_{\max}} K_{\max}^2)$ and a space complexity of $\mathcal{O}(M^2 2^{2K_{\max}} K_{\max})$ where $M$ denotes the number of system levels and $K_{\max}$ the number of time steps within the memory length. Compared with existing methods, our approach requires substantially less memory and computational effort for multilevel systems ($M\geqslant 3$). Numerical experiments are carried out to illustrate the validity and efficiency of our method.
\end{abstract}

\maketitle

\section{Introduction}
\label{sec_intro}
In recent decades, numerical methods for open quantum systems have attracted significant attention due to their broad applications in many fields such as quantum computation \cite{breuer2002theory}, quantum information \cite{nielsen2002quantum}, and chemical physics \cite{weiss2012quantum}.
Any physical system inevitably interacts with its surrounding environment, commonly referred to as the bath in quantum dynamics, and the coupling between the system and the bath modifies the system’s behavior in essential ways.
Because the bath typically involves a large number of degrees of freedom, direct simulation of the combined system is generally intractable.
Therefore, simplified physical models of the bath are often assumed so that its influence on the system can be computed analytically.
This influence transforms the system dynamics into a non-Markovian process, which greatly increases both computational and memory costs.
The intrinsic non-Markovian nature of open quantum systems is the central challenge in their numerical simulation.

System-bath dynamics are often described by master equations.
In the regime of weak system-bath coupling, the dynamics can be approximated by the Markovian Lindblad equation \cite{lindblad1976generators,davies1974markovian,davies1976markovian}, and many numerical methods have been developed within this framework \cite{ziolkowski1995ultrafast,cao2025structure}.
However, in most practical settings, the memory effect cannot be neglected.
In such cases, the system is instead governed by the Nakajima–Zwanzig equation \cite{nakajima1958quantum,zwanzig1960ensemble}, in which memory effects are represented explicitly by a memory kernel.
The transfer tensor method (TTM) \cite{cerrillo2014nonmarkovian,buser2017initial} can be viewed as a numerical discretization of the Nakajima–Zwanzig equation.
Since the memory kernel typically decays in time, the concept of a finite memory length can be introduced, enabling TTM to handle long-time simulations efficiently \cite{sun2024simulation}.

Another major class of approaches relies on path integrals, which provide a powerful framework for simulating open quantum systems.
However, path-integral–based methods often suffer from prohibitive memory requirements.
One classical method in this category is the quasi-adiabatic propagator path integral (QuAPI) \cite{makri1992improved}, where the memory length typically grows exponentially with the simulation time.
Subsequent refinements include the iterative QuAPI (i-QuAPI) \cite{makri1995numerical,makri1998quantum}, which introduces memory truncation to control memory growth, and the blip-summed method \cite{makri2014blip,makri2016blip}, which ignores paths with less important contributions.
Other developments in this direction include the time-evolving matrix product operator (TEMPO) \cite{strathearn2017efficient,strathearn2018efficient},
the differential equation based path integral (DEBPI) \cite{wang2022differential}, and the kink sum method \cite{makri2024kink}.
More recently, the small-matrix decomposition and small-matrix path integral techniques \cite{makri2020smallMatrixPath,makri2021smallMatrixDecomposition,makri2021smallMatrixPathIntegralDriven,makri2021smallMatrixPathIntegralExtended,makri2021smallMatrixModular,kundu2022small,wang2024tree} have significantly reduced memory costs.
Software packages such as \texttt{PathSum} \cite{kundu2023pathsum} further facilitate the application of path-integral–based methods.

Another family of approaches applies Monte Carlo or some numerical quadrature method to deal with the high-dimensional integrals in the Dyson series \cite{dyson1949radiation}.
In general, given a quantum system with Hamiltonian $H = H_0 + W$ with $H_0$ being the unperturbed Hamiltonian and $W$ being the perturbation, the propagator $U(t) = \e^{-\ii H t}$ satisfies
\begin{equation}
\label{eq:Duhammel}
U(t) = \e^{-\ii H_0 t} + \int_0^t \e^{-\ii H_0(t-t_1)} (-\ii W) U(t_1) \,\mathrm{d}t_1.
\end{equation}
One can substitute $U(t_1)$ on the right-hand side of this equation by the equation itself, and obtain an expression with a double integral
\begin{displaymath}
\begin{aligned}
U(t) = \e^{-\ii H_0 t} & + \int_0^t \e^{-\ii H_0(t-t_1)} (-\ii W) \e^{-\ii H_0 t_1} \,\mathrm{d}t_2 \\
& + \int_0^t \int_0^{t_2} \e^{-\ii H_0(t-t_2)} (-\ii W) \e^{-\ii H_0 (t_2-t_1)} (-\ii W) U(t_2) \,\mathrm{d}t_1 \,\mathrm{d}t_2.
\end{aligned}
\end{displaymath}
The Dyson series is obtained by applying this procedure repeatedly, resulting in an infinite series
\begin{equation}
\label{eq:Ut}
\begin{aligned}
U(t) = \e^{-\ii H_0 t} + \sum_{m=1}^{+\infty} (-\ii)^m \int_0^t \!\! \int_0^{t_m} \! \cdots\! \int_0^{t_2} \! \e^{-\ii H_0(t-t_m)} W \e^{-\ii H_0 (t_m-t_{m-1})} W \cdots W \e^{-\ii H_0 t_1} \\
\mathrm{d}t_1 \cdots \,\mathrm{d}t_{m-1} \,\mathrm{d}t_m.
\end{aligned}
\end{equation}
This representation of the propagator can help separate the system operators and bath operators, allowing the influence of the bath to be calculated explicitly.
Following Feynman's method, the reduced density matrix can then be illustrated by the sum of diagrams, enabling the application of the diagrammatic quantum Monte Carlo method (dQMC) in numerical simulations \cite{prokof1998polaron,werner2009diagrammatic}.
In this method, the memory cost is no longer a bottleneck, but it suffers from the numerical sign problem \cite{loh1990sign,cai2023numericalAnalysisInchworm}.
Recently, the inchworm method \cite{chen2017inchwormITheory,chen2017inchwormIIBenchmarks,cai2020inchworm,cai2023numericalAnalysisInchworm} successfully relieves the numerical sign problem by constructing a resummation of the Dyson series with the bold line trick \cite{prokof2007bold,prokof2008bold}.
A bold line is a partial sum of the Dyson series.
By replacing thin line diagrams with bold line diagrams, the inchworm method constructs an equivalent form of the Dyson series, which converges faster and suffers less numerical sign problem \cite{cai2023numericalAnalysisInchworm}.
Some numerical methods are employed to accelerate the computation \cite{yang2021inclusion,cai2022fast}.
By combining the thin lines in the Dyson series and bold lines in the inchworm algorithm in the same diagram, the bold-thin-bold (BTB) method utilizes the shift invariance of some quantities to accelerate the computation of the diagrammatic series \cite{cai2023bold}.

While these methods work well for small quantum systems, extra techniques may be required for larger systems.
For instance, the modular path integral (MPI) method has been successfully applied to various spin-chain models \cite{makri2018modular, kundu2019modular}, and the multisite
tensor network path integral utilizes tensor trains to reduce memory cost \cite{bose2022multisite, bose2022tensor}.
The inchworm method has also been generalized to the spin chain models \cite{wang2023real,sun2024simulation}.
Even in spin-chain models, however, the system space remains finite-dimensional. 
More recently, new methods \cite{wang2025solving,Nishimura2025quantum,zhan2025reducing} have been developed for the simulation of the Caldeira–Leggett model, where the system of interest is typically a single particle living in an infinite-dimensional Hilbert space, which further enriches the toolbox for open quantum system simulations and enables the exploration of genuinely infinite-dimensional models.

In this paper, we derive first-order and second-order discretizations of the Dyson series by closely mimicking the derivation of \eqref{eq:Ut}.
Based on these discretizations, we design a numerically exact iterative method, named the fast resummation of Dyson series (FRODS).
The FRODS avoids direct evaluation of high-dimensional integrals in the continuous-time methods and therefore significantly reduces the computational cost compared with the dQMC and inchworm method.
We rigorously establish the numerical accuracy of FRODS for both the first- and second-order schemes by deriving explicit error bounds for the discretizations.
Additionally, we introduce two techniques to further reduce the computational and memory costs for problems in specific regimes.

In the rest part of the paper, we will introduce the discretization of general Dyson series in \Cref{sec_discrete_Dyson_series}, and then apply the results to open quantum systems in \Cref{sec_first_order}.
The fast iterative method, FRODS to compute the discrete Dyson series for open quantum systems is proposed in \Cref{sec_iterative_scheme}, and it is further accelerated in \Cref{sec_memory_cost_reduction} by considering a finite memory length and a limited number of terms in the Dyson series.
\Cref{sec_proof} is a collection of mathematical proofs for the convergence rates claimed previously, and these results are also verified numerically in \Cref{sec_numerical_result}.
Finally, we draw a conclusion and discuss possible future works in \Cref{sec_conclusion_and_future_work}.

\section{Discrete Dyson series}
\label{sec_discrete_Dyson_series}
Before studying open quantum systems, we first introduce our discretization of the Dyson series \eqref{eq:Ut}.
Assume that a uniform time step $\Delta t$ is used for temporal discretization.
The most naive approximation of \eqref{eq:Ut} is to approximate the integrals in \eqref{eq:Ut} by numerical integration.
However, this approach alone does not provide a viable numerical scheme, due to the existence of the infinite series with respect to $m$.
Here, we will present a different approach which generates a discrete Dyson series with only finite terms.

\subsection{First-order discrete Dyson series}
\label{sec:first_order_discrete_dyson}
Instead of discretizing the final series \eqref{eq:Ut}, we start from the approximation of \eqref{eq:Duhammel}.
Let $U_n$ be the numerical approximation of $U(n\Delta t)$.
A first-order scheme for \eqref{eq:Duhammel} can be developed based on the rectangular rule:
\begin{equation}
\label{eq:Un_1st_orig}
U_n = \e^{-\ii H_0 n \Delta t} + \sum_{k=0}^{n-1} \e^{-\ii H_0 (n-k) \Delta t} (-\ii W \Delta t) U_k.
\end{equation}
For better consistency with the second-order scheme to be introduced later, here we adopt a slightly different quadrature rule:
\begin{equation}
\label{eq:Un_1st}
U_n = \e^{-\ii H_0 n \Delta t} + \sum_{k=0}^{n-1} \e^{-\ii H_0 (n-k-1/2) \Delta t} (-\ii W \Delta t) \e^{-\ii H_0 \Delta t/2} U_k,
\end{equation}
since
\begin{displaymath}
W \e^{-\ii H_0 \Delta t/2} - \e^{-\ii H_0 \Delta t/2} W = \frac{\ii \Delta t}{2} [\e^{-\ii H_0 \eta/2} H_0, W] \sim \mathcal{O}(\Delta t)
\end{displaymath}
for some $\eta \in (0, \Delta t)$, showing that the difference between the right-hand sides of \eqref{eq:Un_1st_orig} and \eqref{eq:Un_1st} is $\mathcal{O}(\Delta t)$.
Mimicking the derivation of the Dyson series \eqref{eq:Ut}, we replace the $U_k$ on the right-hand side of \eqref{eq:Un_1st} with the expression \eqref{eq:Un_1st} itself, yielding
\begin{displaymath}
\begin{aligned}
U_n ={} & \e^{-\ii H_0 n \Delta t} + \sum_{k=0}^{n-1} \e^{-\ii H_0 (n-k-1/2) \Delta t} (-\ii W \Delta t) \e^{-\ii H_0 (k+1/2) \Delta t} \\
&+ \sum_{k=1}^{n-1} \sum_{l=0}^{k-1} \e^{-\ii H_0 (n-k-1/2) \Delta t} (-\ii W \Delta t) \e^{-\ii H_0 (k-l) \Delta t} (-\ii W \Delta t) \e^{-\ii H_0 \Delta t / 2} U_l.
\end{aligned}
\end{displaymath}
Repeating this procedure produces the following discrete Dyson series:
\begin{equation}
\label{eq:discrete_Dyson}
\begin{aligned}
U_n = \e^{-\ii H_0 n\Delta t} + \sum_{m=1}^{n} (-\ii \Delta t)^m \Bigg( \sum_{k_m={m-1}}^{n-1} \sum_{k_{m-1}=m-2}^{k_m-1} \cdots \sum_{k_1=0}^{k_2-1}
\e^{-\ii H_0 (n-k_m-1/2) \Delta t} W \qquad \\ \e^{-\ii H_0 (k_m-k_{m-1}) \Delta t} W \cdots \e^{-\ii H_0 (k_2 - k_1) \Delta t} W \e^{-\ii H_0 (k_1 + 1/2) \Delta t}
\Bigg).
\end{aligned}
\end{equation}
It is obvious that the series is a finite series, and the number of terms is
\begin{displaymath}
1 + n + \frac{n(n-1)}{2} + \frac{n(n-1)(n-2)}{6} + \cdots + 1 = 2^n.
\end{displaymath}

We claim that the series \eqref{eq:discrete_Dyson} can be written equivalently as
\begin{equation}
\label{eq:Dyson_1st}
U_n = \sum_{j_1=0}^1 \cdots \sum_{j_n=0}^1 [\e^{-\ii H_0 \Delta t/2} (-\ii \Delta t W)^{j_n}\e^{-\ii H_0 \Delta t/2}] \cdots [\e^{-\ii H_0 \Delta t/2} (-\ii \Delta t W)^{j_1}\e^{-\ii H_0 \Delta t/2}],
\end{equation}
which also contains $2^n$ terms. This can be observed by noting that the summand in the above sum for indices $j_1, \cdots, j_n$ equals the summand in \eqref{eq:discrete_Dyson} for
\begin{equation}
\label{eq:jtok}
m = j_1 + \cdots + j_n, \qquad k_{\alpha} = \min\{k \mid j_1 + \cdots + j_k \geqslant \alpha \}, \quad \alpha = 1,\cdots,m,
\end{equation}
meaning that $m$ is the total number of ``$1$''s in $j_1, \cdots, j_n$, and $k_{\alpha}$ is the index of the $\alpha$th ``$1$'' in the list.
This form will be used in the further construction of our numerical scheme below.

\subsection{Second-order discrete Dyson series}
To obtain a second-order method, we apply the mid-point rule when discretizing \eqref{eq:Duhammel}:
\begin{equation}
\label{eq:midpoint}
U_n = \e^{-\ii H_0 n\Delta t} + \sum_{k=0}^{n-1} \e^{-\ii H_0(n-k-1/2) \Delta t} (-\ii W \Delta t) U_{k+1/2},
\end{equation}
where $U_{k+1/2}$ is the approximation of $U((k+1/2)\Delta t)$, and we choose it as
\begin{equation}
\label{eq:Uk1/2}
U_{k+1/2} = \e^{-\ii H_0 \Delta t / 2} U_k + \frac{\Delta t}{2} (-\ii W) \e^{-\ii H_0 \Delta t / 2} U_k.
\end{equation}
Inserting \eqref{eq:Uk1/2} into \eqref{eq:midpoint} yields
\begin{equation}
\label{eq:Ut_2nd}
U_n = \e^{-\ii H_0 n\Delta t} + \sum_{k=0}^{n-1} \e^{-\ii H_0(n-k-1/2) \Delta t} \left(-\ii W \Delta t - \frac{1}{2} W^2 \Delta t^2 \right) \e^{-\ii H_0 \Delta t/2} U_k,
\end{equation}
which again be used to replace the $U_k$ on the right-hand side repeatedly to achieve the following discrete Dyson series:
\begin{equation}
\label{eq:discrete_Dyson_2nd}
\begin{aligned}
U_n = \e^{-\ii H_0 n\Delta t} + \sum_{m=1}^{n} (-\ii \Delta t)^m \Bigg( \sum_{k_m=m-1}^{n-1} \sum_{k_{m-1}=m-2}^{k_1-1} \cdots \sum_{k_1=0}^{k_2-1}
\e^{-\ii H_0 (n-k_m-1/2) \Delta t} \widehat{W} \qquad \\ \e^{-\ii H_0 (k_m-k_{m-1}) \Delta t} \widehat{W} \cdots \e^{-\ii H_0 (k_2 - k_1) \Delta t} \widehat{W} \e^{-\ii H_0 (k_1 + 1/2) \Delta t}
\Bigg),
\end{aligned}
\end{equation}
where $\widehat{W} = W - \ii W^2 \Delta t / 2$.

Like \eqref{eq:Dyson_1st}, this series can also be equivalently written in the form of $n$ nested sums, which reads
\begin{equation}
\label{eq:Dyson_2nd}
U_n = \sum_{j_1=0}^2 \cdots \sum_{j_n=0}^2 \left[\e^{-\ii H_0 \Delta t/2} \frac{(-\ii \Delta t W)^{j_n}}{j_n!} \e^{-\ii H_0 \Delta t/2}\right] \cdots \left[\e^{-\ii H_0 \Delta t/2} \frac{(-\ii \Delta t W)^{j_1}}{j_1!} \e^{-\ii H_0 \Delta t/2}\right],
\end{equation}
which contains $3^n$ terms.
In fact, to preserve the second order, we just need to retain $(n+2) \cdot 2^{n-1}$ terms in the summation.
To characterize the result, we define
\begin{equation}
\label{eq:Nl}
N_{\ell}(j_1, \cdots, j_n) = \sum_{k=1}^n \delta_{j_k, \ell}, \qquad \ell = 0,1,2,
\end{equation}
which gives the number of $\ell$'s in the indices $j_1, \cdots, j_n$.
Let
\begin{equation}
\label{eq:Uhat}
\widehat{U}_n = \sum_{\substack{(j_1,\cdots,j_n) \in \{0,1,2\}^n\\ N_2(j_1,\cdots,j_n) \leqslant 1}}
\left[\e^{-\ii H_0 \Delta t/2} \frac{(-\ii \Delta t W)^{j_n}}{j_n!} \e^{-\ii H_0 \Delta t/2}\right] \cdots \left[\e^{-\ii H_0 \Delta t/2} \frac{(-\ii \Delta t W)^{j_1}}{j_1!} \e^{-\ii H_0 \Delta t/2}\right].
\end{equation}
The following theorem implies that $\widehat{U}_n$ also has second-order accuracy:
\begin{theorem}
\label{thm:2nd_dyson}
Assume that $H_0$ is an Hermitian operator and $W$ is bounded. There exists a constant $C$ depending on $T := n\Delta t$ and the norm of $W$, such that $\|U_n - \widehat{U}_n\| \leqslant C \Delta t^2$.
\end{theorem}

\subsection{An alternative method to derive the first- and second-order discrete Dyson series} \label{sec:alternative}
The discrete Dyson series written in the form \eqref{eq:discrete_Dyson} and \eqref{eq:discrete_Dyson_2nd} inspires us to consider an alternative derivation of these formulas.
By definition,
\begin{equation}
\label{eq:Unt}
U(n\Delta t) = \e^{-\ii H n \Delta t} = \underbrace{\e^{-\ii H \Delta t}\e^{-\ii H \Delta t} \cdots\e^{-\ii H \Delta t}}_{n\text{ terms}}.4\end{equation}
For each $\e^{-\ii H \Delta t}$, we apply the Strang splitting:
\begin{equation}
\label{eq:Strang}
\e^{-\ii H \Delta t} \approx \e^{-\ii H_0 \Delta t/2} \e^{-\ii W \Delta t} \e^{-\ii H_0 \Delta t/2},
\end{equation}
which is a second-order approximation.
Before plugging \eqref{eq:Strang} into \eqref{eq:Unt}, we further approximate $\e^{-\ii W \Delta t}$ using Taylor expansion:
\begin{itemize}
\item First order: $\e^{-\ii W \Delta t} \approx I - \ii W \Delta t$,
\item Second order: $\e^{-\ii W \Delta t} \approx I - \ii W \Delta t - \frac{1}{2} W^2 \Delta t^2$.
\end{itemize}
It is not difficult to find that these two approximations will lead to the discrete Dyson series \eqref{eq:discrete_Dyson} and \eqref{eq:discrete_Dyson_2nd}, respectively.

Compared to previous derivations, this approach is more straightforward.
But one advantage of the previous approach is its generalizability to higher-order methods, since the operator splitting method is restricted to second order if we stick to positive time steps.
Nevertheless, in this work, we will only study the first- and second-order schemes, and the two derivations therefore lead to identical results.

\section{Discrete Dyson series in open quantum systems}
\label{sec_first_order}
In this section, we introduce the general structure of open quantum systems and generalize the discretizations in \Cref{sec_discrete_Dyson_series} to the open quantum system settings.

\subsection{Open quantum systems}
\label{subsec_open_quantym_system}\
An open quantum system consists of two parts, the system of interest and a uninteresting environment (bath).
In this paper, we refer to the system of interest as ``system''.
The system and the bath, as a whole, is referred as the ``combined system''.
We consider open quantum systems whose the Hamiltonian can be represented by $H = H_0 + W$ with
\begin{equation}
  H_0 = H_s \otimes \id_b + \id_s \otimes H_b, \qquad W = W_s\otimes W_b.
\end{equation}
In this Hamiltonian, $H_s$ is the Hamiltonian acting barely on system states, and $H_b$ is the one for the bath.
The perturbation $W$ refers to the interaction of the system and the bath, where $W_s$ acts on the system and $W_b$ acts on the bath.
The existence of the interaction term $W$ leads to quantum dissipation and brings difficulty for numerical simulations.

In open quantum systems, the system state can be described by the reduced density matrix
\begin{equation}
\label{eq:rdm}
\rho_s(t) = \tr_b(\rho(t)) = \tr_b\left(U(t) \rho(0) U^\dagger(t)\right).
\end{equation}
Here $\tr_b$ denotes the trace operator on the bath, and $U(t) = \e^{-\ii H t}$ as introduced in \cref{sec_intro}.
Further simplification requires plugging \eqref{eq:Ut} into \eqref{eq:rdm}.
In the harmonic bath model, the bath is modeled by a large number of bosonic quantum harmonic oscillators, and $W_b$ is a linear function of their positions.
Suppose
\begin{displaymath}
\rho(0) = \rho_s(0) \otimes \frac{\e^{\beta H_b}}{\tr_b (\e^{\beta H_b})},
\end{displaymath}
meaning that the system and the bath are initially unentangled, and the initial state of the bath is in a thermal equilibrium with inverse temperature $\beta$.
Then the trace operator $\tr_b$ in \eqref{eq:rdm} can be calculated using Wick's theorem \cite{wick1950evaluation}:
\begin{equation}
\label{eq:Lb}
\begin{aligned}
& \mathcal{L}_b(-t_m, \cdots, -t_1; t_1', \cdots, t'_{m'}) \\
:= {} & \tr_b\left(W_{b,I}(-t_m) \cdots W_{b,I}(-t_1) \rho_b(0)  W_{b,I}^{\dagger}(t_1') \cdots W_{b,I}^{\dagger}(t_{m'}') \right) \\
={} & \left\{ \begin{array}{@{}ll}
  1, & \text{if } m = m' = 0, \\
  0, & \text{if } m + m' \text{ is odd}, \\
  \sum_{\mathfrak{q} \in \mathcal{Q}(-t_m,\cdots,-t_1; t_1', \cdots, t'_{m'})} \prod_{(\tau, \tau') \in \mathfrak{q}} B(\tau,\tau'), & \text{if } m + m' \text{ is even}.
\end{array} \right.
\end{aligned}
\end{equation}
Here, $W_{b,I}(\tau) = \e^{-\ii \tau H_b} W_b \e^{\ii \tau H_b}$, and ``$^\dagger$'' refers to the conjugate transpose.
The function $B(\cdot, \cdot)$ is the bath correlation function depending on the frequencies of quantum oscillators and the coupling intensity, whose precise form will be given in our numerical tests.
In the case of even $m + m'$, the sum is taken over all possible pairings of $\{-t_1, \cdots, -t_m, t_1', \cdots, t'_{m'}\}$, and the product is taken over the pairs contained in $\mathfrak{q}$.
For instance,
\begin{align*}
  \mathcal{L}_b(-t_1; t_1') &= B(-t_1, t_1'), \\
  \mathcal{L}_b(-t_1; t_1', t_2', t_3') &= B(-t_1, t_1') B(t_2', t_3') \\
  & \quad + B(-t_1, t_2') B(t_1', t_3') + B(-t_1, t_3') B(t_1', t_2'), \\
  \mathcal{L}_b(-t_4, -t_3, -t_2, -t_1; t_1', t_2') &= B(-t_4, -t_3) B(-t_2, -t_1) B(t_1', t_2') + B(-t_3, -t_1) B(-t_4, -t_2) B(t_1', t_2') \\
  & \quad + \cdots + B(-t_1, t_2') B(-t_2, t_1') B(-t_4, -t_3), \qquad \text{(15 terms in total)}
\end{align*}
and in general, the expansion of $\mathcal{L}_b$ is the sum of $(m+m'-1)!!$ products.
Using these results, the reduced density matrix can be represented as an infinite series of high-dimensional integrals:
\begin{equation} \label{eq:rho_s}
\begin{aligned}
\rho_s(t) &= \sum_{m=0}^{+\infty} \sum_{m'=0}^{+\infty}(-\ii)^m \ii^{m'} \int_0^t \int_0^{t_m} \cdots \int_0^{t_2}\int_0^t \int_0^{t'_{m'}} \cdots \int_0^{t'_2} \\
& \qquad \e^{-\ii H_s(t-t_m)} W_s \cdots W_s \e^{-\ii H_s t_1} \rho_s(0) \e^{\ii H_s t'_1} W_s \cdots W_s e^{\ii H_s(t-t'_{m'})}
\mathcal{L}_b(-t_m, \cdots, -t_1; t_1', \cdots, t'_{m'}) \\
& \hspace{.55\linewidth} \mathrm{d}t'_1 \cdots \,\mathrm{d}t'_{m'-1} \,\mathrm{d}t'_{m'}
\,\mathrm{d}t_1 \cdots \,\mathrm{d}t_{m-1} \,\mathrm{d}t_m.
\end{aligned}
\end{equation}
For more details, we refer the readers to \cite{chen2017inchwormITheory} and the references therein.

\subsection{First-order scheme}
\label{subsec_1st_order_scheme}
For the purpose of numerical simulation, we will use the discrete Dyson series introduced in \Cref{sec:first_order_discrete_dyson} to replace $U(t)$ in the reduced density matrix \eqref{eq:rdm}.
Let $\rho_{s,n}$ be the numerical approximation of $\rho_s(n\Delta t)$.
Using the first-order approximation \eqref{eq:discrete_Dyson}, we have
\begin{equation}
\label{eq:rdm_discrete}
\begin{aligned}
\rho_{s,n} = \tr_b(U_n\rho(0)U_n^\dagger) = \sum_{\substack{j_{1-},\ldots,j_{n-} \in \{0,1\} \\
j_{1+},\ldots,j_{n+} \in \{0,1\}}}  \mathcal{P}_{s,j_{n-}} \cdots \mathcal{P}_{s,j_{1-}} \rho_s(0) \mathcal{P}_{s,j_{1+}}^{\dagger} \cdots \mathcal{P}_{s,j_{n+}}^{\dagger} \\
\times \tr_b(\mathcal{P}_{b,j_{n-}} \cdots \mathcal{P}_{b,j_{1-}} \rho_b(0) \mathcal{P}_{b,j_{1+}}^{\dagger} \cdots \mathcal{P}_{b,j_{n+}}^{\dagger}),
\end{aligned}
\end{equation}
where we have used $1-, \cdots, n-$ to denote the indices used in the expansion of $U_n$, and $1+, \cdots, n+$ to denote the indices in the expansion of $U_n^{\dagger}$.
The operators $\mathcal{P}_{s,0}$, $\mathcal{P}_{s,1}$, $\mathcal{P}_{b,0}$, and $\mathcal{P}_{b,1}$ are defined by
\begin{equation}
\label{eq_two_types_of_propagators}
\begin{split}
    &\mathcal{P}_{s,0} = \e^{-\ii H_s \dt},\quad
    \mathcal{P}_{s,1} = -\ii \dt \e^{-\ii H_s \dt/2} W_s \e^{-\ii H_s \dt/2}, \\
    &\mathcal{P}_{b,0} = \e^{-\ii H_b \dt}, \quad
    \mathcal{P}_{b,1} = \e^{-\ii H_b \dt/2} W_b \e^{-\ii H_b \dt/2}.
\end{split}
\end{equation}
According to \eqref{eq:jtok} and \eqref{eq:Lb}, the trace $\tr_b\left(\mathcal{P}_{b,j_{n-}} \cdots \mathcal{P}_{b,j_{1-}} \rho_b(0) \mathcal{P}_{b,j_{1+}}^{\dagger} \cdots \mathcal{P}_{b,j_{n+}}^{\dagger}\right)$
can be equivalently written as 
\begin{equation}
\label{eq:Lb_dyson}
\mathcal{L}_b\left(-(k_{m-}^--1/2)\Delta t, \cdots, -(k_1^--1/2)\Delta t; (k_1^+-1/2)\Delta t, \cdots, (k_{m+}^+-1/2)\Delta t\right),
\end{equation}
where $m\pm$ and $k_{\ell}^{\pm}$ are defined as in \eqref{eq:jtok}, so that $m-$ denotes the number of ``1''s in $(j_{1-}, j_{2-}, \cdots, j_{n-})$, and $k_1^-, \cdots, k_{m-}^-$ are the locations of these ``1''s, and $m+, k_1^+, k_2^+, \cdots, k_{m+}^+$ describe the same property for $(j_{1+}, j_{2+}, \cdots, j_{n+})$.

For simplicity, we will use 
$\mathscr{L}_b(j_{n-}, \cdots, j_{2-}, j_{1-}; j_{1+},j_{2+},\cdots, j_{n+})$
to denote the quantity \eqref{eq:Lb_dyson}, and for any $r, r'$ in the index set $\{1-, 1+, 2-, 2+, \cdots\}$, we let
\begin{displaymath}
\mathcal{B}_{r,r'} = \left\{ \begin{array}{@{}ll}
  B\Big({-(|r|-1/2)}\Delta t, -(|r'|-1/2)\Delta t\Big), & \text{if } r,r' \in \{1-, 2-, \cdots \}, \\
  B\Big({-(|r|-1/2)}\Delta t, (|r'|-1/2)\Delta t\Big), & \text{if } r \in \{1-, 2-, \cdots \} \text{ and } r' \in \{1+, 2+, \cdots, \}, \\
  B\Big((|r|-1/2)\Delta t, (|r'|-1/2)\Delta t\Big), & \text{if } r,r' \in \{1+, 2+, \cdots \},
\end{array} \right.
\end{displaymath}
where $|\cdot|$ applied to an index removes the plus or minus sign following the integer.
Below are a few examples with these notations:
\begin{align*}
\mathscr{L}_b(1,0;1,0) &= \mathcal{L}_b\left( -\frac{3}{2} \Delta t; \frac{1}{2} \Delta t\right) = \mathcal{B}_{2-,1+}, \\
\mathscr{L}_b(1,1,1;1,0,1) &= \mathcal{L}_b\left( -\frac{5}{2} \Delta t, -\frac{3}{2} \Delta t, -\frac{1}{2} \Delta t; \frac{1}{2} \Delta t, \frac{5}{2}\Delta t\right) = 0, \\
\mathscr{L}_b(1,0,1;0,1,1) &= \mathcal{L}_b\left( -\frac{5}{2} \Delta t, -\frac{1}{2} \Delta t; \frac{3}{2} \Delta t, \frac{5}{2} \Delta t\right) = \mathcal{B}_{3-,1-} \mathcal{B}_{2+,3+} + \mathcal{B}_{1-, 2+} \mathcal{B}_{3-,3+} + \mathcal{B}_{1-, 3+} \mathcal{B}_{3-,2+}.
\end{align*}

As a result, the reduced density matrix \eqref{eq:rdm_discrete} becomes the sum of  products of $\mathcal{P}$'s and $\mathcal{B}$'s.
For instance,
\begin{align}
\label{eq:1st_order_rho1}
\rho_{s,1} &= \mathcal{P}_{s,0} \rho_s(0) \mathcal{P}_{s,0}^{\dagger} + \mathcal{P}_{s,1} \rho_s(0) \mathcal{P}_{s,1}^{\dagger} \mathcal{B}_{1-,1+}, \\
\label{eq:1st_order_rho2}
\begin{split}
\rho_{s,2} &= \mathcal{P}_{s,0}\mathcal{P}_{s,0}
    \rho_s(0) \mathcal{P}_{s,0}^\dagger\mathcal{P}_{s,0}^\dagger \\
    & \quad +\mathcal{P}_{s,0}\mathcal{P}_{s,0}
    \rho_s(0) \mathcal{P}_{s,1}^\dagger\mathcal{P}_{s,1}^\dagger \Bcal{1+}{2+}
    + \mathcal{P}_{s,0}\mathcal{P}_{s,1}
    \rho_s(0) \mathcal{P}_{s,0}^\dagger\mathcal{P}_{s,1}^\dagger \Bcal{1-}{2+} 
    \\
    & \quad +\mathcal{P}_{s,1} \mathcal{P}_{s,0}
    \rho_s(0) \mathcal{P}_{s,0}^\dagger\mathcal{P}_{s,1}^\dagger \Bcal{2-}{2+} 
    +\mathcal{P}_{s,0}\mathcal{P}_{s,1}
    \rho_s(0) \mathcal{P}_{s,1}^\dagger\mathcal{P}_{s,0}^\dagger \Bcal{1-}{1+}
    \\
    & \quad +\mathcal{P}_{s,1}\mathcal{P}_{s,0}
    \rho_s(0) \mathcal{P}_{s,1}^\dagger\mathcal{P}_{s,0}^\dagger \Bcal{2-}{1+}
    + \mathcal{P}_{s,1}\mathcal{P}_{s,1}
    \rho_s(0) \mathcal{P}_{s,0}^\dagger\mathcal{P}_{s,0}^\dagger \Bcal{2-}{1-}\\
    & \quad +\mathcal{P}_{s,1}\mathcal{P}_{s,1}
    \rho_s(0) \mathcal{P}_{s,1}^\dagger\mathcal{P}_{s,1}^\dagger
    \Bcal{1+}{2+}\Bcal{2-}{1-} \\
    & \quad +\mathcal{P}_{s,1}\mathcal{P}_{s,1}
    \rho_s(0) \mathcal{P}_{s,1}^\dagger\mathcal{P}_{s,1}^\dagger
    \Bcal{1-}{2+}\Bcal{2-}{1+}\\
    & \quad +\mathcal{P}_{s,1}\mathcal{P}_{s,1}
    \rho_s(0) \mathcal{P}_{s,1}^\dagger\mathcal{P}_{s,1}^\dagger
    \Bcal{2-}{2+}\Bcal{1-}{1+}.
\end{split}
\end{align}


Even with the simplified notations, these equations are lengthy and hard to read.
We will therefore follow \cite{chen2017inchwormITheory, cai2020inchworm, yang2021inclusion} and use diagrams to represent these equations.

The diagrammatical representation of equation \eqref{eq:1st_order_rho1} is
\def\fn{1st_order_dyson}
\begin{displaymath}
\dgm{\fn.1} = \dgm{\fn.2} + \dgm{\fn.3}. \end{displaymath}
On the right-hand side, the red cross presents the initial density matrix $\rho_s(0)$.
In the first term \,\includegraphics[scale=.5]{figures/1st_order_dyson.2}, the line segment on the left of the red cross denotes $\mathcal{P}_{s,0}$, and the one on the right denotes $\mathcal{P}_{s,0}^{\dagger}$.
The second term has two white circles representing the coupling operator $-\ii \Delta t W_s$, so that on both sides of the red cross, the line segments stand for $\mathcal{P}_{s,1}$ and $\mathcal{P}_{s,1}^{\dagger}$.
The arc is interpreted as the bath correlation function $B(\cdot,\cdot)$.
Since it connects the first interval on the right side of $\color{red} \boldsymbol{\times}$ and the first interval on the left side of $\color{red} \boldsymbol{\times}$, it corresponds to $\Bcal{1-}{1+}$.
The bold line on the left-hand side of the equation indicates the sum, corresponding to the reduced density matrix $\rho_{s,1}$.

Similarly, the equation \eqref{eq:1st_order_rho2} can be written into a diagrammatic equation as
\begin{equation}
\label{eq:rho2_diagrammatic}
\begin{aligned}
\dgm[.4]{1st_order_dyson.4} = {} &
\dgm[.4]{1st_order_dyson.5} +
\dgm[.4]{1st_order_dyson.6} +
\dgm[.4]{1st_order_dyson.7} \\
{}+{} & \dgm[.4]{1st_order_dyson.8} +
\dgm[.4]{1st_order_dyson.9} +
\dgm[.4]{1st_order_dyson.10} \\
{}+{} & \dgm[.4]{1st_order_dyson.11} +
\dgm[.4]{1st_order_dyson.12} +
\dgm[.4]{1st_order_dyson.13} \\
{}+{} & \dgm[.4]{1st_order_dyson.14}.
\end{aligned}
\end{equation}
In general, every $\rho_{s,n}$ is the sum of all possible diagrams of length $2n$ where any even number of intervals contain circles connected to each other with arcs.
Readers can also refer to the previous works on diagrammatic method for better understanding of the diagrams \cite{cai2020inchworm,cai2022fast,cai2023bold}.

We would like to highlight that the numerical order of this scheme does not directly follow the arguments in \Cref{sec:alternative}, since the coupling operator $W_b$, being a linear combination of the position operators of quantum harmonic oscillators, is unbounded.
As a result, the classical proof for the second-order accuracy of Strang splitting fails.
In this case, a more careful study based on the bounded of the bath correlation function $B(\cdot, \cdot)$ can guarantee the first-order accuracy of our scheme. The result is summarized in the following theorem:
\begin{theorem} \label{thm:1st_order}
Assume that
\begin{itemize}
\item the system Hamiltonian $H_s$ is a Hermitian operator;
\item the initial density matrix $\rho_s(0)$ and the operator $W_s$ are bounded;
\item the bath correlation function $B(\cdot, \cdot)$ is bounded, \text{i.e.} $|B(\tau_1, \tau_2)| \leqslant \bar{B}$ for all $\tau_1$ and $\tau_2$;
\item the bath correlation function $B(\cdot, \cdot)$ has bounded partial derivatives, \text{i.e.} $|\partial_{\tau_1} B(\tau_1, \tau_2)| + |\partial_{\tau_2} B(\tau_1, \tau_2)| \leqslant \bar{B}'$ for all $\tau_1$ and $\tau_2$.
\end{itemize}
There exists a constant $C$ depending on $\|\rho_s(0)\|$, $\|W_s\|$, $\bar{B}$, $\bar{B}'$, and $T = n\Delta t$ such that
\begin{equation} \label{eq:rho_error}
\|\rho_{s,n} - \rho_{s}(n\Delta t)\| \leqslant C \Delta t,
\end{equation}
where $\rho_{s,n}$ and $\rho_s(\cdot)$ are defined in \eqref{eq:rdm_discrete} and \eqref{eq:rho_s}, respectively.
\end{theorem}

The proof of the theorem will be given later in \Cref{sec:proof_1st_order}. Note that the boundedness of $B(\cdot,\cdot)$ holds for most bath configurations.
The readers can refer to \cite{yang2021inclusion} for some examples.

\subsection{Second-order scheme}
\label{subsec_2nd_order_scheme}
The second-order scheme for the density matrix can again be derived by replacing $U(t)$ in \eqref{eq:rdm} with its second-order approximation \eqref{eq:Dyson_2nd}, and applying Wick's theorem \cite{wick1950evaluation}, which yields
\begin{equation}
\label{eq:2nd_order_rho}
\rho_{s,n} = \sum_{\substack{j_{1-},\ldots, j_{n-} \in \{0,1,2\}\\
j_{1+},\ldots, j_{n+} \in \{0,1,2\}}} \mathcal{P}_{s,j_{n-}} \cdots \mathcal{P}_{s,j_{1-}} \rho_s(0) \mathcal{P}_{s,j_{1+}}^{\dagger} \cdots \mathcal{P}_{s,j_{n+}}^{\dagger}
\mathscr{L}_b(j_{n-},\ldots, j_{1-};j_{1+},\ldots, j_{n+}) ,
\end{equation}
where $\mathcal{P}_{s,0}$ and $\mathcal{P}_{s,1}$ are defined in  \eqref{eq_two_types_of_propagators}, and
\begin{equation}
\label{eq_third_type_of_propagators}
\mathcal{P}_{s,2} = -\frac{\Delta t^2}{2} \e^{-\ii H_s \Delta t / 2} W_s^2 \e^{-\ii H_s \Delta t / 2}.
\end{equation}
 The bath influence functional $\mathscr{L}_b(j_{n-},\cdots,j_{1-}; j_{1+}, \cdots j_{n+})$ again represents \eqref{eq:Lb_dyson}.
 When it is represented as a sum of products of $\mathcal{B}_{r,r'}$, an index $r \in \{1-, 1+, 2-, 2+, \cdots\}$ will appear twice in each product if $j_r = 2$.
For instance,
\begin{equation}
\label{eq:Lb_example}
\begin{aligned}
\mathscr{L}_b(1,1;0,2) &= \mathcal{L}_b\left( -\frac{3}{2} \Delta t, -\frac{1}{2} \Delta t; \frac{3}{2} \Delta t, \frac{3}{2} \Delta t\right) = \mathcal{B}_{2-,1-} \mathcal{B}_{2+,2+} + 2\mathcal{B}_{1-,2+} \mathcal{B}_{2-,2+}, \\
\mathscr{L}_b(0,2;1,0) &= \mathcal{L}_b\left( -\frac{1}{2} \Delta t, -\frac{1}{2} \Delta t; \frac{1}{2} \Delta t\right) = 0, \\
\mathscr{L}_b (0,2,1;1,0,2) &= \mathcal{L}_b\left( -\frac{3}{2} \Delta t, -\frac{3}{2} \Delta t, -\frac{1}{2} \Delta t; \frac{1}{2} \Delta t, \frac{5}{2} \Delta t, \frac{5}{2} \Delta t\right) \\
& = 2\mathcal{B}_{2-,1-} \mathcal{B}_{2-,1+} \mathcal{B}_{3+,3+} + 4\mathcal{B}_{2-,1-} \mathcal{B}_{2-,3+} \mathcal{B}_{1+,3+} + \mathcal{B}_{1-,1+} \mathcal{B}_{2-,2-} \mathcal{B}_{3+,3+} \\
& \quad + 2\mathcal{B}_{1-,1+} \mathcal{B}_{2-,3+} \mathcal{B}_{2-,3+} + 2\mathcal{B}_{1-,3+} \mathcal{B}_{2-,2-} \mathcal{B}_{1+,3+} + 4\mathcal{B}_{1-,3+} \mathcal{B}_{2-,1+} \mathcal{B}_{2-,3+}.
\end{aligned}
\end{equation}
In the last example, $\mathscr{L}_b(0,2,1;1,0,2) $ is originally the sum of 15 terms, while the same terms are combined on the right-hand side.

To represent \eqref{eq:2nd_order_rho} diagrammatically, we introduce \dgm{2nd_order_small.1} to denote $\mathcal{P}_{s,2}$, where ``$\circledcirc$'' refers to the square of the coupling operator $W_s^2$.
A ``$\circledcirc$'' may be connected to another $\circ$/$\circledcirc$ with an arc, corresponding to $\mathcal{B}_{\ell,\ell'}$ with $\ell \neq \ell'$, or connected to itself like \dgm{2nd_order_small.2}, standing for $\mathcal{B}_{\ell,\ell}$.
Thus, the diagrammatic representation for the term $\mathcal{P}_{s,0}\mathcal{P}_{s,2}\mathcal{P}_{s,1} \rho_s(0) \mathcal{P}_{s,1}^{\dagger} \mathcal{P}_{s,0}^{\dagger} \mathcal{P}_{s,2}^{\dagger} \mathscr{L}_b(0,2,1;1,0,2)$ is
\def\fn{2nd_order_dyson_example}
\begin{equation} \label{eq:diag_example}
\begin{aligned}
    &2\times \dgm[.4]{\fn.3} + 4\times \dgm[.4]{\fn.4} + \dgm[.4]{\fn.5} \\
    {}+{} & 2\times\dgm[.4]{\fn.6} + 2\times \dgm[.4]{\fn.7} + 4\times\dgm[.4]{\fn.8}.
\end{aligned}
\end{equation}

\def\fn{2nd_order_dyson_example}
We would now like to clarify how to determine of the coefficient in front of each diagram.
In general, given a bath influence functional $\mathcal{L}_b(-t_m, \cdots, -t_1; t_1', \cdots, t_{m'}')$, if $t_k' = t_{k+1}'$ and no other $t_j'$'s are equal to them, meaning that we have some $j_r =2$ in $\mathscr{L}_b$ and a  ``$\circledcirc$'' at $t_k'$ or $t_{k+1}'$ in the corresponding diagram, then in the expansion of $\mathcal{L}_b$ into bath correlation functions like
\eqref{eq_third_type_of_propagators}, 
the coefficients will be affected by coincidence of $t_k'$ and $t_{k+1}'$ in the following way:
\begin{itemize}
\item The two terms $B(\tau_1,t_k') B(t_{k+1}',\tau_2)$ and $B(\tau_1, t_{k+1}') B(t_k',\tau_2)$ with $\tau_1 < t_k' = t_{k+1}' < \tau_2$ are represented by the same diagram.
The same argument applies if $\tau_1 < \tau_2 < t_k' = t_{k+1}'$ or $t_k' = t_{k+1}' < \tau_1 < \tau_2$.
Therefore, in the diagram, if a ``$\circledcirc$'' is connected to two distinct points, a factor $2$ is introduced into the coefficient.
For instance, the diagram \dgm[.4]{\fn.2} should be multiplied by $2$ in the expansion of $\rho_{s,2}$, which can also be seen from $\mathscr{L}_b(1,1;0,2) = \mathcal{B}_{2-,1-} \mathcal{B}_{2+,2+} + 2\mathcal{B}_{1-,2+} \mathcal{B}_{2-,2+}$ as in \eqref{eq:Lb_example}.
\item If there is another pair of coincident points, e.g. $t_j = t_{j+1}$, then $B(-t_j, t_k') B(-t_{j+1}, t_{k+1}')$ and $B(-t_j, t_{k+1}') B(-t_{j+1}, t_k')$ share the same diagram.
This indicates that a pair of ``$\circledcirc$''s linked with a double arc introduces a factor $2$ into the coefficient.
One such example is the fourth diagram in \eqref{eq:diag_example}, corresponding to the term $\Bcal{1-}{1+}\Bcal{2-}{3+}\Bcal{2-}{3+}$ in the expansion of $\mathscr{L}_b(0,2,1;1,0,2)$ (see \eqref{eq:Lb_example}).
\end{itemize}
Note that a ``$\circledcirc$'' connecting to itself does not lead to any additional multiplicative factors.
For instance,
\begin{displaymath}
\mathcal{P}_{s,1} \mathcal{P}_{s,1} \rho_s(0) \mathcal{P}_{s,0} \mathcal{P}_{s,2} \mathscr{L}_b(1,1;0,2) = \dgm[.4]{\fn.1} + 2\times\dgm[.4]{\fn.2},
\end{displaymath}
where the first diagram is not doubled.



Similar to the first-order case, the second-order accuracy of \eqref{eq:2nd_order_rho} also relies on the boundedness of the bath correlation function $B(\cdot, \cdot)$:
\begin{theorem} \label{thm:2nd_order}
Assume that the system Hamiltonian $H_s$, the initial density matrix $\rho_s(0)$, the operator $W_s$, and the bath correlation function satisfy the same hypotheses as in \Cref{thm:1st_order}, and second-order derivatives of the bath correlation function $B(\cdot,\cdot)$ are bounded by $\bar{B}''$.
There exists a constant $C$ depending on $\|\rho_s(0)\|$, $\|W_s\|$, $\bar{B}$, $\bar{B}'$, $\bar{B}''$ and $T = n\Delta t$ such that
\begin{displaymath}
\|\rho_{s,n} - \rho_{s}(n\Delta t)\| \leqslant C \Delta t^2,
\end{displaymath}
where $\rho_{s,n}$ and $\rho_s(\cdot)$ are defined in \eqref{eq:2nd_order_rho} and \eqref{eq:rho_s}, respectively.
\end{theorem}
The proof of this theorem is highly similar to that of \Cref{thm:1st_order}, and we will only provide the sketch of the proof in \Cref{sec:proof_2nd_order}.

As in \eqref{eq:Uhat}, terms in \eqref{eq:2nd_order_rho} with two or more indices being $2$ can be discarded to reduce the computational cost while preserving the second-order accuracy.
Using the notation $N_{\ell}$ defined in \eqref{eq:Nl}, the simplified sum can be written as

\begin{equation}
\label{eq:rhohat}
\hat{\rho}_{s,n} = \sum_{\substack{(j_{1-},\cdots,j_{n-}, j_{1+},\cdots,j_{n+}) \in \{0,1,2\}^{2n}\\ N_2(j_{n-},\cdots,1_{n-}, j_{1+},\cdots,j_{n+}) \leqslant 1}}\mathcal{P}_{s,j_{n-}} \cdots \mathcal{P}_{s,j_{1-}} \rho_s(0) \mathcal{P}_{s,j_{1+}}^{\dagger} \cdots \mathcal{P}_{s,j_{n+}}^{\dagger} \mathscr{L}_b(j_{n-},\cdots,j_{1-}; j_{1+}, \cdots j_{n+}).
\end{equation}
The following theorem confirms its second order of accuracy:
\begin{theorem} \label{thm:2nd_dyson_oqs}
Assume that the system Hamiltonian $H_s$, the initial density matrix $\rho_s(0)$, the operator $W_s$, and the bath correlation function satisfy the same hypotheses as in \Cref{thm:1st_order}.
There exists a constant $C$ depending on $\|\rho_s(0)\|$, $\|W_s\|$, $\bar{B}$ and $T = n\Delta t$ such that
\begin{displaymath}
\|\rho_{s,n} - \hat{\rho}_{s,n}\| \leqslant C \Delta t^2,
\end{displaymath}
where $\rho_{s,n}$ and $\hat{\rho}_{s,n}$ are defined in \eqref{eq:2nd_order_rho} and \eqref{eq:rhohat}, respectively.
\end{theorem}

The proof of the above theorem, to be given in \Cref{sec:proof_2nd_dyson_oqs}, is slightly more involved than \Cref{thm:2nd_dyson}, due to the extra factor $\mathscr{L}_b$ that may grow with $n$ as $\mathcal{O}((2n-1)!!)$ in the worst case. 
The proof also uses the fact that $\mathscr{L}_b$ is zero when $j_{1-} + \cdots + j_{n-} + j_{1+} + \cdots + j_{n+}$ is odd.
Taking this into account, the summation \eqref{eq:2nd_order_rho} contains $(3^{2n}+1)/2$ nonzero terms, while \eqref{eq:rhohat} has only $(n+1) \cdot 2^{2n-1}$ nonzero terms, leading to intrinsic reduction of the computational cost.

\let\oldfn\fn
\def\fn{2nd_order_dyson}
After expanding $\mathscr{L}_b$ into bath correlation functions, the summation \eqref{eq:rhohat} can also be represented diagrammatically.
For example, when $n = 1$, the diagrammatic equation is 
\begin{equation}
\label{eq:rhohat_1}
\hat{\rho}_{s,1} =\dgm{2nd_order_small.3}+\dgm{2nd_order_small.4}+\dgm{\fn.1} + \dgm{\fn.2},
\end{equation}
where ``$\circledcirc$'' stands for the operator $-(\Delta t W_s)^2/2$, and a loop stands for $\Bcal{k}{k}$ with the same $k$.
Thus, the last two diagrams in \eqref{eq:rhohat_1} denote $\mathcal{P}_{s,2}\rho_s(0) \mathcal{P}_{s,0}^{\dagger} \mathcal{B}_{1-,1-}$ and $\mathcal{P}_{s,0}\rho_s(0) \mathcal{P}_{s,2}^{\dagger} \mathcal{B}_{1+,1+}$.
Diagrams such as \dgm[.4]{\fn.3} and \dgm[.4]{\fn.4}, corresponding to $\mathcal{P}_{s,2}\rho_s(0) \mathcal{P}_{s,2}^{\dagger} \mathcal{B}_{1-,1-} \mathcal{B}_{1+,1+}$ and $\mathcal{P}_{s,2}\rho_s(0) \mathcal{P}_{s,2}^{\dagger} \mathcal{B}_{1-,1+} \mathcal{B}_{1-,1+}$, are not included since these terms have more than one $\mathcal{P}_{s,2}$ in their expressions.

The second-order approximation $\hat{\rho}_{s,2}$ includes all diagrams in \eqref{eq:rho2_diagrammatic}, along with the following 28 diagrams:
\begin{equation} \label{eq:hat_rho_2}
\begin{gathered}
\!
\dgm[.4]{\fn.5}, \quad
\dgm[.4]{\fn.12}, \quad
\dgm[.4]{\fn.19}, \quad
\dgm[.4]{\fn.26}, \\
\!
\dgm[.4]{\fn.6}, \quad
\dgm[.4]{\fn.13}, \quad
\dgm[.4]{\fn.20}, \quad
\dgm[.4]{\fn.27}, \\
\!
\dgm[.4]{\fn.7}, \quad
\dgm[.4]{\fn.14}, \quad
\dgm[.4]{\fn.21}, \quad
\dgm[.4]{\fn.28}, \\
\!
\dgm[.4]{\fn.8}, \quad
\dgm[.4]{\fn.15}, \quad
\dgm[.4]{\fn.22}, \quad
\dgm[.4]{\fn.29}, \\
\!
\dgm[.4]{\fn.9}, \quad
\dgm[.4]{\fn.16}, \quad
\dgm[.4]{\fn.23}, \quad
\dgm[.4]{\fn.30}, \\
\!
\dgm[.4]{\fn.10}, \quad
\dgm[.4]{\fn.17}, \quad
\dgm[.4]{\fn.24}, \quad
\dgm[.4]{\fn.31}, \\
\!
\dgm[.4]{\fn.11}, \quad
\dgm[.4]{\fn.18}, \quad
\dgm[.4]{\fn.25}, \quad
\dgm[.4]{\fn.32}.
\end{gathered}
\end{equation}
Note that when adding up the diagrams, the diagrams of the last three rows need to be multiplied by 2.
For instance, the diagrams \dgm[.3]{\fn.27} (2nd row, 4th column) and \dgm[.3]{\fn.30} (5th row, 4th column) originate from the bath influence functional $\mathscr{L}_b(1,1;0,2)$.
The extra factor $2$ for \dgm[.3]{2nd_order_dyson.30} can be observed in the first example of \eqref{eq:Lb_example}.
In general, all diagrams with exactly one ``$\circledcirc$'' but no loops should be doubled when being added.


\section{Fast resummation of Dyson series (FRODS)}
\label{sec_iterative_scheme}
In this section, we will derive an iterative scheme based on the discretization \cref{eq:rdm_discrete} and then accommodate it to the second order discretization.
Since it is an efficient method to compute the discrete Dyson series \cref{eq:Ut}, we call this method ``fast resummation of Dyson series'', or FRODS in short.


\subsection{First-order method}
\label{subsec_iterative_scheme_1st_order}
The basic idea of our fast algorithm is to combine terms that have common factors.
For instance, a more efficient way to compute $\rho_{s,2}$ given in 
\eqref{eq:1st_order_rho2} is the following:
\begin{equation} \label{eq:example_rhos2}
\begin{aligned}
\rho_{s,2} &= \big[\mathcal{P}_{s,0}\mathcal{P}_{s,0} \rho_s(0) \mathcal{P}^{\dagger}_{s,0} + \mathcal{P}_{s,1}\mathcal{P}_{s,1} \rho_s(0) \mathcal{P}^{\dagger}_{s,1} \mathcal{B}_{2-,1-}\\
& \qquad + \mathcal{P}_{s,1}\mathcal{P}_{s,0} \rho_s(0) \mathcal{P}^{\dagger}_{s,1} \mathcal{B}_{2-,1+}+ \mathcal{P}_{s,0}\mathcal{P}_{s,1} \rho_s(0) \mathcal{P}^{\dagger}_{s,1} \mathcal{B}_{1-,1+}\big] \mathcal{P}_{s,0}^{\dagger} \\
& \quad + \big[ \mathcal{P}_{s,1}\mathcal{P}_{s,0} \rho_s(0) \mathcal{P}^{\dagger}_{s,0} + \mathcal{P}_{s,1}\mathcal{P}_{s,1} \rho_s(0) \mathcal{P}^{\dagger}_{s,1} \mathcal{B}_{1-,1+} \big] \mathcal{P}_{s,1}^{\dagger} \mathcal{B}_{2-,2+} \\
& \quad + \big[ \mathcal{P}_{s,0}\mathcal{P}_{s,1} \rho_s(0) \mathcal{P}^{\dagger}_{s,0} + \mathcal{P}_{s,1}\mathcal{P}_{s,1} \rho_s(0) \mathcal{P}^{\dagger}_{s,1} \mathcal{B}_{2-,1+} \big] \mathcal{P}_{s,1}^{\dagger} \mathcal{B}_{1-,2+} \\
& \quad + \big[ \mathcal{P}_{s,0}\mathcal{P}_{s,0} \rho_s(0) \mathcal{P}^{\dagger}_{s,1} + \mathcal{P}_{s,1}\mathcal{P}_{s,1} \rho_s(0) \mathcal{P}^{\dagger}_{s,1} \mathcal{B}_{2-,1-} \big] \mathcal{P}_{s,1}^{\dagger} \mathcal{B}_{1+,2+}.
\end{aligned}
\end{equation}
This regrouping of terms will save a few operator multiplications, leading to lower computational cost.
For better visualization, we will use bold diagrams to denote the sums in the brackets.
For example, the sum 
\begin{equation}
\label{eq_rhos2_1storder_1stterm}
\mathcal{P}_{s,0}\mathcal{P}_{s,0} \rho_s(0) \mathcal{P}^{\dagger}_{s,0} + \mathcal{P}_{s,1}\mathcal{P}_{s,1} \rho_s(0) \mathcal{P}^{\dagger}_{s,1} \mathcal{B}_{2-,1-} + \mathcal{P}_{s,1}\mathcal{P}_{s,0} \rho_s(0) \mathcal{P}^{\dagger}_{s,1} \mathcal{B}_{2-,1+}+ \mathcal{P}_{s,0}\mathcal{P}_{s,1} \rho_s(0) \mathcal{P}^{\dagger}_{s,1} \mathcal{B}_{1-,1+}
\end{equation}
appearing in the first term of \eqref{eq:example_rhos2} will be denoted as a bold line covering three time intervals:
\def\fn{rhos2}
\begin{displaymath}
\begin{aligned}
& \dgm{\fn.1} \\
:= \, & \dgm{\fn.2} + \dgm{\fn.3} + \dgm{\fn.5} + \dgm{\fn.4}.
\end{aligned}
\end{displaymath}
Such a definition of a bold line is consistent with our previous interpretations such as 
\eqref{eq:rho2_diagrammatic}, where thin diagrams with all possible arc connections between time intervals are summed up.
Furthermore, we can extend such definitions to bold diagrams with circles on some intervals.
For instance, the terms inside the brackets in the last three lines of \eqref{eq:example_rhos2} can be represented diagrammatically as
\begin{equation} \label{eq:diag_1circ}
\begin{aligned}
    \dgm{\fn.7} &:= \dgm{\fn.8} + \dgm{\fn.9}, \\
    \dgm{\fn.11} &:= \dgm{\fn.12} + \dgm{\fn.13}, \\
    \dgm{\fn.15} &:= \dgm{\fn.16} + \dgm{\fn.17}.
\end{aligned}
\end{equation}
For a bold diagram with circles on some intervals, it is defined as the sum of all thin diagrams with circles on the same intervals, where these circles are not connected to any other circles, but arc connections between other intervals can be arbitrary.
An interval with an unconnected circle \dgm[.5]{\fn.20} again stands for the operator $\mathcal{P}_{s,1}$ or its conjugate transpose $\mathcal{P}_{s,1}^{\dagger}$.
Thus, the equation \eqref{eq:example_rhos2} can be considered as the sum of right extensions of the four bold diagrams mentioned above:
\begin{equation} \label{eq:rho2_diagram}
\begin{aligned}
\dgm{\fn.19} = & \dgm{\fn.6} + \dgm{\fn.10} \\
+ & \dgm{\fn.14} + \dgm{\fn.18},
\end{aligned}
\end{equation}
where the extended parts, labeled in green, are the factors in \eqref{eq:example_rhos2} multiplied outside the brackets.

We will use the notation $\Lambda(j_{n_l-}, \cdots, j_{1-}; j_{1+}, \cdots, j_{n_r+})$ to denote the bold diagrams with $n_l$ ($n_r$) intervals on the left (right) side of the red cross.
Each $j_r$, $r \in \{1-,1+,2-,2+,\cdots\}$ is 0 or 1, indicating if there is a circle on this interval.
For instance, the diagram \dgm[.5]{\fn.7} corresponds to $\Lambda(1,0;0)$, and \dgm[.5]{\fn.11} corresponds to $\Lambda(0,1;0)$.
Therefore, without using diagrams, we can write the equation \eqref{eq:rho2_diagram} as
\begin{displaymath}
\begin{aligned}
& \Lambda(0,0;0,0) \\
={} & \Lambda(0,0;0) \mathcal{P}_{s,0}^{\dagger} + \Lambda(1,0;0) \mathcal{P}_{s,1}^{\dagger} \Bcal{2-}{2+} + \Lambda(0,1;0) \mathcal{P}_{s,1}^{\dagger} \Bcal{1-}{2+} + \Lambda(0,0;1) \mathcal{P}_{s,1}^{\dagger} \Bcal{1+}{2+},
\end{aligned}
\end{displaymath}
where $\Lambda(\underbrace{0, \cdots, 0}_{n \text{ zeros}}; \underbrace{0, \cdots, 0}_{n \text{ zeros}})$ is equivalent to $\rho_{s,n}$.

In general, like \eqref{eq:rho2_diagram}, any bold diagram can be decomposed into extensions of other bold diagrams that are one interval shorter.
If the rightmost interval of a bold diagram contains a circle, such as
\def\fn{1st_order_scheme}
\begin{equation} \label{eq:bold_with_cirle}
\dgm{\fn.30},
\end{equation}
when it is expanded into thin diagrams by definition, all of them have an unconnected circle on the rightmost interval, while for other intervals, all possible arc connections between intervals without circles must be considered.
Therefore, if we remove the rightmost intervals in all these thin diagrams, their sum should be a bold diagram that excludes the rightmost interval from the original bold diagram.
This indicates that to obtain a bold diagram like \eqref{eq:bold_with_cirle}, we just need to extend the bold diagram without the rightmost interval by adding a segment \dgm[.5]{rhos2.20} at the end.
Thus, the diagram in \eqref{eq:bold_with_cirle} can be computed by
\begin{displaymath}
\dgm{\fn.31}.
\end{displaymath}
The general equation corresponding to this case is
\begin{equation} \label{eq:right_ext_1}
\Lambda(j_{n_l-}, \cdots, j_{1-}; j_{1+}, \cdots, j_{n_r+}, 1) = \Lambda(j_{n_l-}, \cdots, j_{1-}; j_{1+}, \cdots, j_{n_r+}) \mathcal{P}_{s,1}^{\dagger}.
\end{equation}

If a bold diagram does not have a circle on its rightmost interval, e.g.
\begin{equation} \label{eq:no_circ_on_right}
\dgm{\fn.32},
\end{equation}
then in its expansion, the thin diagrams can be categorized into two classes: the first class ends with a bare segment \dgm[.5]{rhos2.21}, and the second ends with a segment including a circle connected to another circle.
Examples are:
\begin{itemize}
\item First class:
\begin{gather*}
\dgm[.5]{\fn.33}, \\
\dgm[.5]{\fn.34}.
\end{gather*}
\item Second class:
\begin{equation} \label{eq:2nd_class}
\begin{gathered}
\dgm[.5]{\fn.35}, \\
\dgm[.5]{\fn.36}, \\
\dgm[.5]{\fn.37}, \\
\dgm[.5]{\fn.38}.
\end{gathered}
\end{equation}
\end{itemize}
Following the same argument as the previous paragraph, one can find that the sum of all diagrams in the first class is the right extension that adds a \dgm[.5]{rhos2.21} to the bold diagram without the rightmost interval.
For the second class, we can group the diagrams in which the rightmost ``$\circ$'' is connected to the same circle.
By this rule, in \eqref{eq:2nd_class}, the first two diagrams belong to the same group, and the last two diagrams belong to another group.
In each group, if we remove the last interval and its connecting arc, the sum of all diagrams will also correspond a bold diagram, which removes the last interval of the original diagram, and adds a circle to the interval that the removed arc connects to.
This means the sum of all diagrams in each group can be written as an extension of a shorter diagram obtained in this manner.
In the example above, the first two diagrams in \eqref{eq:2nd_class} belong to the following extension:
\begin{displaymath}
\dgm{\fn.39},
\end{displaymath}
and the last two diagrams belong to
\begin{displaymath}
\dgm{\fn.40}.
\end{displaymath}
Since the circle on the last segment may connect to four previous intervals, we can summarize our discussion on both classes and obtain
\begin{displaymath}
\begin{aligned}
& \dgm{\fn.32} \\
={}& \dgm{\fn.43} \\
+{}& \dgm{\fn.42} \\
+{}& \dgm{\fn.41} \\
+{}& \dgm{\fn.40} \\
+{}& \dgm{\fn.39}.
\end{aligned}
\end{displaymath}
The general conclusion is
\begin{equation} \label{eq:right_ext_0}
\begin{aligned}
& \Lambda(j_{n_l-}, \cdots, j_{1-}; j_{1+}, \cdots, j_{n_r+}, 0) = \Lambda(j_{n_l-}, \cdots, j_{1-}; j_{1+}, \cdots, j_{n_r+}) \mathcal{P}_{s,0}^{\dagger} \\
& \quad + \sum_{\ell \in \{1-,\cdots,n_l-,1+, \cdots, n_r+\}} \delta_{0,j_{\ell}} \Lambda\left(\hat{j}_{n_l-}^{\ell}, \cdots, \hat{j}_{1-}^{\ell}; \hat{j}_{1+}^{\ell}, \cdots, \hat{j}_{n_r+}^{\ell} \right) \mathcal{P}_{s,1}^{\dagger} \Bcal{\ell}{(n_r+1)+},
\end{aligned}
\end{equation}
where for all $r \in \{1-,\cdots,n_l-,1+, \cdots, n_r+\}$, we define
\begin{equation} \label{eq:hatj}
\hat{j}_r^{\ell} = j_r + \delta_{r\ell} = \left\{ \begin{array}{@{}ll}
j_r & \text{if } r \neq \ell, \\
j_r + 1 & \text{if } r = \ell.
\end{array} \right.
\end{equation}

Similarly, these bold diagrams can also be generated via left extensions.
The formulas are similar to \eqref{eq:right_ext_1} and \eqref{eq:right_ext_0}:

\begin{equation}
\label{eq_1st_order_left}
    \begin{aligned}
& \Lambda(1, j_{n_l-}, \cdots, j_{1-}; j_{1+}, \cdots, j_{n_r+}) = \mathcal{P}_{s,1} \Lambda(j_{n_l-}, \cdots, j_{1-}; j_{1+}, \cdots, j_{n_r+}), \\
& \Lambda(0, j_{n_l-}, \cdots, j_{1-}; j_{1+}, \cdots, j_{n_r+}) = \mathcal{P}_{s,0} \Lambda(j_{n_l-}, \cdots, j_{1-}; j_{1+}, \cdots, j_{n_r+}) \\
& \quad + \sum_{\ell \in \{1-,\cdots,n_l-,1+, \cdots, n_r+\}} \delta_{0,j_{\ell}} \mathcal{P}_{s,1} \Lambda\left(\hat{j}_{n_l-}^{\ell}, \cdots, \hat{j}_{1-}^{\ell}; \hat{j}_{1+}^{\ell}, \cdots, \hat{j}_{n_r+}^{\ell} \right) \Bcal{(n_l+1)-}{\ell}.
\end{aligned}
\end{equation}

These tools allow us to compute any diagram from the initial density matrix $\Lambda(;) = \rho_{s,0}$ via step-by-step extensions.
In our algorithm, we calculate bold diagrams by applying left and right extensions alternately:
\begin{equation}
\label{eq:Lambda_path}
\Lambda(\,;\,) \rightarrow \Lambda(j_{1-}; \, ) \rightarrow \Lambda(j_{1-}; j_{1+}) \rightarrow \Lambda(j_{2-}, j_{1-}; j_{1+}) \rightarrow \Lambda(j_{2-}, j_{1-}; j_{1+}, j_{2+}) \rightarrow \cdots.
\end{equation}
For the $k$th $\Lambda$ in this sequence ($\Lambda(\,;\,)$ is considered the 0th), $2^k$ operators need to be computed.
Since $\rho_{s,1} = \Lambda(0;0)$ and $\rho_{s,2} = \Lambda(0,0;0,0)$, we will regard $\Lambda(\,;\,) \rightarrow \Lambda(j_{1-}; \, ) \rightarrow \Lambda(j_{1-}; j_{1+})$ as the first time step and $\Lambda(j_{1-}; j_{1+}) \rightarrow \Lambda(j_{2-}, j_{1-}; j_{1+}) \rightarrow \Lambda(j_{2-}, j_{1-}; j_{1+}, j_{2+})$ as the second time step.
To better illustrate the procedure, below we write down a few details of the first two time steps using diagrammatic equations:
\begin{itemize}
\item Step 1:
  \begin{itemize}
    \item Left extensions:
      \begin{equation}
      \label{extension_left_1st_order_n1}
          \begin{aligned}
      \dgm{\fn.1} &= \dgm{\fn.2}, \\
      \dgm{\fn.3} &= \dgm{\fn.4}.
      \end{aligned}
      \end{equation}
    \item Right extensions:
      \begin{equation}
      \label{extension_right_1st_order_n1}
          \begin{aligned}
      \dgm{\fn.5} &= \dgm{\fn.6} + \dgm{\fn.7}, \\
      \dgm{\fn.8} &= \dgm{\fn.9}, \\
      \dgm{\fn.10} &= \dgm{\fn.11}, \\
      \dgm{\fn.12} &= \dgm{\fn.13}.
      \end{aligned}
      \end{equation}
  \end{itemize}
\item Step 2:
  \begin{itemize}
    \item Left extensions (8 equations below):
      \begin{equation}
      \label{extension_left_1st_order_n2}
          \begin{aligned}
      \dgm[.43]{\fn.14} &= \dgm[.43]{\fn.15} + \dgm[.43]{\fn.16} + \dgm[.43]{\fn.17}, \\
      \dgm[.43]{\fn.18} &= \dgm[.43]{\fn.19} + \dgm[.43]{\fn.20}, \\
      \vdots \hspace*{35pt} & \hspace*{45pt} \vdots \hspace*{76pt} \vdots
      \end{aligned}
      \end{equation}
    \item Right extensions (16 equations below):
      \begin{equation}
      \label{extension_right_1st_order_n2}
          \begin{aligned}
      \dgm[.43]{\fn.21} &= \dgm[.43]{\fn.22} + \dgm[.43]{\fn.23} \\
      & \quad + \dgm[.43]{\fn.24} + \dgm[.43]{\fn.25}, \\[10pt]
      \dgm[.43]{\fn.26} &= \dgm[.43]{\fn.27} \\
      & \quad + \dgm[.43]{\fn.28} + \dgm[.43]{\fn.29}, \\
      \vdots \hspace*{35pt} & \hspace*{45pt} \vdots \hspace*{76pt} \vdots
      \end{aligned}
      \end{equation}
  \end{itemize}
\end{itemize}

For comprehensiveness, the pseudocode of the first-order algorithm is given in \Cref{algo:1st_order}.
The for-loops in lines 3, 6, 13, and 16 indicate that the computational complexity for the $k$th time step is $\mathcal{O}(2^{2k} k)$.
Hence, the total computational complexity for $N$ time steps is $\mathcal{O}(2^{2N} N)$.
If each operator is denoted by an $M \times M$ matrix, then the computational time can be refined as $\mathcal{O}(2^{2N} N M^3)$, with $M^3$ being the computational cost for one matrix multiplication.
Note that if $\rho_{s,N}$ is computed by directly applying \eqref{eq:rdm_discrete}, the computational cost is
\begin{displaymath}
\underbrace{2^{2N} N M^3}_{\substack{\text{multiplication of} \\\text{system operators}}} + \underbrace{N \sum_{k=0}^N \begin{pmatrix} 2n \\ 2k \end{pmatrix} (2k-1)!!}_{\text{bath influence functional}} \sim \mathcal{O} (2^{2N} N M^3 + 2^N N! \sqrt{N} \mathrm{e}^{\sqrt{2N}}).
\end{displaymath}
In most cases, our iterative method will show remarkably shorter computational time.
One possible disadvatange is the memory cost required by the iterative algorithm.
Since all values of $\Lambda(j_{k-}, \cdots, j_{1-}; j_{1+}, \cdots, j_{k+})$ must be stored for the next-step computation, the storage requirement is $O(M^2 2^{2N})$, which may become unaffordable when $N$ is large.
When the system has a fast decaying memory kernel, the memory usage can be limited by a truncation of the memory length.
This will be further discussed in the next section.

\begin{algorithm}[!ht]
\begin{algorithmic}[1]
\State $\Lambda(\,;\,) \gets \rho_{s,0}$
\For{$k=0,\cdots,N-1$}
\For{$(j_{1-}, j_{1+}, \cdots, j_{k-}, j_{k+}) \in \{0,1\}^{2k}$} \Comment{Left extension}
\State $\Lambda(1,j_{k-}, \cdots, j_{1-}; j_{1+}, \cdots, j_{k+}) \gets \mathcal{P}_{s,1}\Lambda(j_{k-}, \cdots, j_{1-}; j_{1+}, \cdots, j_{k+})$
\State $Op \gets \mathcal{P}_{s,0}\Lambda(j_{k-}, \cdots, j_{1-}; j_{1+}, \cdots, j_{k+})$
\For{$\ell = 1-, 1+, \cdots, k-, k+$ satisfying $j_{\ell} = 0$}
\State $j_{\ell} \gets 1$
\State $Op \gets Op + \mathcal{P}_{s,1} \Bcal{(k+1)-}{\ell} \Lambda(j_{k-}, \cdots, j_{1-}; j_{1+}, \cdots, j_{k+})$
\State $j_{\ell} \gets 0$
\EndFor
\State $\Lambda(0, j_{k-}, \cdots, j_{1-}; j_{1+}, \cdots, j_{k+}) \gets Op$
\EndFor
\medskip
\For{$(j_{1-}, j_{1+}, \cdots, j_{k-}, j_{k+}, j_{(k+1)-}) \in \{0,1\}^{2k+1}$} \Comment{Right extension}
\State $\Lambda(j_{(k+1)-}, \cdots, j_{1-}; j_{1+}, \cdots, j_{k+}, 1) \gets \Lambda(j_{(k+1)-}, \cdots, j_{1-}; j_{1+}, \cdots, j_{k+})\mathcal{P}_{s,1}^{\dagger}$
\State $Op \gets \Lambda(j_{(k+1)-}, \cdots, j_{1-}; j_{1+}, \cdots, j_{k+})\mathcal{P}_{s,0}^{\dagger}$
\For{$\ell = 1-, 1+, \cdots, k-, k+, (k+1)-$ satisfying $j_{\ell} = 0$}
\State $j_{\ell} \gets 1$
\State $Op \gets Op +  \Lambda(j_{(k+1)-}, j_{k-}, \cdots, j_{1-}; j_{1+}, \cdots, j_{k+})\mathcal{P}_{s,1}^{\dagger} \Bcal{\ell}{(k+1)+}$
\State $j_{\ell} \gets 0$
\EndFor
\State $\Lambda(j_{(k+1)-}, j_{k-}, \cdots, j_{1-}; j_{1+}, \cdots, j_{k+}, 0) \gets Op$
\EndFor
\EndFor
\end{algorithmic}
\caption{The first-order scheme for computing $\rho_{s,N}$}
\label{algo:1st_order}
\end{algorithm}

\subsection{Second-order scheme}
The idea to develop the iterative procedure can be extended to the second-order case.
We will again compute quantities $\Lambda(j_{n_l-}, \cdots, j_{1-}; j_{1+}, \cdots, j_{n_r+})$ presented by bold diagrams with circles, and the major difference here is that the indices can also take the value ``$2$'' as in \eqref{eq:2nd_order_rho} and \eqref{eq:rhohat}.
However, for better efficiency, our iterative scheme will not strictly recover the expression  \eqref{eq:2nd_order_rho} or \eqref{eq:rhohat}.
Below, we will demonstrate the process with extensions in the first time step starting with $\rho_{s,0}$ represented by $\color{red} \boldsymbol{\times}$.

Following the first-order scheme, we will first perform a left extension to obtain $\Lambda(j_{1-};)$, where $j_{1-}$ can take the values $0$, $1$ and $2$:
\begin{equation}
\label{eq_2nd_order_eg1_formula}
    \begin{aligned}
        \Lambda(0;) &= \mathcal{P}_{s,0}\rho_s(0)+\mathcal{P}_{s,2}\rho_s(0)\mathcal{B}_{1-,1-}, \\
        \Lambda(1;) &=\mathcal{P}_{s,1}\rho_s(0),\\
        \Lambda(2;) &=2\mathcal{P}_{s,2}\rho_s(0).
    \end{aligned}
\end{equation}
The diagrammatic representations are
\def\fn{2nd_order_original}
\begin{equation}
\label{eq_2nd_order_eg1_diagram}
     \begin{aligned}
         \dgm{\fn.1}&:=\dgm{\fn.2} + \dgm{\fn.3},\\
         \dgm{\fn.4}&:=\dgm{\fn.5},\\
         \dgm{\fn.6}&:=2\times\dgm{\fn.7}.
     \end{aligned}
\end{equation}
\def\fn{2nd_order_others}%
Compared with the equations \eqref{extension_left_1st_order_n1} in the first-order scheme, the first equation in \eqref{eq_2nd_order_eg1_diagram} contains one extra term \dgm{\fn.1}, where the ``$\circledcirc$'' is connected to itself.
This agrees with the general definitions of bold diagrams: if two circles are connected by an arc, they do not appear in the expansion of the bold diagram.
Meanwhile, \eqref{eq_2nd_order_eg1_diagram} contains one more equation for \dgm{2nd_order_original.6}, as a result of the existence of $\mathcal{P}_{s,2}$.
Note that on the right-hand side of this equation, the diagram is multiplied by a factor of $2$.
In general, once the extension adds a ``$\circledcirc$'' to the $r$th segment of the bold diagram, the added double circle is to be connected to other circles to be added in the future, the associated two-point correlation terms $\mathcal{B}_{r,r'}$ or $\mathcal{B}_{r',r}$ will appear twice in the product form of the influence functional $\mathscr{L}_b$ whenever $r \neq r'$, as discussed in \cref{subsec_2nd_order_scheme}.
To ensure consistency in the enumeration of all possible future extensions that may contract with this ``$\circledcirc$'', we manually assign the coefficient to be 2, as shown in the third equation in \eqref{eq_2nd_order_eg1_formula}.
In general, whenever a ``$\circledcirc$'' presents in the extended segment, the diagram must be multiplied by $2$, unless the ``$\circledcirc$'' connects to itself.

\def\fn{2nd_order_original}
  In the right extension followed, we have
\begin{equation}
\label{eq_2nd_order_eg2_diagram}
    \begin{aligned}
        \Lambda(0;0)&=\dgm{\fn.8} = \dgm{\fn.9}+\dgm{\fn.10}+\dgm{\fn.11}\\
        \Lambda(1;0)&=\dgm{\fn.12}=\dgm{\fn.13} + \dgm{\fn.14} + \dgm{\fn.15},\\
        \Lambda(1;1)&=\dgm{\fn.16}=\dgm{\fn.17},\\
        \Lambda(0;1) & = \dgm{\fn.18} =\dgm{\fn.19} + 2 \times \dgm{\fn.20}, \\
        \Lambda(0;2) & = \dgm{\fn.21} = 2\times \dgm{\fn.22}
    \end{aligned}
\end{equation}
\Cref{thm:2nd_dyson_oqs} states that preserving one ``$\circledcirc$'' is sufficient for second-order accuracy. Therefore, in the extensions above, if there has been a ``$\circledcirc$'' in the diagram, we do not introduce new ``$\circledcirc$'' to keep the computational cost low.
For instance, in the right extension of $\Lambda(2;)$, we do not include the terms with $\mathcal{P}_{s,2}^\dagger$ (``$\circledcirc$'' in the diagram). \def\fn{2nd_order_others}
In other words, the diagrams $\dgm{\fn.2}$, $\dgm{\fn.3}$ or $\dgm{\fn.4}$ are excluded from consideration.
However, our method does not guarantee all thin diagrams with two or more ``$\circledcirc$''s are excluded as in \eqref{eq:rhohat}.
\def\fn{2nd_order_original}%
For instance, since $\dgm{\fn.1}$ contains a second-order term $\dgm[.4]{2nd_order_others.1}$, the diagram $\dgm{\fn.8}$ defined by the first equation in \eqref{eq_2nd_order_eg2_diagram} contains the term \def\fn{2nd_order_dyson}
\dgm[.4]{\fn.3}, which is not included in \eqref{eq:rhohat}. 
The iterative scheme, as a result, yields a numerical result different from either \eqref{eq:2nd_order_rho} or \eqref{eq:rhohat}.
We shall later show that it also has second-order accuracy in time.

\def\fn{2nd_order_example}
When the extension continues to $\Lambda(j_{2-},j_{1-}; j_{1+},j_{2+}) \to \Lambda(j_{3-},j_{2-},j_{1-}; j_{1+},j_{2+})$, the bold diagram $\Lambda(0,0,0;0,0)=\dgm{\fn.1}$ contains the following diagrams
\begin{equation} \label{eq:Lambda00000}
    \begin{split}
        &\dgm[.43]{\fn.2}, \quad \dgm[.43]{\fn.4}, \quad \dgm[.43]{\fn.5}\\
        &\dgm[.43]{\fn.6}, \quad \dgm[.43]{\fn.7}, \quad \dgm[.43]{\fn.3}\\
        &\dgm[.43]{\fn.8}, \quad \dgm[.43]{\fn.9}, \quad \dgm[.43]{\fn.10}\\
        &\dgm[.43]{\fn.11}, \quad \dgm[.43]{\fn.12}, \quad \dgm[.43]{\fn.13}
    \end{split}
\end{equation}
The first 5 diagrams also appear in the first-order iterative scheme, while the remaining 7 account for the introduction of new ``$\circledcirc$'' operators.
Here we see a lot more diagrams to be added compared with the first-order case.
In general, when computing $\Lambda(0,j_{n-},\cdots,j_{1-};j_{1+},\cdots,j_{n+})$ with all $j_{1\pm}, \cdots, j_{n\pm}$ being either $0$ or $1$ by left extensions, we need to include all diagrams like the last 6 ones in \eqref{eq:Lambda00000}, which have two arcs connecting the ``$\circledcirc$'' on extended segment with circles on two other segments, corresponding to contributions from the bath influence functionals taking the form $\mathscr{L}_b\big(2, \underbrace{0,\ldots,1,\ldots,1,\ldots,0}_{\text{only two 1s at }\ell_1 \text{ and }\ell_2}\big)$ where $\ell_1 < \ell_2$ and $\ell_1, \ell_2\in\{n-,\cdots,1-,1+,\cdots,n+\}$. By definition of $\mathscr{L}_b$, we have
\begin{equation}
\label{eq_Lb_fourpoints_one_concentratecirc}
    \mathscr{L}_b\big(2, \underbrace{0,\ldots,1,\ldots,1,\ldots,0}_{\text{only two 1s at }\ell_1 \text{ and }\ell_2}\big)=2\Bcal{(n+1)-}{\ell_1}\Bcal{(n+1)-}{\ell_2}+\Bcal{(n+1)-}{(n+1)-}\Bcal{\ell_1}{\ell_2},
\end{equation}
and these diagrams only account for the first term $2\Bcal{(n_l+1)-}{\ell_1}\Bcal{(n_l+1)-}{\ell_2}$.
This analysis shows that 1) all diagrams with two arcs connecting the extended ``$\circledcirc$'' with existing segments should be multiplied by $2$, and 2) the computational cost for a single diagram may scale quadratically with respect to its length, due to the following summation in the second-order correction:
\begin{equation}
\label{eq_2nd_order_eg_4_formula}
    \sum_{\substack{{\ell_1,\ell_2\in \{1-,\cdots,n-,1+,\cdots,n+\}}\\\ell_1<\ell_2}}
    \delta_{0,j_{\ell_1}}\delta_{0,j_{\ell_2}}
    \mathcal{P}_{s,2}
    \Lambda\left(
        \hat{j}_{n-}^{\ell_1\ell_2},
        \cdots,
        \hat{j}_{1-}^{\ell_1\ell_2},
        \hat{j}_{1+}^{\ell_1\ell_2},
        \cdots,
        \hat{j}_{n+}^{\ell_1\ell_2}
    \right)
    2\Bcal{(n+1)-}{\ell_1}\Bcal{(n+1)-}{\ell_2},
\end{equation}
where
\begin{equation} \label{eq:jr_l1l2}
    \hat{j}_{r}^{\ell_1\ell_2} = j_r + \delta_{r\ell_1} + \delta_{r\ell_2}.
\end{equation}
For example, in the case of computing $\Lambda(0,0,0;0,0)$, there are $\displaystyle \binom{4}{2}=6$ such terms to be summed up (the common factor of 2 is omitted):
\begin{equation} \label{eq:Lambda00000_expansion}
    \begin{split}
&\quad\mathcal{P}_{s,2}\Lambda(1,1;0,0)\Bcal{3-}{2-}\Bcal{3-}{1-} + \mathcal{P}_{s,2}\Lambda(1,0;1,0)\Bcal{3-}{2-}\Bcal{3-}{1+}+\mathcal{P}_{s,2}\Lambda(1,0;0,1)\Bcal{3-}{2-}\Bcal{3-}{2+}\\
    & +\mathcal{P}_{s,2}\Lambda(0,1;1,0)\Bcal{3-}{1-}\Bcal{3-}{1+}+\mathcal{P}_{s,2}\Lambda(0,1;0,1)\Bcal{3-}{1-}\Bcal{3-}{2+}+\mathcal{P}_{s,2}\Lambda(0,0;1,1)\Bcal{3-}{1+}\Bcal{3-}{2+}.
    \end{split}
\end{equation}
We will now seek strategies to reduce the computational cost from $O(n^2)$ to $O(n)$ for each diagram.

The reduction of computational complexity can be achieved by further decomposing the second-order correction into two stages, each adding only one arc to the diagram.
Follow this idea, we can factorize $\mathcal{P}_{s,2}$ in \eqref{eq_third_type_of_propagators} into two operators:
\begin{equation}
    \mathcal{P}_{s,2} = \mathcal{G}_{s,2}\mathcal{G}_{s,1}, \quad \mathcal{G}_{s,1} = \frac{\ii}{2}\dt W_s \e^{-\ii\dt H_s /2},\quad \mathcal{G}_{s,2} = \ii \dt \e^{-\ii\dt H_s/2}W_s,
\end{equation}
so that $\mathcal{G}_{s,2} \Bcal{(n+1)-}{\ell_1}$ can be further factored out from \eqref{eq_2nd_order_eg_4_formula}.
Specifically, the summation \eqref{eq:Lambda00000_expansion} is rearranged as:
\begin{equation*}
    \begin{split}
        & \mathcal{G}_{s,2}\Bcal{3-}{2-}\left[\mathcal{G}_{s,1}\Lambda(1,1;0,0)\Bcal{3-}{1-}+\mathcal{G}_{s,1}\Lambda(1,0;1,0)\Bcal{3-}{1+}+\mathcal{G}_{s,1}\Lambda(1,0;0,1)\Bcal{3-}{2+}\right]\\
        {}+{} &\mathcal{G}_{s,2}\Bcal{3-}{1-}\left[\mathcal{G}_{s,1}\Lambda(1,1;0,0)\Bcal{3-}{2-}+\mathcal{G}_{s,1}\Lambda(0,1;1,0)\Bcal{3-}{1+}+\mathcal{G}_{s,1}\Lambda(0,1;0,1)\Bcal{3-}{2+}\right]\\
        {}+{} &\mathcal{G}_{s,2}\Bcal{3-}{1+}\left[\mathcal{G}_{s,1}\Lambda(1,1;0,0)\Bcal{3-}{2-}+\mathcal{G}_{s,1}\Lambda(0,1;1,0)\Bcal{3-}{1+}+\mathcal{G}_{s,1}\Lambda(0,0;1,1)\Bcal{3-}{2+}\right]\\
        {}+{} &\mathcal{G}_{s,2}\Bcal{3-}{2+}\left[\mathcal{G}_{s,1}\Lambda(1,1;0,0)\Bcal{3-}{2-}+\mathcal{G}_{s,1}\Lambda(0,1;1,0)\Bcal{3-}{1-}+\mathcal{G}_{s,1}\Lambda(0,0;1,1)\Bcal{3-}{1+}\right].
    \end{split}
\end{equation*}
\def\fn{2nd_order_example}%
Notably, all terms in the expansion of the above formula appear twice, matching the factor $2$ in \eqref{eq_2nd_order_eg_4_formula}.
For instance, $\mathcal{G}_{s,2}\Bcal{3-}{2-}\mathcal{G}_{s,1}\Lambda(1,1;0,0)\Bcal{3-}{1-}$ in the first line reappears in the second line in the form of $\mathcal{G}_{s,2}\Bcal{3-}{1-}\mathcal{G}_{s,1}\Lambda(1,1;0,0)\Bcal{3-}{2-}$.
In this way, the expressions inside the square brackets can be summed up and stored before multiplied by $\mathcal{G}_{s,2} \Bcal{3-}{\ell}$. 

This intermediate stage computing sums in brackets can be illustrated diagrammatically as:
\begin{equation*}
    \begin{aligned}
        \dgm[.4]{\fn.14}&=\dgm[.4]{\fn.15} + \dgm[.4]{\fn.16} + \dgm[.4]{\fn.17},\\
        \dgm[.4]{\fn.18}&=\dgm[.4]{\fn.19} + \dgm[.4]{\fn.20} + \dgm[.4]{\fn.21},\\
        \dgm[.4]{\fn.22} &=\dgm[.4]{\fn.23}+\dgm[.4]{\fn.24}+\dgm[.4]{\fn.25},\\
        \dgm[.4]{\fn.26} &=\dgm[.4]{\fn.27}+\dgm[.4]{\fn.28}+\dgm[.4]{\fn.29}.
    \end{aligned}
\end{equation*}
The dashed line segments at the leftmost intervals indicate that the diagrams are in an intermediate state, where only a propagator of half of a time step $\e^{-\ii H_s \Delta t/2}$ presented by the green segment and one of the two circles in ``$\circledcirc$'' have been applied, while the propagator of the other half of the time step in $\mathcal{G}_{s,2}$ and the connection to the other circle remain to be completed.
Upon completing these operations, the fully extended diagram \dgm[.4]{\fn.1} can be obtained by
\begin{equation*}
    \dgm[.4]{\fn.30} + \dgm[.4]{\fn.31} + \dgm[.4]{\fn.32}+\dgm[.4]{\fn.33}.
\end{equation*}
We again emphasize that this procedure naturally recovers the original coefficient 2 in \eqref{eq_2nd_order_eg_4_formula}. For example, the intermediate diagram $\dgm[.4]{\fn.14}$ already includes $\dgm[.4]{\fn.15}$, the final contraction $\dgm[.4]{\fn.30}$ consequently contains the term $\dgm[.4]{\fn.34}$, corresponding to  $\mathcal{G}_{s,2}\Bcal{3-}{2-}\mathcal{G}_{s,1}\Lambda(1,1;0,0)\Bcal{3-}{1-}$. Similarly, the intermediate stage $\dgm[.4]{\fn.18}$ includes $\dgm[.4]{\fn.19}$, and so the diagram $\dgm[.4]{\fn.31}$ in turn also yields $\dgm[.4]{\fn.34}$.

This procedure alone does not reduce the computational cost of a single diagram like \dgm[.4]{\fn.1}, but the benefit surfaces when computing all extended diagrams, where each half-extended diagram will be used multiple times when completing the time interval with $\mathcal{G}_{s,2}$.
For instance, the diagram \dgm[.4]{\fn.37} can be applied in the computation of both $\dgm[.4]{\fn.38}$ and \dgm[.4]{\fn.39}.
Such reusability will eventually reduce the average computational cost of each diagram to $O(n)$.
For consistency, we will also introduce the half-extended diagrams with a circle on the extended segment, e.g.,
\begin{displaymath}
\dgm{\fn.40} = \dgm{\fn.41} = \mathcal{G}_{s,1} \Lambda(1,0;1,0),
\end{displaymath}
which will be useful in the computation of diagrams with the leftmost segment being \dgm{\fn.42} or \dgm{\fn.43}.
For instance,
\begin{align*}
\dgm{\fn.44} &= \dgm{\fn.45} + \dgm{\fn.46} \\
& \quad + \dgm{\fn.47} + \dgm{\fn.48}, \\[5pt]
\dgm{\fn.49} &= \dgm{\fn.50}.
\end{align*}
Here the diagram \dgm[.4]{\fn.50} means $\mathcal{G}_{s,2}$ times \dgm[.4]{\fn.40}, and the last two diagrams in the first equation above are similarly interpreted.

To demonstrate the computational cost of this strategy, we will rewrite the diagrammatic equations into mathematical formulas.
Let $L(j_{(n+1)-}, j_{n-}, \cdots, j_{1-}; j_{1+}, \cdots, j_{n+})$ denote the half-extended diagrams, where all indices can only take values $0$ and $1$.
Then these quantities can be computed by
\begin{equation}
\label{eq_2nd_order_left_1st_half}
        \begin{aligned}
            L(0,j_{n-},\cdots,j_{1-};j_{1+},\cdots,j_{n+}) &= \!\!\!\!\!\sum_{\ell\in\{1-,\cdots,n-,1+,\cdots,n+\}}\!\!\!\!\!\delta_{0,j_\ell}\mathcal{G}_{s,1}\Lambda\left(
        \hat{j}_{n-}^{\ell},
        \cdots,
        \hat{j}_{1-}^{\ell},
        \hat{j}_{1+}^{\ell},
        \cdots,
        \hat{j}_{n+}^{\ell}
    \right)\Bcal{(n+1)-}{\ell},\\
    L(1,j_{n-},\cdots,j_{1-};j_{1+},\cdots,j_{n+})&=\mathcal{G}_{s,1}\Lambda\left(
        j_{n-},
        \cdots,
        j_{1-},
        j_{1+},
        \cdots,
        j_{n+}
    \right).
        \end{aligned}
    \end{equation}
This completes the first stage of the left extension, amounting to a time complexity $O(2^n n)$.
In the second stage, we apply $\mathcal{G}_{s,2}$ and complete the connection with the second circle.
The expression \eqref{eq_2nd_order_eg_4_formula} thus becomes:
\begin{equation}
\label{eq_2nd_order_left_2nd_half}
    \sum_{\ell'\in\{1-,\cdots,n_l-,1+,\cdots,n_r+\}}\delta_{0,j_{\ell'}}\mathcal{G}_{s,2} L(0,\hat{j}_{n_l-}^{\ell'},
        \cdots,
        \hat{j}_{1-}^{\ell'},
        \hat{j}_{1+}^{\ell'},
        \cdots,
        \hat{j}_{n_r+}^{\ell'})\Bcal{(n_l+1)-}{\ell'}
\end{equation}
The overall time complexity remains at $O(2^n n)$.
Averagely, the computational cost for each diagram reduces to $O(n)$, which effectively removes the quadratic growth as in \eqref{eq:Lambda00000_expansion}.

Below, we write down complete formulas for the second stage of the left extension.
For simplicity, we assume $L(j_{(n+1)-},j_{n-},\cdots,j_{1-};j_{1+},\cdots,j_{n+}) = 0$ when any of these indices equals $2$:
\begin{align}
\label{eq_2nd_order_left_0}
\begin{split}
    & \Lambda(0, j_{n-}, \cdots, j_{1-}; j_{1+}, \cdots, j_{n+}) = \mathcal{P}_{s,0} \Lambda(j_{n-}, \cdots, j_{1-}; j_{1+}, \cdots, j_{n+}) \\
& \qquad \qquad + \sum_{\ell \in \{1-,\cdots,n-,1+, \cdots, n+\}} \delta_{0,j_{\ell}} \mathcal{P}_{s,1} \Lambda\left(\hat{j}_{n-}^{\ell}, \cdots, \hat{j}_{1-}^{\ell}; \hat{j}_{1+}^{\ell}, \cdots, \hat{j}_{n+}^{\ell} \right) \Bcal{(n+1)-}{\ell}\\
&\qquad \qquad + \sum_{\ell'\in\{1-,\cdots,n-,1+,\cdots,n+\}}\delta_{0,j_{\ell'}}\mathcal{G}_{s,2} L(0,\hat{j}_{n-}^{\ell'},
        \cdots,
        \hat{j}_{1-}^{\ell'},
        \hat{j}_{1+}^{\ell'},
        \cdots,
        \hat{j}_{n+}^{\ell'})\Bcal{(n+1)-}{\ell'}\\
&\qquad \qquad + \mathcal{G}_{s,2} L(1,j_{n-}, \cdots, j_{1-}; j_{1+}, \cdots, j_{n+})\Bcal{(n+1)-}{(n+1)-},
\end{split} \\[5pt]
  \label{eq_2nd_order_left_1}
\begin{split}
& \Lambda(1, j_{n_l-}, \cdots, j_{1-}; j_{1+}, \cdots, j_{n+}) = \mathcal{P}_{s,1} \Lambda(j_{n-}, \cdots, j_{1-}; j_{1+}, \cdots, j_{n+})\\
&\qquad \qquad +\sum_{\ell \in \{1-,\cdots,n-,1+, \cdots, n+\}}
\delta_{0,j_\ell} \mathcal{G}_{s,2} L\left(1,\hat{j}_{n-}^{\ell}, \cdots, \hat{j}_{1-}^{\ell}; \hat{j}_{1+}^{\ell}, \cdots, \hat{j}_{n+}^{\ell}\right)\Bcal{(n+1)-}{\ell} \\
&\qquad \qquad + \mathcal{G}_{s,2} L\left(0,j_{n-}, \cdots, j_{1-}; j_{1+}, \cdots, j_{n+}\right),
\end{split} \\[5pt]
    \label{eq_2nd_order_left_2}
& \Lambda(2, j_{n-}, \cdots, j_{1-}; j_{1+}, \cdots, j_{n+})=2\mathcal{G}_{s,2} L(1,j_{n-}, \cdots, j_{1-}; j_{1+}, \cdots, j_{n+}).
\end{align}
We would like to remark that the two-stage strategy only applies when multiplying the second-order correction term $\mathcal{P}_{s,2}$.
Extensions with $\mathcal{P}_{s,0}$ or $\mathcal{P}_{s,1}$ are completed within one stage as in the first-order scheme.

In the computation of $\Lambda(1,j_{n_l-}, \cdots, j_{1-}; j_{1+}, \cdots, j_{n_r+})$ as described in \eqref{eq_2nd_order_left_1}, the contributions involving the introduction of a ``$\circledcirc$’’ (i.e., the second and third lines where one ``$\circ$'' in ``$\circledcirc$'' has already been connected and the other $\circ$ remains to be connected) naturally recover the coefficient 2 in front of $\mathcal{P}_{s,2}$, since a given diagram can appear in both terms depending on the contraction order. When the remaining ``$\circ$'' is subsequently connected in future extensions, the factor 2 is inherited, ensuring consistency with the expression in \eqref{eq_Lb_fourpoints_one_concentratecirc}.

We have illustrated the two-stage strategy in the left extension.
The same idea can be also applied to the right extension, where we extend the diagrams with $\mathcal{G}_{s,1}^\dagger$ and $\mathcal{G}_{s,2}^\dagger$ consecutively.
As an example, we will show the extensions in the first time step below:
\def\fn{2nd_order_scheme_final}
\begin{itemize}
    \item Step 1:
    \begin{itemize}
        \item Left extension:
        \begin{itemize}
            \item First stage:
                \begin{equation*}
                    \begin{aligned}
                        L(0;\,) = \dgm{\fn.1} &=0\\
                        L(1;\,) = \dgm{\fn.2} &=\dgm{\fn.3}
                    \end{aligned}
                \end{equation*}
            \item Second stage:
            \begin{equation*}
                \begin{aligned}
                    \Lambda(0;\,) = \dgm{\fn.4}&=\dgm{\fn.5}+\dgm{\fn.6}\\
                    \Lambda(1;\,) = \dgm{\fn.7}&=\dgm{\fn.8} + \dgm{\fn.9}\\
                    \Lambda(2;\,) = \dgm{\fn.10}&=2\times\dgm{\fn.11}
                \end{aligned}
            \end{equation*}
        \end{itemize}
        \item Right extension:
        \begin{itemize}
            \item First stage:
                \begin{equation*}
                    \begin{aligned}
                        R(0;0) = \dgm{\fn.12} &= \dgm{\fn.13}\\
                        R(0;1) = \dgm{\fn.14} &=\dgm{\fn.15}\\
                        R(1,0) = \dgm{\fn.16} &= 0\\
                        R(1,1) = \dgm{\fn.17} &=\dgm{\fn.18}
                    \end{aligned}
                \end{equation*}
            \item Second stage:
            \begin{equation*}
                \begin{aligned}
                    \Lambda(0;0) = \dgm{\fn.19}&=\dgm{\fn.20} + \dgm{\fn.21} + \dgm{\fn.22}  \\
                    \Lambda(1;0) = \dgm{\fn.23} &=\dgm{\fn.24} + \dgm{\fn.25} + \dgm{\fn.26}\\
                    \Lambda(2;0) = \dgm{\fn.27}&=\dgm{\fn.28}\\
                    \Lambda(0,1) = \dgm{\fn.32}&=\dgm{\fn.33}+\dgm{\fn.34}+\dgm{\fn.35}\\
                    \Lambda(1,1) = \dgm{\fn.36}&=\dgm{\fn.37}+\dgm{\fn.38}\\
                    \Lambda(2,1) = \dgm{\fn.39} &=\dgm{\fn.40}\\
                    \Lambda(0,2) = \dgm{\fn.41} &=2\times\dgm{\fn.42}\\
                    \Lambda(1,2) = \dgm{\fn.43} &=2\times\dgm{\fn.44}
                \end{aligned}
            \end{equation*}
        \end{itemize}
    \end{itemize}
\end{itemize}
Here one can observed that diagrams with half intervals can also be used multiple times in future computations.
For instance, the diagram $R(1,1) = \dgm{\fn.17}$ is used for computing $\Lambda(0,1)$ and $\Lambda(1,2)$, and $R(0,1) = \dgm{\fn.14}$ is needed for computing $\Lambda(0,0)$ and $\Lambda(0,2)$. For references, some diagrammatic equations in the second time step are also provided:
\begin{itemize}
    \item Step 2:
    \begin{itemize}
        \item Left extension:
        \begin{itemize}
            \item First-stage extension:
            \begin{equation*}
                \begin{aligned}
                    \dgm{\fn.45}&=\dgm{\fn.46} + \dgm{\fn.47}\\
                    \dgm{\fn.48}&=\dgm{\fn.49}\\
                    \dgm{\fn.50}&=\dgm{\fn.51}\\
                    \dgm{\fn.52}&=\dgm{\fn.53}\\
                    \dgm{\fn.54}&=\dgm{\fn.55}\\
                    \dgm{\fn.56}&=\dgm{\fn.57}\\
                    \dgm{\fn.58}&=\dgm{\fn.59}
                \end{aligned}
            \end{equation*}
            \item Second-stage extension (35 equations below):
            \begin{equation*}
                \begin{aligned}
                \dgm[0.4]{\fn.60}&=\dgm[0.4]{\fn.61}+\dgm[0.4]{\fn.62}+\dgm[0.4]{\fn.63}+\dgm[0.4]{\fn.64}\\
                &+\dgm[0.4]{\fn.65}+\dgm[0.4]{\fn.66}\\
                \dgm[0.4]{\fn.67}&=\dgm[0.4]{\fn.68}+\dgm[0.4]{\fn.69}+\dgm[0.4]{\fn.70}+\dgm[0.4]{\fn.71}\\
                \dgm[0.4]{\fn.72}&= \dgm[0.4]{\fn.73}+ \dgm[0.4]{\fn.74}+ \dgm[0.4]{\fn.75}+ \dgm[0.4]{\fn.76}\\
                \dgm[0.4]{\fn.77}&= \dgm[0.4]{\fn.78}+ \dgm[0.4]{\fn.79}+ \dgm[0.4]{\fn.80}+ \dgm[0.4]{\fn.81}\\
                \dgm[0.4]{\fn.82}&=\dgm[0.4]{\fn.83}+\dgm[0.4]{\fn.84}\\
                \dgm[0.4]{\fn.85}&=\dgm[0.4]{\fn.86}+\dgm[0.4]{\fn.86}\\
                \dgm[0.4]{\fn.88}&=\dgm[0.4]{\fn.89}+\dgm[0.4]{\fn.90}+\dgm[0.4]{\fn.91}+\dgm[0.4]{\fn.92}\\
                \dgm[0.4]{\fn.93}&=2\times\dgm[0.4]{\fn.94}\\
                \vdots \hspace*{20pt} & \hspace*{70pt} \vdots
                \end{aligned}
            \end{equation*}
        \end{itemize}
        \item Right extension:
        \begin{itemize}
            \item First-stage extension (15 equations below):
            \begin{equation*}
                \begin{aligned}
                    \dgm[0.4]{\fn.95}&=\dgm[0.4]{\fn.96} +\dgm[0.4]{\fn.97}+\dgm[0.4]{\fn.98}\\
                    \dgm[0.4]{\fn.99}&=\dgm[0.4]{\fn.100} + \dgm[0.4]{\fn.101}\\
                    \dgm[0.4]{\fn.102}&=\dgm[0.4]{\fn.103}+\dgm[0.4]{\fn.104}\\
                    \dgm[0.4]{\fn.105}&=\dgm[0.4]{\fn.106}+\dgm[0.4]{\fn.107}\\
                    \vdots \hspace*{35pt} & \hspace*{53pt} \vdots \hspace*{90pt} \vdots
                \end{aligned}
            \end{equation*}
            \item Second-stage extension (47 equations below):
            \begin{equation*}
                \begin{aligned}
                    &\dgm[0.4]{\fn.108} \\
                    =&\dgm[0.4]{\fn.109}
                    +\dgm[0.4]{\fn.110}
                    +\dgm[0.4]{\fn.111}
                    +\dgm[0.4]{\fn.112}\\
                    +&\dgm[0.4]{\fn.113}
                    +\dgm[0.4]{\fn.114}
                    +\dgm[0.4]{\fn.115}
                    +\dgm[0.4]{\fn.116}\\
                    & \hspace*{53pt} \vdots \hspace*{90pt} \vdots\hspace*{90pt} \vdots\hspace*{90pt} \vdots
                \end{aligned}
            \end{equation*}
        \end{itemize}
    \end{itemize}
\end{itemize}
The detailed algorithm is attached in appendix (see \cref{algo:2nd_order}).
We let $\Tilde{\rho}_{s,n}$ denote the output of the second-order iterative scheme at step $n$, i.e., $\Tilde{\rho}_{s,n} :=\Lambda(\underbrace{0,\ldots,0}_{n\text{~zeors}}; \underbrace{0,\ldots,0}_{n \text{ zeros}})$. We show in the following theorem that $\Tilde{\rho}_{s,n}$ admits a second-order accuracy:

\begin{theorem}
\label{thm:2nd_order_iterative}
   Assume the hypotheses of \cref{thm:2nd_order} and \cref{thm:2nd_dyson_oqs} hold, then there exists a constant $C$, depending only on $\|\rho_s(0)\|$, $\|W_s\|$, $\bar{B}$ and $T=n\dt$, such that
    \begin{displaymath}
        \|\Tilde{\rho}_{s,n}-\rho_s(n\dt) \|\leqslant C\dt^2.
    \end{displaymath}
\end{theorem}

Recall that $\hat{\rho}_{s,n}$ corresponds to the truncated second-order expansion \cref{eq:rhohat} which allows at most one ``$\circledcirc$'' in each summand, while $\rho_{s,n}$ denotes the full second–order expansion \cref{eq:2nd_order_rho}. The quantity $\Tilde{\rho}_{s,n}$ is constructed by including more diagrams than $\hat{\rho}_{s,n}$ but fewer than those in $\rho_{s,n}$. In this sense, it can be regarded as an intermediate approximation lying strictly between $\hat{\rho}_{s,n}$ and $\rho_{s,n}$, yielding its second-order accuracy by the squeeze theorem.
The complete proof of the theorem can be found in \Cref{sec_proof_scheme}.

Lastly, we would like to discuss the memory and computational cost of the second-order method.
We first estimate the number of diagrams in the $k$th step, \emph{i.e.}, the quantities of the form $\Lambda(j_{k-},\cdots,j_{1}-;j_{1}+,\cdots,j_{k+})$.
The number will be
\begin{equation}
    \underbrace{2^{2k}}_{\text{diagrams without $\circledcirc$}} + \underbrace{2k\cdot 2^{2k-1}}_{\text{diagrams with one $\circledcirc$}} = (k+1)2^{2k}.
\end{equation}
Therefore, the memory cost for $N$ time steps is $\mathcal{O}(M^2N2^{2N})$ in the second-order iterative scheme where $M^2$ comes from the size of the matrix.
To evaluate each of these quantities, we need to traverse all previous the time intervals, leading to a total computational cost $\mathcal{O}(M^3N^2 2^{2N})$ where the extra $N$ comes from the traversal and $M^3$ stands for the matrix multiplication.
Although these costs are already much smaller than the ones directly using \cref{eq:rhohat}, they grows exponentially with the time step $N$.
In the following section, we will further discuss the probabilities to keep memory and computational costs under control.

\section{Memory cost reduction}
\label{sec_memory_cost_reduction}
For both the first-order and second-order methods in the previous section, a common issue is the fast exponential growth of the memory cost with respect to the number of time steps, which puts a severe restriction on the simulation time, usually within 20 time steps.
Such a restriction can be alleviated in the cases of short memory length and weak system-bath coupling, which will be studied in the following subsections.
\subsection{Truncation of the bath correlation function}
\label{sec:truncation}
For most bath correlation functions $B(\tau_1, \tau_2)$, the modulus of the function value depends only on the difference $\Delta \tau := \big||\tau_1| - |\tau_2|\big|$, and $|B(\tau_1, \tau_2)|$ often decays as $\Delta \tau$ increases.
Such a decay agrees with the physical intuition that the older history has weaker impact on the presence and has also been observed in previous research works \cite[Figure 7.1]{cai2020inchworm}\cite[Figure 3]{cai2023bold}.
Therefore, a reasonable approximation of the bath correlation function is the following truncated function:
\begin{equation} \label{eq:tilde_B}
\tilde{B}(\tau_1, \tau_2) = \left\{ \begin{array}{@{}ll}
B(\tau_1, \tau_2), & \text{if } \big||\tau_1| - |\tau_2|\big| < T_{{\max}}, \\
0, & \text{otherwise}.
\end{array} \right.
\end{equation}
For simplicity, we will omit the ``\,$\tilde{~}$\,'' and still use $B$ to represent the truncated function.
As a result, in the discrete Dyson series, $\Bcal{\ell}{\ell'}$ becomes zero when $\big||\ell'| - |\ell|\big| > K_{{\max}} := \lfloor T_{{\max}} / \Delta t \rfloor$, where $\ell$ and $\ell'$ are indices in the index set $\{1\pm, 2\pm, \cdots \}$, and $|\ell|$ removes the plus or minus sign that follows the numeral. 

This assumption allows us to remove many diagrams from the summation.
For instance, if we use the first-order method to compute $\rho_{s,5}$, in the case $K_{{\max}} = 2$, the last step is
\begin{displaymath}
\begin{aligned}
\rho_{s,5} &= \Lambda(0,0,0,0,0; 0,0,0,0,0) \\
&= \Lambda(0,0,0,0,0; 0,0,0,0) \mathcal{P}_{s,0}^{\dagger} +\Lambda(1,0,0,0,0; 0,0,0,0) \mathcal{P}_{s,1}^{\dagger} \Bcal{5-}{5+} \\
& \quad + \Lambda(0,1,0,0,0; 0,0,0,0) \mathcal{P}_{s,0}^{\dagger} \Bcal{4-}{5+} +\Lambda(0,0,1,0,0; 0,0,0,0) \mathcal{P}_{s,1}^{\dagger} \Bcal{3-}{5+} \\
& \quad + \Lambda(0,0,0,1,0; 0,0,0,0) \mathcal{P}_{s,0}^{\dagger} \Bcal{2-}{5+} +\Lambda(0,0,0,0,1; 0,0,0,0) \mathcal{P}_{s,1}^{\dagger} \Bcal{1-}{5+} \\
& \quad + \Lambda(0,0,0,0,0; 1,0,0,0) \mathcal{P}_{s,0}^{\dagger} \Bcal{1+}{5+} +\Lambda(0,0,0,0,0; 0,1,0,0) \mathcal{P}_{s,1}^{\dagger} \Bcal{2+}{5+} \\
& \quad + \Lambda(0,0,0,0,0; 0,0,1,0) \mathcal{P}_{s,0}^{\dagger} \Bcal{3+}{5+} +\Lambda(0,0,0,0,0; 0,0,0,1) \mathcal{P}_{s,1}^{\dagger} \Bcal{4+}{5+} \\
&= \Lambda(0,0,0,0,0; 0,0,0,0) \mathcal{P}_{s,0}^{\dagger} +\Lambda(1,0,0,0,0; 0,0,0,0) \mathcal{P}_{s,1}^{\dagger} \Bcal{5-}{5+} \\
& \quad + \Lambda(0,1,0,0,0; 0,0,0,0) \mathcal{P}_{s,0}^{\dagger} \Bcal{4-}{5+} +\Lambda(0,0,1,0,0; 0,0,0,0) \mathcal{P}_{s,1}^{\dagger} \Bcal{3-}{5+} \\
& \quad + \Lambda(0,0,0,0,0; 0,0,1,0) \mathcal{P}_{s,0}^{\dagger} \Bcal{3+}{5+} +\Lambda(0,0,0,0,0; 0,0,0,1) \mathcal{P}_{s,1}^{\dagger} \Bcal{4+}{5+},
\end{aligned}
\end{displaymath}
whose diagrammatic representation is
\def\fn{cost_reduction}
\begin{equation} \label{eq:rho5}
\begin{aligned}
& \dgm{\fn.1} \\
={} & \dgm{\fn.2} \\
& + \dgm{\fn.3} \\
& + \dgm{\fn.4} \\
& + \dgm{\fn.5} \\
& + \dgm{\fn.6} \\
& + \dgm{\fn.7}.
\end{aligned}
\end{equation}
Assume that the red cross is the origin and the length of each segment is $\Delta t$.
The equation \eqref{eq:rho5} shows that diagrams with circles in the interval $[-2\Delta t, 2\Delta t]$, or $\Lambda(j_{5-}, \cdots, j_{1-}; j_{1+}, \cdots, j_{4+})$ with $j_{1\pm}$ or $j_{2\pm}$ equal to 1, are not required when computing $\rho_{s,5}$.
Following the same idea, during the computation of any $\Lambda(j_{n_l-}, \cdots, j_{1-}; j_{1+}, \cdots, j_{n_r+})$ with $n_l > K_{{\max}} + 1$ and $n_l - n_r = 0$ or $1$,
if
\begin{displaymath}
j_{1-} = j_{1+} = j_{2-} = j_{2+} = \cdots = j_{(n_l-K_{{\max}}-1)-} = j_{(n_l-K_{{\max}}-1)+} = 0,
\end{displaymath}
meaning that the corresponding diagram does not contain any circles in $[-(n_l-K_{{\max}}-1)\Delta t, (n_l-K_{{\max}}-1)\Delta t]$,
then the computation will not involve any diagrams with circles in the same interval $[-(n_l-K_{{\max}}-1)\Delta t, (n_l-K_{{\max}}-1)\Delta t]$.
Hence, all diagrams with circles in $[-k\Delta t, k \Delta t]$ become useless after the $(k+K_{{\max}})$th time step.

By this principle, all the $2^{2K_{{\max}}}$ diagrams $\Lambda(j_{K_{{\max}}-}, \cdots, j_{1-}; j_{1+}, \cdots, j_{K_{{\max}}+})$ still need to be computed, but in the $(K_{{\max}}+1)$th step, the quantities
\begin{equation} \label{eq:useless}
\begin{gathered}
\Lambda(j_{(K_{{\max}}+1)-}, \cdots, j_{2-}, 1; 0, j_{2+}, \cdots, j_{K_{{\max}}+}), \\
\Lambda(j_{(K_{{\max}}+1)-}, \cdots, j_{2-}, 0; 1, j_{2+}, \cdots, j_{K_{{\max}}+}), \\
\Lambda(j_{(K_{{\max}}+1)-}, \cdots, j_{2-}, 1; 1, j_{2+}, \cdots, j_{K_{{\max}}+}),
\end{gathered}
\end{equation}
if computed, will become useless in future time steps.
Therefore, it is unnecessary to calculate the diagrams \eqref{eq:useless} in the $(K_{{\max}}+1)$th time step, and we only need to store the results of $\Lambda(j_{(K_{{\max}}+1)-}, \cdots, j_{2-}, 0; 0, j_{2+}, \cdots, j_{(K_{{\max}}+1)+})$, which amounts to $M^2 2^{2K_{{\max}}}$ complex numbers.
Moving on, we just need to compute $\Lambda(j_{(K_{{\max}}+2)-}, \cdots, j_{3-}, 0, 0; 0, 0, j_{2+}, \cdots, j_{(K_{{\max}}+2)+})$ in the $(K_{{\max}}+2)$th time step, and the memory cost stays the same.
In general, in the $k$th time step with $k > K_{{\max}}$, all quantities we need to compute have the form
\begin{equation} \label{eq:Lambda0}
\Lambda(j_{k-}, \cdots, j_{(k-K_{{\max}}+1)-}, \underbrace{0, \cdots, 0}_{(k-K_{{\max}}) \text{ zeros}}; \underbrace{0, \cdots, 0,}_{(k-K_{{\max}}) \text{ zeros}} j_{(k-K_{{\max}}+1)+}, \cdots, j_{k+}),
\end{equation}
so that the memory cost no longer grows.
Similarly, in the second-order scheme, the same truncation requires us to compute \eqref{eq:Lambda0} after the $K_{{\max}}$th time step, but one of the indices $j_{(k-K_{{\max}}+1)\pm}, \cdots, j_{k\pm}$ can take the value $2$.
Thus, we need to store $M^2 (K_{{\max}}+1) 2^{2K_{{\max}}}$ complex numbers, which is again independent of $k$.

The truncation of the bath correlation function also leads to a reduction of the computational cost.
For instance, in the first-order scheme like \eqref{eq:rho5}, when $k > K_{{\max}}$, the computations of diagrams by extensions become
\begin{itemize}
\item Left extension:
\begin{align*}
& \Lambda(1, j_{k-}, \cdots, j_{1-}; j_{1+}, \cdots, j_{k+}) = \mathcal{P}_{s,0}\Lambda(j_{k-}, \cdots, j_{1-}; j_{1+}, \cdots, j_{k+}), \\
& \Lambda(0, j_{k-}, \cdots, j_{1-}; j_{1+}, \cdots, j_{k+}) = \mathcal{P}_{s,1}\Lambda(j_{k-}, \cdots, j_{1-}; j_{1+}, \cdots, j_{k+}) \\
& \quad + \sum_{\ell \in \{(k-K_{{\max}}+1)\pm, \cdots, k\pm\}} \delta_{0,j_{\ell}} \mathcal{P}_{s,1} \mathcal{B}_{(k+1)-, \ell} \Lambda\left( \hat{j}_{k-}^{\ell}, \cdots, \hat{j}_{1-}^{\ell}; \hat{j}_{1+}^{\ell}, \cdots, \hat{j}_{k+}^{\ell} \right).
\end{align*}
\item Right extension:
\begin{align*}
& \Lambda(j_{(k+1)-}, j_{k-}, \cdots, j_{1-}; j_{1+}, \cdots, j_{k+}, 1) = \Lambda(j_{(k+1)-}, j_{k-}, \cdots, j_{1-}; j_{1+}, \cdots, j_{k+})\mathcal{P}_{s,1}^{\dagger}, \\
& \Lambda(j_{(k+1)-}, j_{k-}, \cdots, j_{1-}; j_{1+}, \cdots, j_{k+}, 0) = \Lambda(j_{(k+1)-}, j_{k-}, \cdots, j_{1-}; j_{1+}, \cdots, j_{k+}) \mathcal{P}_{s,0}^{\dagger} \\
& \quad + \sum_{\ell \in \{(k-K_{{\max}}+1)\pm, \cdots, k\pm, (k+1)-\}} \delta_{0,j_{\ell}} \mathcal{B}_{\ell, (k+1)+} \Lambda\left( \hat{j}_{(k+1)-}^{\ell}, \hat{j}_{k-}^{\ell}, \cdots, \hat{j}_{1-}^{\ell}; \hat{j}_{1+}^{\ell}, \cdots, \hat{j}_{k+}^{\ell} \right) \mathcal{P}_{s,1}^{\dagger}.
\end{align*}
\end{itemize}
We refer the readers to \eqref{eq:hatj} for the definition of $\hat{j}_r^{\ell}$.
The computational cost of the $k$th time step now becomes $O(M^3 2^{2K_{{\max}}} K_{{\max}})$, which no longer grows with $k$.
For the second-order scheme, we again just need to change the range of $\ell$ and $\ell'$ in  \eqref{eq_2nd_order_left_1st_half}\eqref{eq_2nd_order_left_0}\eqref{eq_2nd_order_left_1} (and the corresponding equations in right extensions) to 
\begin{displaymath}
\{n_l-, \cdots, (n_r-K_{{\max}}+1)-; (n_r-K_{{\max}}+1)+, \cdots, n_r+\},
\end{displaymath}
where we assume $0 \leqslant n_l-n_r \leqslant 1$ and $n_r > K_{{\max}}$.

\begin{remark}
The truncation of the bath correlation function is inspired by the i-QuAPI method \cite{makri1995numerical,makri1995tensorITheory,makri1995tensorIINumerical,makri1998quantum}, which assumes a finite memory length $K_{{\max}}$ to avoid unlimited growth of the memory cost.
In the i-QuAPI method, when the difference of two time points is larger than $K_{{\max}}\dt$, their correlation is small enough to be neglected.
Despite their similar ideas, the time and space complexities are not alike, as shown in the following table:
\begin{center}
\medskip
\begin{tabular}{c|cc}
& i-QuAPI \cite{kundu2023pathsum} & 2nd-order FRODS \\ \hline
\rule{0pt}{10pt}Time complexity for each time step & $O(M^{2K_{{\max}}+2})$ & $O(M^3 2^{2K_{{\max}}} K_{{\max}}^2)$ \\[2pt]
Space complexity for each time step & $O(M^{2K_{{\max}}})$ & $O(M^2 2^{2K_{{\max}}} K_{{\max}})$
\end{tabular}
\medskip
\end{center}
When $M=2$, the i-QuAPI method outperforms the FRODS in both time and space complexities. 
However, when $M \geqslant 3$, the FRODS shows its advantages as the number of system levels does not affect the base of the exponent in both complexities.
For a three-level system with memory length $K_{{\max}}=15$, the time complexity of i-QuAPI is $M^{2 K_{{\max}}+2} \approx 1.8\times 10^{15} $ while for FRODS, this number is $M^3 2^{K_{{\max}}} K_{{\max}}^2 \approx 6.5 \times 10^{12}$.
For space complexity, we have $M^{2K_{\max}}\approx 2\times 10^{14}$ while $M^2 2^{2K_{{\max}}} K_{\max} \approx 1.4 \times 10^{11}$.
For systems with more levels, we can expect the gap between the two methods to widen even further, since the exponent in i-QuAPI scales directly with the system levels $M$, 
while in FRODS the dependence on $M$ is only polynomial. 
These estimations demonstrate the superiority of the FRODS method, especially in multi-level systems.

However, for fixed $K_{\max}$ and $\Delta t$, the i-QuAPI method is usually more accurate, despite the second-order accuracy of both schemes.
The reason is that the i-QuAPI method does not make the approximation $\e^{-\ii W \Delta t} \approx I - \ii W \Delta t - \frac{1}{2} W^2 \Delta t^2$ as we did in \Cref{sec:alternative}.
\end{remark}

\subsection{Restriction on the number of coupling operators}
One classical method for computing the reduced density matrix is to plug the continuous form of the Dyson series \eqref{eq:Ut} into the definition of $\rho_s(t)$ \eqref{eq:rdm}, and then apply the Monte Carlo method to compute the sum of integrals.
This approach, as well as some of its variations like the inchworm Monte Carlo method \cite{chen2017inchwormITheory}, usually requires a truncation of the infinite series in \eqref{eq:Ut}, and one can preserve less terms for weaker coupling between the system and the bath \cite{chen2017inchwormIIBenchmarks}, since the $m$th term in the infinite series \eqref{eq:Ut} has the order of magnitude $O(\|W\|^m)$.
The idea of reducing the computational cost by applying a limit to the number of coupling operators in each term has also been applied in the DEBPI method \cite{wang2022differential} and the kink sum SMatPI method \cite{makri2024kink}.
In this section, we are going to adopt the same strategy to further accelerate our algorithm.

Instead of truncating the discrete Dyson series, we will simply remove some bold diagrams during our computation.
Precisely speaking, for a chosen positive integer $D_{{\max}}$, all $\Lambda(j_{n_l-}, \cdots, j_{1-}; j_{1+}, \cdots, j_{n_r+})$ and $L(j_{n_l-}, \cdots, j_{1-}; j_{1+}, \cdots, j_{n_r+})$ with $j_{n_l-} + \cdots + j_{1-} + j_{1+} + \cdots j_{n_r+} > D_{{\max}}$ are regarded as zero, where the sum of all $j$'s denotes the number of circles in the bold diagram.
By definition, when we expand the bold diagram with $d$ circles into thin diagrams, each of them will include at least $d$ circles (see e.g. \eqref{eq:diag_1circ} and \eqref{eq_2nd_order_eg1_diagram}), indicating that all these diagrams contain at least $d$ operators $W_s$ inside $\mathcal{P}_{s,1}$ or $\mathcal{P}_{s,2}$ when written as products like the terms in \eqref{eq:1st_order_rho2}.
Hence, if we are only interested in terms with less than or equal to $D_{{\max}}$ coupling operators in the discrete Dyson series, it is unnecessary to compute any bold diagrams with more than $D_{{\max}}$ circles.

For instance, in the first-order method, when $D_{{\max}} = 3$, during the left extension in the 3rd time step, the following diagrams will not be computed:
\def\fn{weak_coupling}
\begin{gather*}
\dgm{\fn.1}, \quad
\dgm{\fn.2}, \quad
\dgm{\fn.3}, \\
\dgm{\fn.4}, \quad
\dgm{\fn.5}, \quad
\dgm{\fn.6},
\end{gather*}
since the numbers of circles in these diagrams exceed $D_{{\max}}$.
For diagrams with exactly $D_{{\max}}$ circles, if no circles lie on the leftmost segment, the computation can be simplified, e.g.:
\begin{equation} \label{eq:example_weak_coupling}
\begin{aligned}
\dgm{\fn.7} & = \dgm{\fn.8} + \dgm{\fn.9} \\
& = \dgm{\fn.8}.
\end{aligned}
\end{equation}
In general, the following rules apply to both first- and second-order schemes:
\begin{itemize}
\item All bold diagrams with more than $D_{{\max}}$ circles are not computed.
\item If the diagram to be computed has exactly $D_{{\max}}$ circles, any extension with arcs, such as the diagram \dgm[.43]{\fn.9} in \eqref{eq:example_weak_coupling}, can be dropped.
\item The computation of other diagrams are not affected.
\end{itemize}
In the second case of $D_{{\max}}$ circles, below are some examples with $D_{{\max}} = 3$ for the second-order scheme, showing that no arcs are used in the extensions:
\begin{align*}
\dgm{\fn.10} &= \dgm{\fn.11}, \\
\dgm{\fn.12} &= \dgm{\fn.13}, \\
\dgm{\fn.14} &= \dgm{\fn.15}, \\
\dgm{\fn.16} &= \dgm{\fn.17} + \dgm{\fn.18}, \\
\dgm{\fn.19} &= \dgm{\fn.20}.
\end{align*}

\begin{figure}[!ht]
\centering
\subfloat[$N_{\mathrm{diag}}(D_{{\max}}, K_{{\max}})$]{
\includegraphics[height=.29\textwidth]{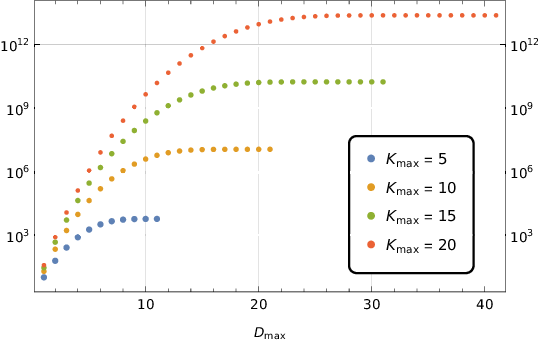}
\label{fig:Ndiag}
} \quad
\subfloat[$\dfrac{N_{\mathrm{diag}}(D_{{\max}}, K_{{\max}})}{2^{2K_{{\max}}}(1+K_{{\max}})}$]{
\includegraphics[height=.29\textwidth]{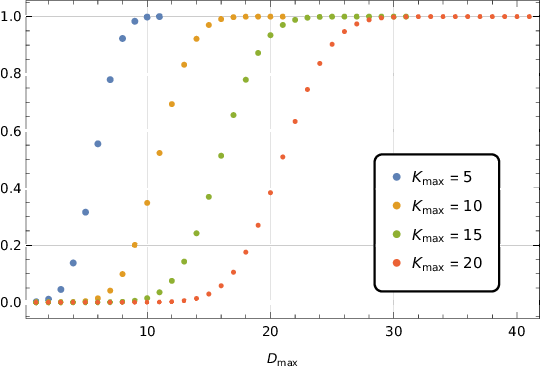}
\label{fig:ratio}
}
\caption{Numbers of diagrams for various $D_{{\max}}$ and $K_{{\max}}$.}
\label{fig:ndiag}
\end{figure}

For the second-order scheme, the value of $D_{{\max}}$ can be chosen as any number between $1$ and $2K_{{\max}}+1$.
Given a fixed $D_{{\max}}$, the number of bold diagrams now reduces to
\begin{displaymath}
N_{\mathrm{diag}}(D_{{\max}}, K_{{\max}}) = \underbrace{\sum_{k=0}^{\min(2K_{{\max}}, D_{{\max}})} \begin{pmatrix} 2K_{{\max}} \\ k \end{pmatrix}}_{\text{diagrams without ``$\circledcirc$''}} +
\underbrace{\sum_{k=2}^{D_{{\max}}} 2K_{{\max}} \begin{pmatrix} 2K_{{\max}}-1 \\ k-2 \end{pmatrix}}_{\text{diagrams with one ``$\circledcirc$''}}.
\end{displaymath}
\Cref{fig:Ndiag} shows how this number grows with $D_{\max}$ and $K_{\max}$.
To demonstrate how much memory can be saved by introducing an upper bound for the number of circles, we plot the ratio $N_{\mathrm{diag}}(D_{{\max}}, K_{{\max}}) / N_{\mathrm{diag}}(2K_{{\max}}+1, K_{{\max}})$ in Figure \ref{fig:ratio}, where $N_{\mathrm{diag}}(2K_{{\max}}+1, K_{{\max}}) = 2^{2K_{{\max}}} (1+K_{{\max}})$ is the number of bold diagrams without bounds for circles.
It shows that for a fixed $D_{{\max}}$, more savings can be achieved for larger $K_{{\max}}$.
Later in our numerical tests, we will see that this technique can save a significant amount of computational and memory cost without sacrificing much accuracy.

\section{Proofs of convergence rates}
\label{sec_proof}
In previous sections, the numerical accuracies of various formulae has been stated in several theorems.
Here we will present proofs of these theorems.

\subsection{Proof of \Cref{thm:2nd_dyson}}
\begin{proof}
    Since $H_0$ is a Hermitian operator and $W$ is bounded, each term in $U_n$ defined by \eqref{eq:Dyson_2nd} has a norm bounded by
    \begin{equation}
        \frac{(\|W\| \Delta t)^{N_1+2N_2}}{2^{N_2}}
    \end{equation}
    where $N_1$ and $N_2$ are defined by \cref{eq:Nl}, representing the numbers of ``1''s and ``2''s in the indices $j_1, \cdots, j_n$.
    Recall that the difference between $U_n$ and $\widehat{U}_n$ is all the terms with $N_2 \geqslant 2$.
    Therefore,
    \begin{equation} \label{eq:Udiff}
    \begin{aligned}
        & \Vert U_n - \widehat{U}_n \Vert\\
        \leqslant{} & \sum_{N_1=0}^{n-2} \sum_{N_2=2}^{n-N_1} \binom{n}{N_2} \binom{n-N_2}{N_1} \frac{(\|W\|\Delta t)^{N_1+2N_2}}{2^{N_2}} \\
        ={}& (1+ \|W\|\Delta t)^{n-1}
        \left(
                \frac{(1+\|W\|\dt + \frac{1}{2}\|W\|^2 \dt^2)^n}{(1+\|W\|\dt)^{n-1}}
            -\frac{n}{2} \|W\|^2 \dt^2 - \|W\| \dt-1
        \right) \\
        ={} & \frac{\|W\|^4 T^2 \e^{\|W\|T} }{8} \dt^2 + {O}(\dt^3)
    \end{aligned}
    \end{equation}
    where we fixed the simulation time $T=n\dt$.
    In the inequality, the two binomial coefficients come from different combinations of choosing $N_1$ ``1''s and $N_2$ ``2''s in the indices $j_1,\cdots,j_n$.
\end{proof}

\subsection{Proof of \Cref{thm:1st_order}}
\label{sec:proof_1st_order}
Due to the non-Markovian dynamics of open quantum systems, the proof of the numerical order becomes significantly more involved.
To facilitate the proof, below we will introduce some notations that are used only in this section and \Cref{sec:proof_2nd_order}.

We define
\begin{displaymath}
  G_s(\tau, \tau') = \left\{ \begin{array}{@{}ll}
    \e^{\ii H_s (\tau - \tau')} & \text{if } \tau < \tau' < 0, \\
    \e^{\ii H_s (\tau' - \tau)} & \text{if } 0 < \tau < \tau', \\
    \e^{\ii H_s \tau} \rho_s(0) \e^{\ii H_s \tau'} & \text{if } \tau < 0 < \tau',
  \end{array} \right.
\end{displaymath}
and
\begin{displaymath}
  \mathcal{U}_s(\tau_1, \tau_2, \cdots, \tau_{m-1}, \tau_m) =
    \left( \prod_{i=2}^{m-1} \ii \operatorname{sgn} \tau_i \right)
    G_s(\tau_1, \tau_2) W_s G_s(\tau_2, \tau_3) W_s \cdots W_s G_s(\tau_{m-1}, \tau_m).
\end{displaymath}
Then \eqref{eq:rho_s} can be reformulated as
\begin{equation} \label{eq:rho_s_expanded}
  \rho_s(t) = \sum_{m=0}^{+\infty} \int_{-t}^t \int_{-t}^{s_m} \cdots \int_{-t}^{s_2}
    \mathcal{U}_s(-t, s_1, \cdots, s_m, t) \mathcal{L}_b(s_1, \cdots, s_m)
    \,\mathrm{d}s_1 \cdots \,\mathrm{d}s_{m-1} \mathrm{d}s_m.
\end{equation}
When $m = 0$, the summand is regarded as $U_s(-t,t) = G_s(-t,t)$.
For further simplification, when $t_1 < t_2$, we let
\begin{displaymath}
  S_m(t_1,t_2) = \{ (s_1, s_2, \cdots, s_m) \mid t_1 < s_1 < s_2 < \cdots < s_m < t_2 \}.
\end{displaymath}
Then the equation \eqref{eq:rho_s_expanded} can be simplified to
\begin{equation}
  \rho_s(t) = \sum_{m=0}^{+\infty} \int_{S_m(-t,t)}
    \mathcal{U}_s(-t, \boldsymbol{s}, t) \mathcal{L}_b(\boldsymbol{s})
    \,\mathrm{d}\boldsymbol{s}.
\end{equation}
As an extension, for $t > 0$, $t' > 0$ and $\boldsymbol{\tau} \in \overline{S_m(-t,t')}$, we let
\begin{equation} \label{eq:rhos_tt'}
  \rho_s(t, t'; \boldsymbol{\tau}) =     \sum_{m=0}^{+\infty}
    \int_{S_m(-t,t')}
  \mathcal{U}_s(-t, \sigma[\boldsymbol{s}, \boldsymbol{\tau}], t')
    \mathcal{L}_b(\boldsymbol{s}) \,\mathrm{d}\boldsymbol{s},
\end{equation}
where $\sigma[\boldsymbol{s},\boldsymbol{\tau}]$ refers to the sorted vector including all components of $\boldsymbol{s}$ and $\boldsymbol{\tau}$.
For instance,
\begin{displaymath}
\sigma[(4,2), (1,3)] = (1,2,3,4), \qquad
\sigma[(-1,5,2), (1.6, -4)] = (-4, -1, 1.6, 2, 5).
\end{displaymath}
Here we also allow $\boldsymbol{\tau}$ to be an empty vector, and in this case, $\rho_s(-t,t;\boldsymbol{\tau})$ equals $\rho_s(t)$.

The quantities $\rho_s(t,t'; \boldsymbol{\tau})$ can be viewed as the continuous version of bold diagrams with circles such as \eqref{eq:bold_with_cirle} and \eqref{eq:no_circ_on_right}, and these circles are located at time points specified by $\boldsymbol{\tau}$.
With the help of this definition, we will show how diagram extensions \eqref{eq:right_ext_1}\eqref{eq:right_ext_0}\eqref{eq_1st_order_left} are formulated for these continuous-time quantities, and compare the results with our iterative scheme, which leads to the proof of the convergence rate.
Specifically, the proof of \Cref{thm:1st_order} will be conducted in the following steps:
\begin{itemize}
\item Step 1: Prove a continuous version of the extension formulas \eqref{eq_1st_order_left}.
\item Step 2: Discretize the continuous formulas to develop an iterative scheme for computing $\rho_s(t,t';\boldsymbol{\tau})$. This scheme will include more terms than our iterative method introduced in \Cref{subsec_iterative_scheme_1st_order}.
\item Step 3: Prove that the new numerical scheme has first-order accuracy.
\item Step 4: Prove that the difference between the new scheme and the scheme in \Cref{subsec_iterative_scheme_1st_order} is of first order.
\end{itemize}
These steps will be carried out in the following subsections.

\subsubsection{Extensions of $\rho_s(t,t';\boldsymbol{\tau})$}
\label{sec:extension}
In this part, we will prove a lemma showing how we can approximate $\rho_s(t+\Delta t, t'; \cdot)$ using $\rho_s(t,t'; \cdot)$, which can be regarded as left extensions of diagrams:
\begin{lemma} \label{lem:rho_s_extension}
  Assume that the bath correlation function $B(\cdot, \cdot)$ has an upper bound $\bar{B}$.
  Given $t>0$, $t'>0$, and $\Delta t > 0$, for any $\boldsymbol{\tau} \in S_n(-t,t')$ and $\boldsymbol{\tau}' \in S_{n'}(-t-\Delta t,-t)$, let 
  \begin{equation} \label{eq:rho_s1}
    \rho_{s,l}^{(1)}(t,t',\Delta t; \boldsymbol{\tau}', \boldsymbol{\tau}) = \int_{-t}^{t'} \int_{-t-\Delta t}^t 
      \mathcal{U}_s(-t-\Delta t, \sigma[\boldsymbol{\tau}', s'], t) \rho_s(t,t'; \sigma[\boldsymbol{\tau},s]) B(s',s) \,\mathrm{d}s' \,\mathrm{d}s.
  \end{equation}
  Then
  \begin{equation} \label{eq:left_extension}
  \begin{aligned}
    \|\rho_s(t+\Delta t,t';\sigma[\boldsymbol{\tau}', \boldsymbol{\tau}]) - \mathcal{U}_s(-t-\Delta t, \boldsymbol{\tau}', t) \rho_s(t,t';\boldsymbol{\tau}) -
      \rho_{s,l}^{(1)}(t,t',\Delta t; \boldsymbol{\tau}', \boldsymbol{\tau})\| \qquad \\
    \leqslant C_{\rho,n+n'}(t+t') A^{(1)}(t+t', \Delta t) \Delta t^2,
  \end{aligned}
  \end{equation}
  where
  \begin{equation} \label{eq:Crho}
  C_{\rho,n}(T) = \|\rho_s(0)\| \|W_s\|^n \exp(T^2\|W_s\|^2 \bar{B}/2),
  \end{equation}
  and $A^{(1)}(\cdot,\cdot)$ is a positive function that increases monotonically with respect to both parameters.
\end{lemma}

This lemma indicates that $\mathcal{U}_s(-t-\Delta t, \boldsymbol{\tau}', t) \rho_s(t,t';\boldsymbol{\tau}) + \rho_{s,l}^{(1)}(t,t',\Delta t; \boldsymbol{\tau}', \boldsymbol{\tau})$ can serve as a second-order approximation of $\rho_s(t+\Delta t,t';\sigma[\boldsymbol{\tau}', \boldsymbol{\tau}])$, mimicking the one-sided extension from $t$ to $t+\Delta t$ as in \eqref{extension_left_1st_order_n2}.
The first term $\mathcal{U}_s(-t-\Delta t, \boldsymbol{\tau}', t) \rho_s(t,t';\boldsymbol{\tau})$ is the extension without introducing new arcs (bath correlation functions).
The second-term $\rho_{s,l}^{(1)}(t,t',\Delta t; \boldsymbol{\tau}', \boldsymbol{\tau})$ introduces one arc connecting the original interval $(-t,t')$ and the extended part $(-t-\Delta t, -t)$, and the integrals in \eqref{eq:rho_s1} imply that all such connections are taken into account.

The proof of its second-order accuracy is given as follows:

\begin{proof}[Proof of \Cref{lem:rho_s_extension}]
By straightforward calculation using \eqref{eq:rho_s},
\begin{equation} \label{eq:rho_tdt}
  \rho_s(t+\Delta t,t'; \sigma[\boldsymbol{\tau}', \boldsymbol{\tau}]) = \mathcal{I}_0(t,t',\Delta t; \boldsymbol{\tau}', \boldsymbol{\tau}; []) + \sum_{k = 1}^{+\infty} \int_{S_k(-t-\Delta t,t')} 
      \mathcal{I}_k(t, t', \Delta t; \boldsymbol{\tau}', \boldsymbol{\tau}; \boldsymbol{s}')
    \,\mathrm{d}\boldsymbol{s}',
\end{equation}
where
\begin{displaymath}
  \mathcal{I}_k(t, t', \Delta t; \boldsymbol{\tau}', \boldsymbol{\tau}; \boldsymbol{s}') =
    \sum_{m=0}^{+\infty} \int_{S_m(-t,t')}
        \mathcal{U}_s(-t-\Delta t, \sigma[\boldsymbol{s}', \boldsymbol{\tau}', \boldsymbol{s}, \boldsymbol{\tau}], t')
        \mathcal{L}_b(\boldsymbol{s}, \boldsymbol{s}')
      \,\mathrm{d}\boldsymbol{s},
\end{displaymath}
and when $k = 0$, $\boldsymbol{s}'$ is an empty vector.
In the integrand, the bath influence functional $\mathcal{L}_b(\boldsymbol{s}, \boldsymbol{s}')$ equals zero if $m + k$ is odd, and when $m + k$ is even, according to \eqref{eq:Lb}, its absolute value can be bounded by $(m+k-1)!! \bar{B}^{(m+k)/2}$.
Therefore, $\mathcal{I}_k(t, t', \Delta t; \boldsymbol{\tau}', \boldsymbol{\tau}; \boldsymbol{s})$ can be bounded by
\begin{displaymath}
  \begin{aligned}
    & \|\mathcal{I}_k(t, t', \Delta t; \boldsymbol{\tau}', \boldsymbol{\tau}; \boldsymbol{s}')\| \\
    \leqslant{} & \|\rho_s(0)\| \sum_{\substack{m=0\\m+k \text{ even}}}^{+\infty}
      \frac{(t+t')^m}{m!} \|W_s\|^{m + k + n + n'} (m + k - 1)!! \bar{B}^{(m + k)/2} \\
    ={} & C_{\rho,n+n'}(t'+t) \sum_{j=0}^{\lfloor k/2 \rfloor}
      \frac{k!}{j! (k -2j)!} \frac{(t+t')^{k-2j}(\|W_s\|^2 \bar{B})^{k-j}}{2^j} \\
    \leqslant{} & C_{\rho,n+n'}(t+t') \sum_{j=0}^{\lfloor k/2 \rfloor}
      \frac{k!}{\lfloor k/2 \rfloor!} \frac{\lfloor k/2 \rfloor!}{j! \, (\lfloor k/2 \rfloor - j)!}
      \frac{[(t+t') \|W_s\|^2 \bar{B}]^k}{[2(t+t')^2 \|W_s\|^2 \bar{B}]^j} \\
    \leqslant{} & C_{\rho,n+n'}(t+t') \frac{k!}{\lfloor k/2 \rfloor!} [(t+t')\|W_s\|^2 \bar{B}]^k
    \left( 1 + \frac{1}{2(t+t')^2\|W_s\|^2\bar{B}} \right)^{\lfloor k/2 \rfloor}
  \end{aligned}
\end{displaymath}
Note that
\begin{gather*}
\mathcal{I}_0(t,t',\Delta t;\boldsymbol{\tau}', \boldsymbol{\tau}; []) = \mathcal{U}_s(-t-\Delta t, \boldsymbol{\tau}', t) \rho_s(t,t';\boldsymbol{\tau}), \\
\int_{S_1(-t-\Delta t,t)} \mathcal{I}_1(t,t',\Delta t; \boldsymbol{\tau}', \boldsymbol{\tau}; \boldsymbol{s}) \,\mathrm{d}\boldsymbol{s} = \rho_{s,l}^{(1)}(t,t',\Delta t; \boldsymbol{\tau}', \boldsymbol{\tau}).
\end{gather*}
One can estimate the left-hand side of \eqref{eq:left_extension} by
\begin{displaymath}
\begin{aligned}
  & \|\rho_s(t+\Delta t,t';\sigma[\boldsymbol{\tau}', \boldsymbol{\tau}]) - \mathcal{U}_s(-t-\Delta t, \boldsymbol{\tau}', t) \rho_s(t,t';\boldsymbol{\tau}) - \rho_{s,l}^{(1)}(t,t',\Delta t; \boldsymbol{\tau}', \boldsymbol{\tau})\| \\
  ={} & \left\| \sum_{k=2}^{+\infty} \int_{S_k(t-\Delta t,t)} \mathcal{I}_k(t,t',\Delta t;\boldsymbol{\tau}', \boldsymbol{\tau}; \boldsymbol{s}) \,\mathrm{d}\boldsymbol{s} \right\| \\
  \leqslant {} & C_{\rho,n+n'}(t+t') \sum_{k=2}^{+\infty} \frac{(\Delta t)^k}{k!} \cdot \frac{k!}{\lfloor k/2 \rfloor!} [(t+t')\|W_s\|^2 \bar{B}]^k
    \left( 1 + \frac{1}{2(t+t')^2\|W_s\|^2\bar{B}} \right)^{\lfloor k/2 \rfloor} \\
  ={} & C_{\rho,n+n'}(t+t') \left(1 + (t+t')\|W_s\|^2 \bar{B} \Delta t \right) \left[ \e^{\frac{1}{2} \Delta t^2 \|W_s\|^2 \bar{B} (1 + 2(t+t')^2 \|W_s\|^2 \bar{B})} - 1\right].
\end{aligned}
\end{displaymath}
Using $\e^x-1 \leqslant x \e^x$, we can find that \eqref{eq:left_extension} holds for
\begin{displaymath}
A^{(1)}(t+t', \Delta t) = \frac{1}{2} \|W_s\|^2 \bar{B} (1 + 2(t+t')^2 \|W_s\|^2 \bar{B}) \left(1 + (t+t')\|W_s\|^2 \bar{B} \Delta t \right) \e^{\frac{1}{2} \Delta t^2 \|W_s\|^2 \bar{B} (1 + 2(t+t')^2 \|W_s\|^2 \bar{B})}.
\end{displaymath}
\end{proof}

\subsubsection{Construction of a first-order method}
The purpose of this section is to develop a numerical method to approximate $\rho_s(t,t';\boldsymbol{\tau})$.
However, we are not interested in $\rho_s(t,t';\boldsymbol{\tau})$ for any $\boldsymbol{\tau}$.
Like the quantities $\Lambda(j_{n-}, \cdots, j_{1-}; j_{1+}, \cdots, j_{n+})$, only $\boldsymbol{\tau}$'s with all components in the middle of discrete cells are taken into account.
For the sake of simplicity, given positive integers $n$ and $n'$, for any sequence of nonnegative integers $\boldsymbol{j} = (j_{n-}, \cdots, j_{1-}; j_{1+}, \cdots, j_{n'+})$, we define
\begin{displaymath}
\rho_{s,n,n'}^{\Delta t}(\boldsymbol{j}) = \rho_s(n\Delta t, n'\Delta t; \boldsymbol{\tau}_{\boldsymbol{j}}),
\end{displaymath}
where
\begin{equation} \label{eq:tau_j}
\begin{aligned}
  \boldsymbol{\tau}_{\boldsymbol{j}} = \Big( & \underbrace{-(n-1/2)\Delta t, \cdots, -(n-1/2) \Delta t}_{j_{n-} \text{ times}}, \cdots, \underbrace{-\Delta t/2, \cdots, -\Delta t/2}_{j_{1-} \text{ times}}, \\
  & \underbrace{\Delta t/2, \cdots, \Delta t/2}_{j_{1+} \text{ times}}, \cdots, \underbrace{(n'-1/2)\Delta t, \cdots, (n'-1/2) \Delta t}_{j_{n'+} \text{ times}} \Big).
\end{aligned}
\end{equation}
In this equation, if $j_{k\pm} = 0$, then the corresponding midpoint $\pm(k-1/2) \Delta t$ does not appear in the list.
Furthermore, we will introduce the short-hand notations
\begin{displaymath}
|\boldsymbol{j}| = j_{n-} + \cdots + j_{1-} + j_{1+} + \cdots j_{n'+}, \qquad
\mathcal{V}_{s,r}^{\Delta t} = \e^{-\ii H_s \Delta t/2} (-\ii W_s)^r \e^{-\ii H_s \Delta t/2}.
\end{displaymath}

To approximate $\rho_{s,n,n'}^{\Delta t}(\boldsymbol{j})$ based on \Cref{lem:rho_s_extension}, we still need to discretize the integral \eqref{eq:rho_s1}.
Here we apply the midpoint quadrature rule, and the result is given in the following lemma:
\begin{lemma} \label{lem:quadrature}
Let $\boldsymbol{\tau}' = \big( \underbrace{-(n+1/2)\Delta t, \cdots, -(n+1/2)\Delta t}_{r \text{ times}} \big)$, $\boldsymbol{j} = (j_{n-}, \cdots, j_{1-}, j_{1+}, \cdots, j_{n'+})$, and
\begin{equation} \label{eq:Q1}
Q_{s,l}^{(1)}(n, n'; r, \boldsymbol{j}) = \Delta t^2 \sum_{\ell \in \{1-,\cdots,n-,1+,\cdots,n'+\}} \mathcal{V}_{s,r+1}^{\Delta t} \rho_{s,n,n'}^{\Delta t}(\boldsymbol{j} + \boldsymbol{e}_{\ell} \big) \Bcal{(n+1)-}{\ell},
\end{equation}
where $\boldsymbol{j} + \boldsymbol{e}_{\ell}$ adds the $\ell$th element of $\boldsymbol{j}$ by $1$.
Suppose the bath correlation function $B(\cdot,\cdot)$ is bounded by $\bar{B}$, and both $\partial_{\tau_1} B(\tau_1,\tau_2)$ and $\partial_{\tau_2} B(\tau_1,\tau_2)$ are bounded by $\bar{B}'$.
Then
\begin{equation} \label{eq:rhoQ}
\Big\|\rho_{s,l}^{(1)}\big(n\Delta t, n'\Delta t, \Delta t; \boldsymbol{\tau}', \boldsymbol{\tau}_{\boldsymbol{j}} \big) - Q_{s,l}^{(1)}(n,n'; r,\boldsymbol{j}) \Big\| \leqslant C_{\rho,|\boldsymbol{j}|+r}((n+n')\Delta t) A^{(2)}((n+n')\Delta t) \Delta t^2,
\end{equation}
where $A^{(2)}(\cdot)$ is a positive and monotonically increasing function.
\end{lemma}

With $Q_{s,l}^{(1)}(n,n';r,\boldsymbol{j})$ approximating the integral $\rho_{s,l}^{(1)}(n\Delta t, n'\Delta t, \Delta t; \boldsymbol{\tau}', \boldsymbol{\tau}_{\boldsymbol{j}})$, the result \eqref{eq:left_extension} can be naturally converted to an iterative numerical method that computes $\rho_{s,n+1,n'}^{\Delta t}(\cdot)$ from $\rho_{s,n,n'}^{\Delta t}(\cdot)$, and similarly, we can also perform right extension to obtain $\rho_{s,n,n'+1}^{\Delta t}(\cdot)$ from $\rho_{s,n,n'}^{\Delta t}(\cdot)$.
Since our aim is to find $\rho_s(n\Delta t) \approx \rho_{s,n,n}^{\Delta t}(0,\cdots,0; 0,\cdots,0)$, we can further restrict our focus to the quantities $\rho_{s,n+1,n}^{\Delta t}$ and $\rho_{s,n,n}^{\Delta t}$ like \eqref{eq:Lambda_path}, so that $\rho_{s,n,n}^{\Delta t}$ can always be obtained by alternate left and right extensions.
Below, we will use $\Sigma(j_{(n+1)-}, j_{n-}, \cdots, j_{1-}; j_{1+}, \cdots, j_{n+})$ and $\Sigma(j_{n-}, \cdots, j_{1-}; j_{1+}, \cdots, j_{n+})$ to denote the numerical approximations of $\rho_{s,n+1,n}^{\Delta t}$ and $\rho_{s,n,n}^{\Delta t}$, respectively.

Applying the results in \Cref{lem:rho_s_extension} and \Cref{lem:quadrature}, the scheme reads
\begin{align*}
& \Sigma(\,;\,) = \rho_s(0), \\
& \Sigma(j_{(n+1)-}, j_{n-}, \cdots, j_{1-}; j_{1+}, \cdots, j_{n+}) =
\mathcal{V}_{s,j_{(n+1)-}}^{\Delta t} \Sigma(\hat{j}_{n-}, \cdots, \hat{j}_{1-}; \hat{j}_{1+}, \cdots, \hat{j}_{n+}) \\
& \qquad + \Delta t^2 \sum_{\ell \in \{1-,\cdots,n-,1+,\cdots,n+\}} \mathcal{V}_{s,1+j_{(n+1)-}}^{\Delta t} \Sigma(\hat{j}^{\ell}_{n-}, \cdots, \hat{j}^{\ell}_{1-}; \hat{j}^{\ell}_{1+}, \cdots, \hat{j}^{\ell}_{n+}) \Bcal{(n+1)-}{\ell}, \\
& \Sigma(j_{n-}, \cdots, j_{1-}; j_{1+}, \cdots, j_{n+}) = \Sigma(j_{n-}, \cdots, j_{1-}; j_{1+}, \cdots, j_{(n-1)+})(\mathcal{V}_{s,j_{n+}}^{\Delta t})^{\dagger} \\
& \qquad + \Delta t^2 \sum_{\ell \in \{1-,\cdots,n-,1+,\cdots,(n-1)+\}}  \Sigma(\hat{j}^{\ell}_{n-}, \cdots, \hat{j}^{\ell}_{1-}; \hat{j}^{\ell}_{1+}, \cdots, \hat{j}^{\ell}_{(n-1)+}) (\mathcal{V}_{s,1+j_{n+}}^{\Delta t})^{\dagger} \Bcal{\ell}{n+},
\end{align*}
where $\hat{j}_k^{\ell} = j_k + \delta_{k\ell}$ as defined in \eqref{eq:hatj}.

We will show in the next part that the scheme has first-order.
Here, we provide the proof of \Cref{lem:quadrature}, which requires the following result:

\begin{lemma} \label{lem:derivative}
Suppose the bath correlation function $B(\cdot,\cdot)$ is bounded by $\bar{B}$.
For any $t > 0$, $t' > 0$, and $\boldsymbol{\tau} \in S_n(-t,t')$, it holds that
\begin{displaymath}
\|\rho_s(t, t'; \boldsymbol{\tau})\| \leqslant C_{\rho,n}(t+t').
\end{displaymath}
When the $k$th component of $\boldsymbol{\tau}$, denoted by $\tau_k$, is nonzero,
\begin{displaymath}
\|\partial_{\tau_k} \rho_s(t,t';\boldsymbol{\tau})\| \leqslant 2\|H_s\| C_{\rho,n}(t+t').
\end{displaymath}
\end{lemma}

\begin{proof}
The norm of $\rho_s(t,t';\boldsymbol{\tau})$ can be directly estimated using the triangle inequality:
\begin{equation} \label{eq:rho_estimate}
\begin{aligned}
\|\rho_s(t,t';\boldsymbol{\tau})\| & \leqslant \|\rho_s(0)\|\sum_{\substack{m = 0\\ m\text{ even}}}^{+\infty} \frac{(t+t')^m}{m!} \|W_s\|^{m+n} (m-1)!! \bar{B}^{m/2} \\
&= \|\rho_s(0)\| \sum_{m=0}^{+\infty} \frac{(t+t')^{2m}}{2^m m!} \|W_s\|^{2m+n} \bar{B}^m
= \|\rho_s(0)\| \|W_s\|^n \exp((t+t')^2\|W_s\|^2 \bar{B}/2).
\end{aligned}
\end{equation}
The right-hand side equals $C_{\rho,n}(t+t')$ by definition \eqref{eq:Crho}.

When $\tau_k \neq 0$, we write $\boldsymbol{\tau}$ as $\boldsymbol{\tau} = (\boldsymbol{\tau}^{(1)}, \tau_k, \boldsymbol{\tau}^{(2)})$.
Then the derivative of $\rho_s(t,t';\boldsymbol{\tau})$ is
\begin{displaymath}
\begin{aligned}
\partial_{\tau_k} \rho_s(t,t';\boldsymbol{\tau}) &= \sum_{m=0}^{+\infty} \sum_{m'=0}^{+\infty} \int_{S_m(-t,\tau_k)} \int_{S_{m'}(\tau_k,t')} \\
& \qquad (\ii \operatorname{sgn} \tau_k) \, \mathcal{U}_s(-t,\sigma[\boldsymbol{s},\boldsymbol{\tau}^{(1)}],\tau) [H_s, W_s] \mathcal{U}_s(\tau,\sigma[\boldsymbol{s}', \boldsymbol{\tau}^{(2)}], t') \mathcal{L}_b(\boldsymbol{s}, \boldsymbol{s}') \,\mathrm{d}\boldsymbol{s}' \,\mathrm{d}\boldsymbol{s},
\end{aligned}
\end{displaymath}
which can be estimated by
\begin{equation} \label{eq:derivative_estimation}
\begin{aligned}
\|\partial_{\tau_k}\rho_s(t,t';\boldsymbol{\tau})\| & \leqslant
  \|\rho_s(0)\| \sum_{\substack{m,m'=0\\m+m' \text{ even}}}^{+\infty} \frac{(t+\tau_k)^m}{m!} \frac{(t'-\tau_k)^{m'}}{m'!} 2\|H_s\| \|W_s\|^{n+m+m'} (m+m'-1)!! \bar{B}^{(m+m')/2} \\
& = \|\rho_s(0)\| \sum_{\substack{m=0\\m \text{ even}}}^{+\infty} \frac{(t+t')^m}{m!} 2\|H_s\| \|W_s\|^{n+m} (m-1)!! \bar{B}^{m/2}.
\end{aligned}
\end{equation}
The right-hand side can be calculated in the same way as \eqref{eq:rho_estimate}.
\end{proof}

We now present the proof of \Cref{lem:quadrature}:
\begin{proof}[Proof of \Cref{lem:quadrature}]
Let
\begin{displaymath}
K_{s,l}^{(1)}(n, n'; r, \boldsymbol{j}) = \Delta t \int_{-n\Delta t}^{n'\Delta t} \mathcal{V}_{s,r+1}^{\Delta t} \rho_s(n\Delta t, n'\Delta t; \sigma[\boldsymbol{\tau}_{\boldsymbol{j}},s]) B(-(n+1/2)\Delta t, s)\,\mathrm{d}s.
\end{displaymath}
We will show that both $\|\rho_s^{(1)}(n\Delta t, n'\Delta t, \Delta t; \boldsymbol{\tau}', \boldsymbol{\tau}_{\boldsymbol{j}}) - K_{s,l}^{(1)}(n,n'; r, \boldsymbol{j})\|$ and $\|K_{s,l}^{(1)}(n,n'; r, \boldsymbol{j}) - Q_{s,l}^{(1)}(n,n'; r, \boldsymbol{j})\|$ are second order in $\Delta t$.

By definition,
\begin{displaymath}
\begin{aligned}
& \mathcal{U}_s\big({-(n+1)\Delta t}, \sigma[\boldsymbol{\tau}', s'], -n\Delta t \big) \\
={} & \left\{ \begin{array}{@{}ll}
  \e^{-\ii H_s (s' + (n+1)\Delta t)} (-\ii W_s) \e^{-\ii H_s (n\Delta t - s')} (-\ii W_s)^r \e^{-\ii H_s \Delta t/2}, & \text{if } s' < -(n+1/2)\Delta t, \\
  \e^{-\ii H_s \Delta t/2} (-\ii W_s)^r \e^{-\ii H_s (s' + (n+1/2)\Delta t)} (-\ii W_s) \e^{-\ii H_s (n\Delta t - s')}, & \text{if } s' > -(n+1/2)\Delta t.
\end{array}
\right.
\end{aligned}
\end{displaymath}
Therefore, $\mathcal{U}_s\big({-(n+1)\Delta t}, \sigma[\boldsymbol{\tau}', s'], -n\Delta t \big)$ is continuous with respect to $s'$ in $[-(n+1)\Delta t, -n\Delta t]$, and in particular, when $s'=-(n+1/2)\Delta t$, it equals $\mathcal{V}_{s,r+1}^{\Delta t}$. However, its derivative is discontinuous at $s' = -(n+1/2)\Delta t$.
By straightforward calculation, we can bound its derivative by
\begin{displaymath}
\left\|\frac{\partial}{\partial s'}\mathcal{U}_s\big({-(n+1)\Delta t}, \sigma[\boldsymbol{\tau}', s'], -n\Delta t \big) \right\| \leqslant 2\|H_s\| \|W_s\|^{r+1},
\end{displaymath}
which holds for both left- and right-derivatives at $-(n+1/2)\Delta t$.
Taylor expansion at $s' = -(n+1/2) \Delta t$ yields the following bound:
\begin{displaymath}
\begin{aligned}
& \left\|\int_{t-\Delta t}^t \mathcal{U}_s\big({-(n+1)\Delta t}, \sigma[\boldsymbol{\tau}', s'], -n\Delta t \big) B(s', s) \,\mathrm{d}s' - \Delta t\,\mathcal{V}_{s,r+1}^{\Delta t} B(-(n+1/2)\Delta t, s)\right\| \\
\leqslant{} & \frac{\Delta t^2}{4} (\|W_s\|^{r+1} \bar{B}' + 2\|H_s\| \|W_s\|^{r+1} \bar{B}).
\end{aligned}
\end{displaymath}
Consequently,
\begin{equation} \label{eq:rhoK}
\begin{aligned}
& \|\rho_s^{(1)}(n\Delta t, n'\Delta t, \Delta t; \boldsymbol{\tau}', \boldsymbol{\tau}_{\boldsymbol{j}}) -   K_{s,l}^{(1)}(n, n'; r, \boldsymbol{j})\| \\
\leqslant{} &
  \frac{(n+n')\Delta t^3}{4} (\|W_s\|^{r+1} \bar{B}' + 2\|H_s\| \|W_s\|^{r+1} \bar{B}) C_{\rho, |\boldsymbol{j}|+1}((n+n')\Delta t).
\end{aligned}
\end{equation}

The estimation of $\|K_{s,l}^{(1)}(n,n'; r, \boldsymbol{j}) - Q_{s,l}^{(1)}(n,n'; r, \boldsymbol{j})\|$ is similar.
Note that $Q_{s,l}^{(1)}(n,n'; r, \boldsymbol{j})$ approximates the integral in $K_{s,l}^{(1)}(n,n'; r, \boldsymbol{j})$ with the midpoint rule, but the derivative of $\rho_s(-n\Delta t, n'\Delta t; \sigma[\boldsymbol{\tau}_{\boldsymbol{j}},s])$ is discontinuous on all the midpoints in $\boldsymbol{\tau}_{\boldsymbol{j}}$.
Therefore, we can apply \Cref{lem:derivative} to obtain
\begin{equation} \label{eq:KQ}
\|K_{s,l}^{(1)}(n,n'; r, \boldsymbol{j}) - Q_{s,l}^{(1)}(n,n'; r, \boldsymbol{j})\| \leqslant  \frac{(n+n') \Delta t^3}{4} \|W_s\|^{r+1} C_{\rho,|\boldsymbol{j}|+1}((n+n')\Delta t) (\bar{B}' + 2\|H_s\| \bar{B}).
\end{equation}
The conclusion \eqref{eq:rhoQ} can then be drawn by combining \eqref{eq:rhoK} and \eqref{eq:KQ}, which gives $A^{(2)}(t) = (\bar{B}' + 2\|H_s\| \bar{B})\|W_s\|^2 t/2$.
\end{proof}

\subsubsection{Convergence order of the new scheme}
This part shows the first-order accuracy of $\Sigma(j_{n-}, \cdots, j_{1-}; j_{1+}, \cdots, j_{n+})$ as an approximation of $\rho_{s,n,n}^{\Delta t}(j_{n-}, \cdots, j_{1-}; j_{1+}, \cdots, j_{n+})$.
The result is given in the following lemma:
\begin{lemma} \label{lem:Sigma_order}
Assume that the bath correlation function $B(\cdot, \cdot)$ is bounded by $\bar{B}$, and its derivatives are bounded by $\bar{B}'$.
Let $\boldsymbol{j} = (j_{n-}, \cdots, j_{1-}; j_{1+}, \cdots, j_{n+})$. Then
\begin{equation} \label{eq:Sigma}
\|\Sigma(j_{n-}, \cdots, j_{1-}; j_{1+}, \cdots, j_{n+}) - \rho_{s,n,n}^{\Delta t}(\boldsymbol{j})\| \leqslant C_1(2n\Delta t, \Delta t) \|W_s\|^{|\boldsymbol{j}|} \Delta t,
\end{equation}
where $C_1(\cdot,\cdot)$ increases monotonically with respect to both parameters.
\end{lemma}

The proof of this lemma generally follows the classical strategy used in the analysis of numerical methods for ordinary differential equations.
We will first establish an estimation of the location truncation error:
\begin{lemma} \label{lem:scheme}
Assume that the bath correlation function $B(\cdot, \cdot)$ is bounded by $\bar{B}$, and its derivatives are bounded by $\bar{B}'$.
Given $\boldsymbol{j} = (j_{n-}, \cdots, j_{1-}; j_{1+}, \cdots, j_{n'+})$, let $[r,\boldsymbol{j}] = (r, j_{n-}, \cdots, j_{1-}; j_{1+}, \cdots, j_{n'+})$. Then
\begin{equation} \label{eq:left_ext_est}
\|\rho_{s,n+1,n'}^{\Delta t}([r,\boldsymbol{j}]) - \mathcal{V}_{s,r}^{\Delta t} \rho_{s,n,n'}^{\Delta t}(\boldsymbol{j}) - Q_{s,l}^{(1)}(n,n';r,\boldsymbol{j})\| \leqslant \|W_s\|^{r+|\boldsymbol{j}|} A((n+n')\Delta t, \Delta t) \Delta t^2,
\end{equation}
where $A(\cdot,\cdot)$ increases monotonically with respect to both parameters.

Similarly, for $[\boldsymbol{j}, r] = (j_{n-}, \cdots, j_{1-}; j_{1+}, \cdots, j_{n'+}, r)$, it holds that
\begin{equation} \label{eq:right_ext_est}
\|\rho_{s,n,n'+1}^{\Delta t}([\boldsymbol{j}, r]) - \rho_{s,n,n'}^{\Delta t}(\boldsymbol{j}) (\mathcal{V}_{s,r}^{\Delta t})^{\dagger} - Q_{s,r}^{(1)}(n,n';r,\boldsymbol{j})\| \leqslant \|W_s\|^{r+|\boldsymbol{j}|} A((n+n')\Delta t, \Delta t)\Delta t^2,
\end{equation}
where $[\boldsymbol{j},r] = (j_{n-}, \cdots, j_{1-}, j_{1+}, \cdots, j_{n'+}, r)$, and
\begin{displaymath}
Q_{s,r}^{(1)}(n, n'; r, \boldsymbol{j}) = \Delta t^2 \sum_{ \ell \in \{1-,\cdots,n-,1+,\cdots,n'+\}} \rho_{s,n,n'}^{\Delta t}(\boldsymbol{j} + \boldsymbol{e}_{\ell} \big)(\mathcal{V}_{s,r}^{\Delta t})^{\dagger} \Bcal{\ell}{(n+1)+}.
\end{displaymath}
\end{lemma}
\begin{proof}
The estimate \eqref{eq:left_ext_est} is a corollary of \Cref{lem:rho_s_extension} and \Cref{lem:quadrature}.
In \eqref{eq:left_extension}, we set $t = n\Delta t$, $t' = n'\Delta t$, $\boldsymbol{\tau} = \boldsymbol{\tau}_{\boldsymbol{j}}$, and $\boldsymbol{\tau}' = \big( \underbrace{-(n+1/2)\Delta t, \cdots, -(n+1/2)\Delta t}_{r \text{ times}} \big)$, which gives
\begin{displaymath}
\left\|\rho_{s,n+1,n'}^{\Delta t}([r,\boldsymbol{j}]) - \mathcal{V}_{s,r}^{\Delta t} \rho_{s,n,n'}^{\Delta t}(\boldsymbol{j}) - \rho_{s,l}^{(1)}\left(n\Delta t, n'\Delta t, \Delta t; \boldsymbol{\tau}', \boldsymbol{\tau}_{\boldsymbol{j}}\right) \right\| \leqslant C_{\rho,r+|\boldsymbol{j}|} A^{(1)}((n+n')\Delta t, \Delta t) \Delta t^2.
\end{displaymath}
Hence, by \eqref{eq:rhoQ}, we obtain
\begin{displaymath}
\begin{aligned}
& \|\rho_{s,n+1,n'}^{\Delta t}([r,\boldsymbol{j}]) - \mathcal{V}_{s,r}^{\Delta t} \rho_{s,n,n'}^{\Delta t}(\boldsymbol{j}) - Q_{s,l}^{(1)}(n,n';r,\boldsymbol{j})\| \\
\leqslant{} & \!\left\|\rho_{s,n+1,n'}^{\Delta t}([r,\boldsymbol{j}]) - \mathcal{V}_{s,r}^{\Delta t} \rho_{s,n,n'}^{\Delta t}(\boldsymbol{j}) - \rho_{s,l}^{(1)}\left(n\Delta t, n'\Delta t, \Delta t; \boldsymbol{\tau}', \boldsymbol{\tau}_{\boldsymbol{j}}\right) \right\| \\
& \quad + \Big\|\rho_{s,l}^{(1)}\big(n\Delta t, n'\Delta t, \Delta t; \boldsymbol{\tau}', \boldsymbol{\tau}_{\boldsymbol{j}} \big) - Q_{s,l}^{(1)}(n,n'; r,\boldsymbol{j}) \Big\| \\
\leqslant{} & C_{\rho,r+|\boldsymbol{j}|}((n+n')\Delta t) A^{(1)}((n+n')\Delta t, \Delta t) \Delta t^2 + C_{\rho,r+|\boldsymbol{j}|}((n+n')\Delta t) A^{(2)}((n+n')\Delta t) \Delta t^2.
\end{aligned}
\end{displaymath}
The definition \eqref{eq:Crho} shows that \eqref{eq:left_ext_est} holds with
\begin{displaymath}
A(t,\Delta t) = \|\rho_s(0)\| \exp(t^2 \|W_s\|^2 \bar{B}/2) [A^{(1)}(t,\Delta t) + A^{(2)}(t)],
\end{displaymath}
whose monotonicity follows the monotonicity of $A^{(1)}$ and $A^{(2)}$.

To show \eqref{eq:right_ext_est}, 
one needs to repeat the proofs of \Cref{lem:rho_s_extension} and \Cref{lem:quadrature}, but change the left extension to right extension.
Since the steps are nearly identical, we omit the details here.
\end{proof}

\Cref{lem:Sigma_order} is then the result of recursion:
\begin{proof}[Proof of \Cref{lem:Sigma_order}]
Let
\begin{align*}
e(j_{n-}, \cdots, j_{1-}, j_{1+}, \cdots, j_{n+}) &= \|\Sigma(j_{n-}, \cdots, j_{1-}; j_{1+}, \cdots, j_{n+}) - \rho_{s,n,n}^{\Delta t}(\boldsymbol{j})\|, \\
e(j_{(n+1)-}, \cdots, j_{1-}, j_{1+}, \cdots, j_{n+}) &= \|\Sigma(j_{(n+1)-}, \cdots, j_{1-}; j_{1+}, \cdots, j_{n+}) - \rho_{s,n,n}^{\Delta t}([j_{(n+1)-}, \boldsymbol{j}])\|.
\end{align*}
By the recursive relation of $\Sigma$, we have
\begin{align*}
& e(\,;\,) = 0, \\
& e(j_{(n+1)-}, j_{n-}, \cdots, j_{1-}; j_{1+}, \cdots, j_{n+}) \leqslant 
\|W_s\|^{j_{(n+1)-}} e(\hat{j}_{n-}, \cdots, \hat{j}_{1-}; \hat{j}_{1+}, \cdots, \hat{j}_{n+}) \\
& \qquad + \Delta t^2 \sum_{\ell \in \{1-,\cdots,n-,1+,\cdots,n+\}} \|W_s\|^{1 + j_{(n+1)-}} \bar{B} e(\hat{j}^{\ell}_{n-}, \cdots, \hat{j}^{\ell}_{1-}; \hat{j}^{\ell}_{1+}, \cdots, \hat{j}^{\ell}_{n+}) \\
& \qquad + A(2n\Delta t, \Delta t) \|W_s\|^{j_{(n+1)-} + |\boldsymbol{j}|} \Delta t^2, \\
& e(j_{n-}, \cdots, j_{1-}; j_{1+}, \cdots, j_{n+}) \leqslant \|W_s\|^{j_{n+}}e(j_{n-}, \cdots, j_{1-}; j_{1+}, \cdots, j_{(n-1)+}) \\
& \qquad + \Delta t^2 \sum_{\ell \in \{1-,\cdots,n-,1+,\cdots,(n-1)+\}} \|W_s\|^{1 + j_{n+}} \bar{B}  e(\hat{j}^{\ell}_{n-}, \cdots, \hat{j}^{\ell}_{1-}; \hat{j}^{\ell}_{1+}, \cdots, \hat{j}^{\ell}_{(n-1)+}) \\
& \qquad + A((2n-1)\Delta t, \Delta t) \|W_s\|^{|\boldsymbol{j}|} \Delta t^2.
\end{align*}
Let
\begin{align*}
E_{n+1,n} &= \max_{\substack{j_{1-}, \cdots, j_{(n+1)-}\\j_{1+}, \cdots, j_{n+}}} \frac{e(j_{(n+1)-}, \cdots, j_{1-}; j_{1+}, \cdots, j_{n+})}{\|W_s\|^{|\boldsymbol{j}|+j_{(n+1)-}}}, \\
E_{n,n} &= \max_{\substack{j_{1-}, \cdots, j_{n-}\\j_{1+}, \cdots, j_{n+}}} \frac{e(j_{n-}, \cdots, j_{1-}; j_{1+}, \cdots, j_{n+})}{\|W_s\|^{|\boldsymbol{j}|}}.
\end{align*}
Then
\begin{align*}
E_{0,0} &= 0, \\
E_{n+1,n} &\leqslant (1 + \Delta t^2 \cdot (2n)\|W_s\|^2 \bar{B}) E_{n,n} + A(2n\Delta t, \Delta t) \Delta t^2, \\
E_{n,n} &\leqslant (1 + \Delta t^2 \cdot (2n-1)\|W_s\|^2 \bar{B}) E_{n,n-1} + A((2n-1)\Delta t, \Delta t) \Delta t^2.
\end{align*}
By recursion and the monotonicity of $A(\cdot, \Delta t)$, we obtain
\begin{displaymath}
E_{n,n} \leqslant \frac{A(2n\Delta t, \Delta t)}{2n\Delta t \|W_s\|^2 \bar{B}} [\e^{(2n\Delta t)^2 \|W_s\|^2 \bar{B}} - 1] \Delta t.
\end{displaymath}
Hence, the estimate \eqref{eq:Sigma} holds for
\begin{displaymath}
C_1(t, \Delta t) = \frac{A(t, \Delta t)}{t \|W_s\|^2 \bar{B}} [\e^{t^2 \|W_s\|^2 \bar{B}} - 1].
\end{displaymath}
\end{proof}

\subsubsection{Difference between two numerical schemes}
Finally, we will show that the difference between $\Lambda(0, \cdots, 0; 0, \cdots, 0)$ and $\Sigma(0, \cdots, 0; 0, \cdots, 0)$ has order $O(\Delta t)$.
By setting all components of $\boldsymbol{j}$ to zero, we see that $\Sigma(0, \cdots, 0; 0, \cdots, 0)$ is a first-order approximation of $\rho_s(n\Delta t)$, so that $\Lambda(0,\cdots,0; 0,\cdots,0)$ is also first-order accurate.
\begin{lemma} \label{lem:LambdaSigma}
Suppose the bath correlation function $B(\cdot, \cdot)$ is bounded by $\bar{B}$.
For any $n > 0$,
\begin{equation} \label{eq:LambdaSigma}
\|\Lambda(\underbrace{0, \cdots, 0}_{n \text{ zeros}}; \underbrace{0, \cdots, 0}_{n\text{ zeros}}) - \Sigma(\underbrace{0, \cdots, 0}_{n \text{ zeros}}; \underbrace{0, \cdots, 0}_{n\text{ zeros}})\| \leqslant C_2(2n\Delta t, \Delta t) \Delta t,
\end{equation}
where $C_2(\cdot,\cdot)$ is a positive function that increases with respect to both parameters.
\end{lemma}
\begin{proof}
By construction, $\Sigma(0, \cdots, 0; 0, \cdots, 0)$ is obtained by alternating left and right extensions starting from $\rho_s(0)$, and each extension multiplies calculated operators by $\mathcal{V}_{s,r}^{\Delta t}$ or its conjugate transpose.
As a result, $\Sigma(0, \cdots, 0; 0, \cdots, 0)$ can be written as a linear combination of
\begin{equation} \label{eq:product}
\mathcal{V}_{s,j_{n-}}^{\Delta t} \cdots \mathcal{V}_{s,j_{1-}}^{\Delta t} \rho_s(0) (\mathcal{V}_{s,j_{1+}}^{\Delta t})^{\dagger} \cdots (\mathcal{V}_{s,j_{n+}}^{\Delta t})^{\dagger}
\end{equation}
for all $\boldsymbol{j} = (j_{n-}, \cdots, j_{1-}, j_{1+}, \cdots, j_{n+})$ satisfying $|\boldsymbol{j}| \leqslant 4n$ and $|\boldsymbol{j}|$ is even.
The reason for the above constraint is that each extension may introduce a bath correlation function $B(\cdot,\cdot)$ in the coefficient, corresponding to two $W_s$'s in the operator, and obtaining $\Sigma(0, \cdots, 0; 0, \cdots, 0)$ needs  $2n$ extensions without extra $W_s$'s, amounting to at most $4n$ coupling operators.

Similarly, the expansion of $\Lambda(0,\cdots,0; 0,\cdots,0)$ is also a linear combination of \eqref{eq:product}, but indices $j_{n-}, \cdots, j_{1-}$ and $j_{1+}, \cdots, j_{n+}$ can only take values $0$ or $1$.
In this case, the coefficients of \eqref{eq:product} in the expansions of $\Lambda(0,\cdots,0; 0,\cdots,0)$ and $\Sigma(0,\cdots,0; 0,\cdots,0)$ are identical, both equal to $\mathcal{L}_b(\boldsymbol{\tau}_{\boldsymbol{j}})$ (see \eqref{eq:tau_j}).
This implies that their difference comes from the contribution of \eqref{eq:product} where any component of $\boldsymbol{j}$ is greater than equal to $2$.
This condition will be denoted by ``$\max \boldsymbol{j} \geqslant 2$'' below.

We now estimate the difference as follows:
\begin{displaymath}
\|\Lambda(\underbrace{0, \cdots, 0}_{n \text{ zeros}}; \underbrace{0, \cdots, 0}_{n\text{ zeros}}) - \Sigma(\underbrace{0, \cdots, 0}_{n \text{ zeros}}; \underbrace{0, \cdots, 0}_{n\text{ zeros}})\| \leqslant
\sum_{m=1}^{2n} \sum_{\substack{|\boldsymbol{j}|=2m\\ \max \boldsymbol{j} \geqslant 2}} (\Delta t\|W_s\|)^{2m} (2m-1)!! \bar{B}^m = \mathrm{I} + \mathrm{II}.
\end{displaymath}
where
\begin{align*}
\mathrm{I} &= \sum_{m=1}^{n} \left[ \binom{2n+2m-1}{2m} - \binom{2n}{2m} \right] (\Delta t\|W_s\|)^{2m} (2m-1)!! \bar{B}^m, \\
\mathrm{II} &= \sum_{m=n+1}^{2n} \binom{2n+2m-1}{2n-1}(\Delta t\|W_s\|)^{2m} (2m-1)!! \bar{B}^m.
\end{align*}
We are going to show that $\mathrm{I}$ has order $O(\Delta t)$ and $\mathrm{II}$ has a higher order.

To estimate $\mathrm{I}$, we let
\begin{displaymath}
f_m(x) = \binom{2x}{2m} = \frac{(2x)(2x-1)\cdots(2x-2m+1)}{(2m)!},
\end{displaymath}
so that when $x \geqslant m$,
\begin{displaymath}
|f_m'(x)| = \left| \frac{2}{(2m)!} \sum_{k=0}^{2m-1} \prod_{\substack{j=0\\j \neq k}}^{2m-1} (2x-k) \right| \leqslant \frac{2}{(2m-1)!} (2x)^{2m-1}.
\end{displaymath}
By the mean value theorem, there exists $\xi \in (n, n+m-1/2)$ such that
\begin{displaymath}
\begin{aligned}
\mathrm{I} &\leqslant \sum_{m=1}^n (m-1/2) \cdot \frac{2}{(2m-1)!} (2\xi)^{2m-1} (\Delta t \|W_s\|)^{2m} (2m-1)!! \bar{B}^m \\
&\leqslant \sum_{m=1}^n \frac{2m-1}{2^{m-1} (m-1)!} (4n)^{2m-1} (\Delta t \|W_s\|)^{2m} \bar{B}^m \\
&\leqslant \Delta t \|W_s\| \sum_{m=0}^{+\infty} \frac{2(m+1)}{2^m m!} (4n\Delta t \|W_s\|)^{2m+1} \bar{B}^{m+1} \\
& = 4n\Delta t^2 \|W_s\|^2 \bar{B} (1 + 16 n^2 \Delta t^2 \|W_s\|^2 \bar{B}) \e^{8n^2\Delta t^2 \|W_s\|^2 \bar{B}}.
\end{aligned}
\end{displaymath}

The other term $\mathrm{II}$ can be estimated as follows:
\begin{displaymath}
\begin{aligned}
\mathrm{II} &= \sum_{m=n+1}^{2n} \frac{(2n)(2n+1)\cdots (2n+2m-1)}{2^m m!} (\Delta t \|W_s\|)^{2m} \bar{B}^m \\
& \leqslant \sum_{m=n+1}^{2n} \frac{(2n+2)\cdots (2n+2m-1)}{2^{m-2} (m-2)!} (\Delta t \|W_s\|)^{2m} \bar{B}^m \\
& \leqslant \sum_{m=n+1}^{2n} \frac{(6n)^{2m-2}}{2^{m-2} (m-2)!} (\Delta t \|W_s\|)^{2m} \bar{B}^m \\
& \leqslant \Delta t^2 \|W_s\|^2 \bar{B} \sum_{m=0}^{+\infty} \frac{(2m)(6n)^{2m}}{2^m m!} (\Delta t \|W_s\|)^{2m} \bar{B}^m \\
& = 36\Delta t^2 (n\Delta t)^2 \|W_s\|^4 \bar{B}^2 \e^{18n^2\Delta t^2 \|W_s\|^2 \bar{B}}.
\end{aligned}
\end{displaymath}
Adding up the estimates of $\mathrm{I}$ and $\mathrm{II}$, we see that \eqref{eq:LambdaSigma} holds for
\begin{displaymath}
C_2(t,\Delta t) = 2t \|W_s\|^2 \bar{B} (1 + 4t^2 \|W_s\|^2 \bar{B}) \e^{2t^2 \|W_s\|^2 \bar{B}} + 36 t \Delta t^2 \|W_s\|^4 \bar{B}^2 \e^{9t^2\|W_s\|^2\bar{B}/2}.
\end{displaymath}
\end{proof}

\begin{proof}[Proof of \Cref{thm:1st_order}]
Note that
\begin{displaymath}
    \rho_{s,n} = \Lambda(\underbrace{0,\cdots,0}_{n \text{ zeros}};\underbrace{0,\cdots,0}_{n \text{ zeros}}), \quad \rho(n\Delta t) = \rho_{s,n,n}^{\Delta t}(\underbrace{0,\cdots,0}_{n \text{ zeros}};\underbrace{0,\cdots,0}_{n \text{ zeros}}).
\end{displaymath}
By triangle inequality,
\begin{displaymath}
\begin{aligned}
& \|\rho_{s,n} - \rho(n\Delta t)\| \\
\leqslant{} & \|\Lambda(\underbrace{0,\cdots,0}_{n \text{ zeros}};\underbrace{0,\cdots,0}_{n \text{ zeros}}) - \Sigma(\underbrace{0,\cdots,0}_{n \text{ zeros}};\underbrace{0,\cdots,0}_{n \text{ zeros}})\| + \|\Sigma(\underbrace{0,\cdots,0}_{n \text{ zeros}};\underbrace{0,\cdots,0}_{n \text{ zeros}}) - \rho_{s,n,n}^{\Delta t}(\underbrace{0,\cdots,0}_{n \text{ zeros}};\underbrace{0,\cdots,0}_{n \text{ zeros}})\| \\
\leqslant {} & [C_1(2n\Delta t, \Delta t) + C_2(2n\Delta t, \Delta t)] \Delta t.
\end{aligned}
\end{displaymath}
The conclusion of the theorem \eqref{eq:rho_error} holds for $C = C_1(2n\Delta t, n\Delta t) + C_2(2n\Delta t, n\Delta t)$ due to the monotonicity of $C_1$ and $C_2$.
\end{proof}

\subsection{Sketch of the proof for \Cref{thm:2nd_order}}
\label{sec:proof_2nd_order}
The general procedure to show the second-order accuracy of 
\eqref{eq:2nd_order_rho} follows the same structure as the steps listed before \Cref{sec:extension}.
However, due to the additional second-order terms, the proof becomes a lot more tedious compared with the first-order case.
Instead of presenting the complete proof, below we will only highlight the difference between the proofs in the first-order and second-order cases.

\paragraph{Step 1: Extension of $\rho_s(t,t';\boldsymbol{\tau})$}
We will again use the definition of $\rho_s(t,t';\boldsymbol{\tau})$ in \eqref{eq:rhos_tt'} and study its extension as in \eqref{eq:left_extension}.
To prove one higher order of accuracy, the approximation of $\rho_s(t+\Delta, t'; \sigma[\boldsymbol{\tau}',\boldsymbol{\tau}])$ requires an extra term $\rho_{s,l}^{(2)}(t,t',\Delta t; \boldsymbol{\tau}', \boldsymbol{\tau})$, defined as
\begin{equation} \label{eq:2nd_order_approx}
\begin{aligned}
& \rho_{s,l}^{(2)}(t,t',\Delta t; \boldsymbol{\tau}', \boldsymbol{\tau}) = \\
& \quad \frac{1}{2} \int_{-t-\Delta t}^{-t} \int_{-t-\Delta t}^{-t}
    \mathcal{U}_s(-t-\Delta t, \sigma[\boldsymbol{\tau}',s_1',s_2'], t) \rho_s(t,t'; \boldsymbol{\tau}) B(s_1',s_2')\,\mathrm{d}s_1' \,\mathrm{d}s_2' + {} \\
& \qquad \frac{1}{2} \int_{-t}^{t'} \int_{-t}^{t'} \int_{-t-\Delta t}^{-t} \int_{-t-\Delta t}^{-t}
    \mathcal{U}_s(-t-\Delta t, \sigma[\boldsymbol{\tau}', s_1', s_2'], t) \rho_s(t,t'; \sigma[\boldsymbol{\tau},s_1, s_2]) B(s_1',s_1) B(s_2',s_2)\,\mathrm{d}s_1' \,\mathrm{d}s_2' \,\mathrm{d}s_1 \,\mathrm{d}s_2.
\end{aligned}
\end{equation}
Thus, one can prove that
\begin{displaymath}
\mathcal{U}_s(-t-\Delta t, \boldsymbol{\tau}',t) \rho_s(t,t';\boldsymbol{\tau}) + \rho_{s,l}^{(1)}(t,t',\Delta t; \boldsymbol{\tau}', \boldsymbol{\tau}) + \rho_{s,l}^{(2)}(t,t',\Delta t; \boldsymbol{\tau}', \boldsymbol{\tau})
\end{displaymath}
is a third-order approximation of $\rho_s(t+\Delta t, t'; \sigma[\boldsymbol{\tau}', \boldsymbol{\tau}])$.

\paragraph{Step 2: Construction of a new second-order scheme}
Based on the approximation \eqref{eq:2nd_order_approx}, developing a second-order scheme requires discretization of $\rho_{s,l}^{(1)}(n\Delta t, n'\Delta t, \Delta t; \boldsymbol{\tau}', \boldsymbol{\tau})$ and $\rho_{s,l}^{(2)}(n\Delta t, n'\Delta t, \Delta t; \boldsymbol{\tau}', \boldsymbol{\tau})$ for $\boldsymbol{\tau}$ given by \eqref{eq:tau_j} and $\boldsymbol{\tau}' = (-(n+1/2)\Delta t, \cdots, -(n+1/2)\Delta t)$, and the desired error is $O(\Delta t^3)$ for both quantities.
For $\rho_{s,l}^{(1)}(n\Delta t, n'\Delta t, \Delta t; \boldsymbol{\tau}', \boldsymbol{\tau})$, we can use the following approximation:
\begin{displaymath}
\rho_{s,l}^{(1)}(n\Delta t, n'\Delta t, \Delta t; \boldsymbol{\tau}', \boldsymbol{\tau}) = \left\{ \begin{array}{@{}ll}
  Q_{s,l}^{(1)}(n,n';0,\boldsymbol{j}), & \text{if } \boldsymbol{\tau}' \text{ is an empty vector}, \\
  Q_{s,l}^{(1)}(n,n';r,\boldsymbol{j}) + R_{s,l}^{(1)}(n,n'; r,\boldsymbol{j}), & \text{if } \boldsymbol{\tau}' \in \mathbb{R}^r,
\end{array} \right.
\end{displaymath}
where $Q_{s,l}^{(1)}(n,n';r,\boldsymbol{j})$ is the second-order approximation defined in \eqref{eq:Q1}, and
\begin{equation} \label{eq:R1}
R_{s,l}^{(1)}(n,n';r,\boldsymbol{j}) \approx \frac{\Delta t^3}{8} \sum_{\ell \in \{1-,\cdots,n-,1+,\cdots,n'+\}} \e^{-\ii H_s \Delta t/2} [[H_s, W_s], (-\ii W_s)^r] \e^{-\ii H_s \Delta t/2} \rho_{s,n,n'}^{\Delta t}(\boldsymbol{j} + \boldsymbol{e}_{\ell} \big) \Bcal{(n+1)-}{\ell}
\end{equation}
is the correction term to achieve one higher order.
Note that when $r = 0$, the commutator $[[H_s, W_s], (-\ii W_s)^r]$ is zero.
Therefore, the two cases in \eqref{eq:R1} are consistent.
The reason why an extra term is needed for a nonempty $\boldsymbol{\tau}$ lies in the discontinuity of $\mathcal{U}_s(-t-\Delta t, \sigma[\boldsymbol{\tau}', s'], -t)$ with respect $s'$ at $s' = -(n+1/2)\Delta t$.
The other term $\rho_{s,l}^{(2)}$ can be simply approximated by the midpoint method:
\begin{displaymath}
\begin{aligned}
\rho_{s,l}^{(2)}(n\Delta t, n'\Delta t, \Delta t; \boldsymbol{\tau}', \boldsymbol{\tau}) & \approx
\frac{\Delta t^2}{2} \mathcal{V}_{s,r+2}^{\Delta t} \rho_{s,n,n'}^{\Delta t}(\boldsymbol{j}) \Bcal{(n+1)-}{(n+1)-} \\
& \quad + \frac{\Delta t^4}{2} \sum_{\substack{\ell_1 \in \{1-,\cdots,n-,1+,\cdots,n'+\} \\ \ell_2 \in \{1-,\cdots,n-,1+,\cdots,n'+\}}} \mathcal{V}_{s,r+2}^{\Delta t} \rho_{s,n,n'}^{\Delta t}\big(\boldsymbol{j} + \boldsymbol{e}_{\ell_1} + \boldsymbol{e}_{\ell_2} \big) \Bcal{(n+1)-}{\ell_1} \Bcal{(n+1)-}{\ell_2}.
\end{aligned}
\end{displaymath}

In the numerical scheme, we again use $\Sigma(j_{(n+1)-}, j_{n-}, \cdots, j_{1-}; j_{1+}, \cdots, j_{n+})$ and $\Sigma(j_{n-}, \cdots, j_{1-}; j_{1+}, \cdots, j_{n+})$ to approximate $\rho_{s,n+1,n}^{\Delta t}$ and $\rho_{s,n,n}^{\Delta t}$.
The analysis above inspires the following construction of the left extension in the numerical scheme:
\begin{equation} \label{eq:Sigma_ext}
\begin{aligned}
& \Sigma(j_{(n+1)-}, j_{n-}, \cdots, j_{1-}; j_{1+}, \cdots, j_{n+}) \\
={} &
\left( \mathcal{V}_{s,j_{(n+1)-}}^{\Delta t} + \frac{\Delta t^2}{2} \mathcal{V}_{s,2+j_{(n+1)-}}^{\Delta t} \Bcal{(n+1)-}{(n+1)-} \right) \Sigma(j_{n-}, \cdots, j_{1-}; j_{1+}, \cdots, j_{n+}) \\
& + \Delta t^2 \sum_{\ell \in \{1-,\cdots,n-,1+,\cdots,n+\}} \left(\mathcal{V}_{s,1+j_{(n+1)-}}^{\Delta t} + \Delta t \mathcal{W}_{s,1+j_{(n+1)-}}^{\Delta t} \right) \Sigma(\hat{j}^{\ell}_{n-}, \cdots, \hat{j}^{\ell}_{1-}; \hat{j}^{\ell}_{1+}, \cdots, \hat{j}^{\ell}_{n+}) \Bcal{(n+1)-}{\ell} \\
& + \frac{\Delta t^4}{2} \sum_{\substack{\ell_1 \in \{1-,\cdots,n-,1+,\cdots,n+\}\\\ell_2 \in \{1-,\cdots,n-,1+,\cdots,n+\}}} \mathcal{V}_{s,2+j_{(n+1)-}}^{\Delta t} \Sigma(\hat{j}^{\ell_1 \ell_2}_{n-}, \cdots, \hat{j}^{\ell_1 \ell_2}_{1-}; \hat{j}^{\ell_1 \ell_2}_{1+}, \cdots, \hat{j}^{\ell_1 \ell_2}_{n+}) \Bcal{(n+1)-}{\ell_1}\Bcal{(n+1)-}{\ell_2},
\end{aligned}
\end{equation}
where 
\begin{displaymath}
\mathcal{W}_{s,r}^{\Delta t} = \frac{1}{8} \e^{-\ii H_s \Delta t/2} [[H_s, W_s], (-\ii W_s)^{r-1}] \e^{-\ii H_s \Delta t/2},
\end{displaymath}
and the definitions of $\hat{j}_r$ and $\hat{j}_r^{\ell_1 \ell_2}$ can be found in \eqref{eq:hatj}\eqref{eq:jr_l1l2}.
The scheme for the right extension is similar and will be omitted.

\paragraph{Step 3: Convergence order of the new scheme}
This step is generally standard.
We first conclude that the local truncation error has order $O(\Delta t^3)$ by construction, and then use the recursion inequality of the error to show the second-order accuracy of $\Sigma$'s.

\paragraph{Step 4: Difference between two numerical schemes}
As in the proof of \Cref{lem:Sigma_order}, we can also expand $\Sigma(0,\cdots,0; 0,\cdots,0)$ by its recurrence formula, and the result is a linear combination of terms in the following form:
\begin{equation} \label{eq:Z_prod}
\mathcal{Z}_{s,j_{n-}}^{\Delta t} \cdots \mathcal{Z}_{s,j_{1-}}^{\Delta t} \rho_s(0) (\mathcal{Z}_{s,j_{1+}}^{\Delta t})^{\dagger} \cdots (\mathcal{Z}_{s,j_{n+}}^{\Delta t})^{\dagger},
\end{equation}
where $\mathcal{Z}$ can be either $\mathcal{V}$ or $\mathcal{W}$.
Here, some terms do not appear in the expansion of $\rho_{s,n}$ defined by \eqref{eq:2nd_order_rho}, and they can be categorized into two types:
\begin{itemize}
\item Type 1: All $\mathcal{Z}$'s are chosen as $\mathcal{V}$, and $\max \boldsymbol{j} \geqslant 3$.
\item Type 2: At least one of the $\mathcal{Z}$'s is chosen as $\mathcal{W}$.
\end{itemize}
For Type-1 terms, the coefficient of \eqref{eq:Z_prod} can be bounded by $\Delta t^{|\boldsymbol{j}|} (|\boldsymbol{j}|-1)!! \bar{B}^{|\boldsymbol{j}|/2}$.
Following the proof of \Cref{lem:LambdaSigma}, one can show that the sum of all Type-1 terms can be bounded by $O(\Delta t^2)$, and the main difference is that the condition $\max \boldsymbol{j} \geqslant 2$ becomes $\max \boldsymbol{j} \geqslant 3$.

For Type-2 terms, the coefficient of \eqref{eq:Z_prod} can be bounded by $\Delta t^{|\boldsymbol{j}| + N_{\mathcal{W}}} (|\boldsymbol{j}|-1)!! \bar{B}^{|\boldsymbol{j}|/2}$, where $N_{\mathcal{W}}$ is the number of $\mathcal{W}$'s in the product.
For these operators $\mathcal{W}_{s,j_k}^{\Delta t}$, we let $m$ be the sum of all these $j_k$'s.
Then $m$ must be greater than or equal to $2N_{\mathcal{W}}$ since each $\mathcal{W}_{s,j_k}^{\Delta t}$ is nonzero only when $j_k \geqslant 2$.
Comparing the sums of indices on both sides of \eqref{eq:Sigma_ext}, one can find that the right-hand side can have at most 4 more $W_s$'s than the left-hand side, but when $\mathcal{W}$ appears, the right-hand side can have only 2 more.
This implies that in the expansion of $\Sigma(0,\cdots,0; 0,\cdots,0)$, for a given $N_{\mathcal{W}}$, the maximum value of $m$ is $4(2n-N_{\mathcal{W}}) + 2N_{\mathcal{W}} = 8n-2N_{\mathcal{W}}$.
Using $m'$ to denote the sum of rest of $j_k$'s, we can then bound the sum of all Type-2 terms by
\begin{displaymath}
\begin{aligned}
\|\rho_s(0)\| \sum_{N_{\mathcal{W}}=1}^{2n} \sum_{m=2N_{\mathcal{W}}}^{8n-2N_{\mathcal{W}}} & \sum_{\substack{m'=0\\ m'+m \text{ even}}}^{8n-2N_{\mathcal{W}}-m} \binom{2n}{N_{\mathcal{W}}} \binom{m-N_{\mathcal{W}}-1}{N_{\mathcal{W}}-1} \binom{2n-N_{\mathcal{W}}+m'-1}{m'} \\
& \times (\Delta t \|H_s\|)^{N_{\mathcal{W}}} (\Delta t\|W_s\|)^{m+m'} (m+m'-1)!! \bar{B}^{(m+m')/2},
\end{aligned}
\end{displaymath}
which can also be shown to have order $O(\Delta t^2)$.

With these results, we conclude that the difference between $\Sigma(0,\cdots,0; 0,\cdots,0)$ and $\rho_{s,n}$ given in \eqref{eq:2nd_order_rho} is $O(\Delta t^2)$, thus demonstrating the second-order accuracy of $\rho_{s,n}$.

\subsection{Proof of \Cref{thm:2nd_dyson_oqs}}
\label{sec:proof_2nd_dyson_oqs}
\begin{proof}
For a given $\boldsymbol{j} = (j_{n-}, \cdots, j_{1-}; j_{1+}, \cdots, j_{n+})$, let $|\boldsymbol{j}| = j_{n-} + \cdots + j_{1-} + j_{1+} + \cdots + j_{n+}$.
Then when $|\boldsymbol{j}|$ is even, each term in the sum \eqref{eq:2nd_order_rho} or \eqref{eq:rhohat} can be bounded by
\begin{displaymath}
\frac{(\|W_s\|^2 \Delta t^2 \bar{B})^{|\boldsymbol{j}|}/2}{2^{N_2}} \|\rho_s(0)\| (|\boldsymbol{j}|-1)!!,
\end{displaymath}
where $N_2$ is the number of ``2''s in the sequence $j_{n-}, \cdots, j_{1-}, j_{1+}, \cdots, j_{n+}$.
It is obvious that $N_1 + 2N_2 = |\boldsymbol{j}|$ for $N_1$ being the number of ``1''s in $j_{n-}, \cdots, j_{1-}, j_{1+}, \cdots, j_{n+}$.

As in the proof of \Cref{thm:2nd_dyson}, the difference between $\rho_{s,n}$ and $\hat{\rho}_{s,n}$ contains all terms with $N_2 \geqslant 2$.
Since we only need to consider terms with even $|\boldsymbol{j}|$, we let $N = |\boldsymbol{j}|/2$, so that the difference between $\rho_{s,n}$ and $\hat{\rho}_{s,n}$ can be bounded by
\begin{equation}
\label{eq:rho_diff}
\|\rho_{s,n} - \hat{\rho}_{s,n}\| \leqslant \sum_{N=2}^{2n} \sum_{N_2 = \max(2,2N-2n)}^N \binom{2n}{N_2} \binom{2n-N_2}{2N-2N_2}
\frac{(\|W_s\|^2 \Delta t^2 \bar{B})^N}{2^{N_2}} \|\rho_s(0)\| (2N-1)!!,
\end{equation}
where the lower bound of $N_2$ comes from the $N_1 = 2N-2N_2$ and $N_1 + N_2 \leqslant 2n$.
The two binomials have the same meanings as in \eqref{eq:Udiff}, except that $2n$ indices are considered when studying the approximation of density matrices.

To proceed, we need the following inequality:
\begin{displaymath}
\begin{aligned}
\frac{1}{2^{N_2}} \binom{2n}{N_2} \binom{2n-N_2}{2N-2N_2} &= \frac{(2n)!}{2^{N_2} N_2! (2n-2N+N_2)! (2N-2N_2)!} \\
& = \frac{1}{N_2! (2N-N_2)!} \cdot [(2n)(2n-1) \cdots (2n-2N+N_2+1)] \\
& \hspace{68pt} \cdot \frac{(2N-N_2)(2N-N_2-1)\cdots (2N-2N_2+1)}{2^{N_2}} \\
& \leqslant \frac{1}{N_2! (2N-N_2)!} \cdot (2n)^{2N-N_2} \cdot \frac{(2N)^{N_2}}{2^{N_2}} \\
& = \frac{(2n)^{2N-N_2} N^{N_2}}{N_2! (2N-N_2)!} = \binom{2N}{N_2} \frac{(2n)^{2N-N_2} N^{N_2}}{(2N)!},
\end{aligned}
\end{displaymath}
which can be plugged in to \eqref{eq:rho_diff} to obtain
\begin{displaymath}
\begin{aligned}
& \|\rho_{s,n} - \hat{\rho}_{s,n}\| \\
\leqslant{} & \sum_{N=2}^{2n} \sum_{N_2 = \max(2,2N-2n)}^N \binom{2N}{N_2} \frac{(2n)^{2N-N_2} N^{N_2}}{(2N)!} (\|W_s\|^2 \Delta t^2 \bar{B})^N \|\rho_s(0)\| (2N-1)!! \\
\leqslant{} & \sum_{N=2}^{2n} \sum_{N_2 = \max(2,2N-2n)}^N \binom{2N}{N_2} \frac{(2n)^{2N-2} N^2}{(2N)!!} (\|W_s\|^2 \Delta t^2 \bar{B})^N \|\rho_s(0)\|,
\end{aligned}
\end{displaymath}
where the second ``$\leqslant$'' is the result of $(2n)^{2N-N_2} N^{N_2} \leqslant (2n)^{2N-2} N^2$, due to $N \leqslant 2n$ and $N_2 \geqslant 2$.
We now use $T = n \Delta t$ to make further simplifications:
\begin{displaymath}
\begin{aligned}
\|\rho_{s,n} - \hat{\rho}_{s,n}\| & \leqslant \Delta t^2 \sum_{N=2}^{2n} \sum_{N_2 = 0}^{2N} \binom{2N}{N_2} \frac{(2T)^{2N-2} N^2}{(2N)!!} (\|W_s\|^2 \bar{B})^N \|\rho_s(0)\| \\
& = 4 \Delta t^2 \|\rho_s(0)\| \sum_{N=2}^{2n} \frac{(4T)^{2N-2} N^2}{(2N)!!} (\|W_s\|^2 \bar{B})^N \\
& \leqslant 4 \Delta t^2 \|\rho_s(0)\| \sum_{N=0}^{+\infty} \frac{(4T)^{2N-2} N^2}{(2N)!!} (\|W_s\|^2 \bar{B})^N \\
& = 2 \|\rho_s(0)\| \|W_s\|^2 \Delta t^2 \bar{B} \Delta t \e^{8 \|W_s\|^2 T^2 \bar{B}} [1 + 8 \|W_s\|^2 T^2 \bar{B}].
\end{aligned}
\end{displaymath}
\end{proof}

\subsection{Proof of \Cref{thm:2nd_order_iterative}}
\label{sec_proof_scheme}
\begin{proof}
%
We define the sets of distinct thin diagrams considered in the expansions of $\rho_{s,n}$, $\hat{\rho}_{s,n}$ and $\Tilde{\rho}_{s,n}$ as
\[
\mathcal{J}_{\mathrm{full}},\qquad \mathcal{J}_{\mathrm{hat}},\qquad \mathcal{J}_{\mathrm{it}},
\]
respectively. Each $D\in \mathcal{J}_{\mathrm{full}}$ is a distinct diagram built by the $2n$ segments on the time axis with or without markers ``$\circ$''/$``\circledcirc$'' and arc connections representing bath correlation functions consistent with the full second-order expansion defined in \eqref{eq:2nd_order_rho}.
For instance, when $n = 1$,
\def\fn{2nd_order_others}%
\begin{align*}
\mathcal{J}_{\mathrm{hat}} &= \Big\{ \dgm{\fn.5},\, \dgm{\fn.6},\, \dgm{\fn.7},\, \dgm{\fn.8}\Big\}, \\
\mathcal{J}_{\mathrm{it}} &= \Big\{ \dgm{\fn.5},\, \dgm{\fn.6},\, \dgm{\fn.7},\, \dgm{\fn.8},\, \dgm{\fn.9}\Big\}, \\
\mathcal{J}_{\mathrm{full}} &= \Big\{ \dgm{\fn.5},\, \dgm{\fn.6},\, \dgm{\fn.7},\, \dgm{\fn.8},\, \dgm{\fn.9},\, \dgm{\fn.10}\Big\}.
\end{align*}
Let $w_{\mathrm{full}}(D)$ be the scalar coefficient of $D$ induced by ``$\circledcirc$''s.
For instance, $w_{\mathrm{full}}(D) = 2$ for any diagram $D$ in the last three rows of \eqref{eq:hat_rho_2}.
Define $w_{\mathrm{hat}}(D)$ and $w_{\mathrm{it}}(D)$ analogously for $\hat{\rho}_{s,n}$ and $\Tilde{\rho}_{s,n}$. Then these three approximations of the density matrix can be written equivalently as
\begin{displaymath}
    \rho_{s,n} = \sum_{D\in \mathcal{J}_{\mathrm{full}}}w_{\mathrm{full}}(D)D,\quad \hat{\rho}_{s,n}=\sum_{D\in \mathcal{J}_{\mathrm{hat}}}w_{\mathrm{hat}}(D)D,\quad \Tilde{\rho}_{s,n} = \sum_{D\in \mathcal{J}_{\mathrm{it}}}w_{\mathrm{it}}(D)D,\quad
\end{displaymath}

We first claim that
\begin{equation}
\label{eq:strict_inclusion}
      \mathcal{J}_{\mathrm{hat}}\subset \mathcal{J}_{\mathrm{it}}\subset \mathcal{J}_{\mathrm{full}}.
\end{equation}
It is obvious that both $\mathcal{J}_{\mathrm{hat}}$ and $\mathcal{J}_{\mathrm{it}}$ are subsets of $\mathcal{J}_{\mathrm{full}}$, since $\mathcal{J}_{\mathrm{full}}$ has included all possible diagrams with at most two circles on each segment.
The reason why $\mathcal{J}_{\mathrm{it}}$ is only a subset of $\mathcal{J}_{\mathrm{full}}$ is that ``$\circledcirc$'' will not be added during extension once the bold diagram already includes a ``$\circledcirc$''.
Therefore, all diagrams in $\mathcal{J}_{\mathrm{full}} \backslash \mathcal{J}_{\mathrm{it}}$ must included at least two ``$\circledcirc$''s.
Despite this, $\mathcal{J}_{\mathrm{it}}$ may still include thin diagrams with two ``$\circledcirc$''s as explain in the text after  \eqref{eq_2nd_order_eg2_diagram}, indicating that $\mathcal{J}_{\mathrm{hat}}$, allowing only one ``$\circledcirc$'' in thin diagrams, is a proper subset of $\mathcal{J}_{\mathrm{it}}$.

Next, we will show
\begin{equation}
\label{eq:matched_coeff}
    \begin{cases}
w_{\mathrm{it}}(D)=w_{\mathrm{full}}(D)&\text{for all }D\in \mathcal{J}_{\mathrm{it}},\\
    w_{\mathrm{hat}}(D)=w_{\mathrm{it}}(D)=w_{\mathrm{full}}(D)&\text{for all }D\in \mathcal{J}_{\mathrm{hat}}.
    \end{cases}
\end{equation}
As we introduced in \Cref{subsec_2nd_order_scheme}, for any $D\in \mathcal{J}_{\mathrm{full}}$, its coefficient $w_{\mathrm{full}}(D)$ is $2^{m_1 + m_2}$ if it contains $m_1$ pairs of ``$\circledcirc$''s interconnected by a double arc and $m_2$ ``$\circledcirc$''s connecting to two different nodes.
If $D \in \mathcal{J}_{\mathrm{hat}}$ and contains one ``$\circledcirc$'', by the same argument, the corresponding diagram should be doubled.

It remains to prove the coefficient $w_{\mathrm{it}}(D)$ obtained by the iterative process also follows this rule.
Our algorithm does not allow two ``$\circledcirc$''s to be connected by a double arc, and whenever a ``$\circledcirc$'' is added, we introduce a factor $2$ unless it connects to itself.
Therefore, for all $D \in \mathcal{J}_{\mathrm{it}}$, the coefficient $w_{\mathrm{it}}(D)$ always equals $w_{\mathrm{full}}(D)$, which also equals $w_{\mathrm{hat}}(D)$ if it contains only one ``$\circledcirc$''.

We now prove the second-order accuracy of our algorithm.
Using the conclusions \eqref{eq:strict_inclusion} and \eqref{eq:matched_coeff}, one has
\begin{displaymath}
    \rho_{s,n}-\Tilde{\rho}_{s,n} = \sum_{D\in \mathcal{J}_{\mathrm{full}}\setminus \mathcal{J}_{\mathrm{it}}} w_{\mathrm{full}}(D)\,D,\qquad \rho_{s,n}-\hat{\rho}_{s,n}
=\sum_{D\in \mathcal{J}_{\mathrm{full}}\setminus \mathcal{J}_{\mathrm{hat}}} w_{\mathrm{full}}(D)\,D,
\end{displaymath}
and $\mathcal{J}_{\mathrm{full}}\setminus \mathcal{J}_{\mathrm{it}}\subset \mathcal{J}_{\mathrm{full}}\setminus \mathcal{J}_{\mathrm{hat}}$ shows that the first difference is a \emph{subseries} of the second, and the upper bound of each term in the sum is estimated in the proof of \cref{thm:2nd_dyson_oqs}.
As a result, there exists a constant $C_1$ depending on $\|\rho_s(0)\|$, $\|W_s\|$, $\bar{B}$ and $n\dt$ such that,
\begin{equation}
    \|\rho_{s,n}-\Tilde{\rho}_{s,n}\| \leqslant\sum_{D\in \mathcal{J}_{\mathrm{full}}\setminus \mathcal{J}_{\mathrm{it}}} w_{\mathrm{full}}(D)\,\|D\| \leqslant\sum_{D\in \mathcal{J}_{\mathrm{full}}\setminus \mathcal{J}_{\mathrm{hat}}} w_{\mathrm{full}}(D)\|D\|\leqslant C_1\dt^2,
\end{equation}
According to \cref{thm:2nd_order}, there exists a constant $C_2$ depending on $\|\rho_s(0)\|$, $\|W_s\|$, $\bar{B}$, $\bar{B}'$, $\bar{B}''$ and $n\dt$ such that,
\begin{equation}
    \|\rho_{s,n}-\rho_s(n\dt)\|\leqslant C_2\dt^2.
\end{equation}
Therefore 
\begin{equation}
    \|\Tilde{\rho}_{s,n}-\rho_s(n\dt)\|\leqslant\|\rho_{s,n}-\Tilde{\rho}_{s,n}\| +\|\rho_{s,n}-\rho_s(n\dt)\| \leqslant(C_1+C_2)\dt^2.
\end{equation}
\end{proof}

\section{Numerical results}
\label{sec_numerical_result}
In this section, we will carry out numerical tests to validate our method.
The tests will focus on the numerical accuracy, computational time, and memory cost of our numerical schemes. 
To begin with, we will introduce our settings for the thermal bath, which defines the bath correlation function $B(\cdot, \cdot)$

\subsection{Bosonic bath and Ohmic spectral density}
The FRODS method is designed based on Wick's theorem, where the bath influence functional can be written in the form of \cref{eq:Lb}.
Such a form can be achieved by modeling the environment by a set of harmonic oscillators with frequencies $\omega_l$, whose Hamiltonian is
\begin{equation}
    H_b = \sum_{l=1}^L \frac{1}{2} \left(\hat{p}_l^2 + \omega_l^2\hat{q}_l^2\right),
\end{equation}
where $L$ is the number of quantum harmonic oscillators in the bath, and the operators $\hat{p}_l$ and $\hat{q}_l$ are the momentum and position operators for the $l$th harmonic oscillator, respectively.
This assumption is physically motivated as many realistic environments, such as phonons in a solid and photons in a cavity.
The interaction between the system and the bath is given by
\begin{equation}
    W = W_s \otimes W_b = W_s\otimes \left( \sum_{l=1}^L c_l \hat{q}_l\right),
\end{equation}
and $c_l$ is the coupling intensity between the system and the $l$th harmonic oscillator.
In our experiments, we choose the frequencies $\omega_j$ and the coupling coefficient $c_j$ by the Ohmic spectral density \cite{makri1999linear}
\begin{equation}
\begin{split}
    &\omega_j = -\omega_c \ln \left(
        1 - \frac{j}{L}(1-\exp(-\omega_{\max}/\omega_c))
    \right), \\
    & c_j = \omega_j \sqrt{
        \frac{\xi\omega_c}{L}
        (1-\exp(-\omega_{\max}/\omega_c))
    },\qquad
    j = 1,\cdots,L,
\end{split}
\end{equation}
where the frequencies $\omega_j$ are distributed in $[0,\omega_{\max}]$ following the Poisson distribution, and $\xi$ denotes the Kondo parameter indicating the coupling intensity between the system and the bath.

The two-point correlation function $B(\tau_1,\tau_2)$ mentioned in \cref{sec_first_order} is given analytically by
\begin{equation}
    B(\tau_1,\tau_2) = 
    \frac{1}{2}\sum_j \frac{c_j^2}{\omega_j}
    \left(
        \coth\left(\frac{\beta \omega_j}{2}\right)
        \cos(\omega_j \Delta \tau)
        -\ii \sin(\omega_j \Delta \tau)
    \right)
\end{equation}
with $\Delta \tau  = | \tau_1 | - | \tau_2 |$.
The parameter $\beta$ is the inverse temperature mentioned in \cref{subsec_open_quantym_system}.
This expression shows that the derivatives of $B(\cdot,\cdot)$ are bounded, meaning that the conditions in \cref{thm:2nd_order_iterative} are satisfied in our test cases.

\subsection{Convergence Tests}
In this section, we carried out the convergence test of the FRODS in the setting of spin-boson model, where the system of interest is a single spin with two eigenstates, also known as a two-level system.
The system Hamiltonian is given by
\begin{equation}
    H_s = \epsilon\hat{\sigma}_z + \Delta \hat{\sigma}_x
\end{equation}
where $\hat{\sigma}_x,\hat{\sigma}_z$ are Pauli matrices. The parameter $\epsilon$ is the energy difference of two states and the parameter $\Delta$ denotes the frequency of spin flipping.
The system interaction operator $W_s$ is set to $\hat{\sigma}_z$ in our experiments. The initial bath is in a thermal state

\begin{equation}
\rho_b = Z^{-1}\operatorname{exp}(-\beta H_b)
    \label{eq:thermal_state}
\end{equation}
where $Z$ is a normalizing factor satisfying $\operatorname{tr}(\rho_b)=1$.
The experiments will based on the following physical parameters.
\begin{equation}
    \epsilon = 0,\quad \Delta = 1, \quad \beta = 5, \quad \omega_c = 2.5. 
\end{equation}
The set of parameters is consistent with the one in \cite{wang2022differential}.
Following most literature for the spin-boson model \cite{forsythe1999dissipative,tang2015extended,cai2020inchworm,wang2022differential}, we will focus on the observable $O_s = \hat{\sigma}_z$, whose expected value is given by:
\begin{displaymath}
\langle O_s(t) \rangle = \tr_s(\rho_s(t) O_s).
\end{displaymath}
Since a spin is a two-level system ($M = 2$), the i-QuAPI method actually has lower memory and computational costs.
The numerical experiments in this section are only for the purpose of validation rather than showing the superiority of our method.
\begin{figure}[!ht]
    \centering
    \begin{subfigure}[b]{0.45\textwidth} 
        \centering
\includegraphics[width=1.\textwidth]{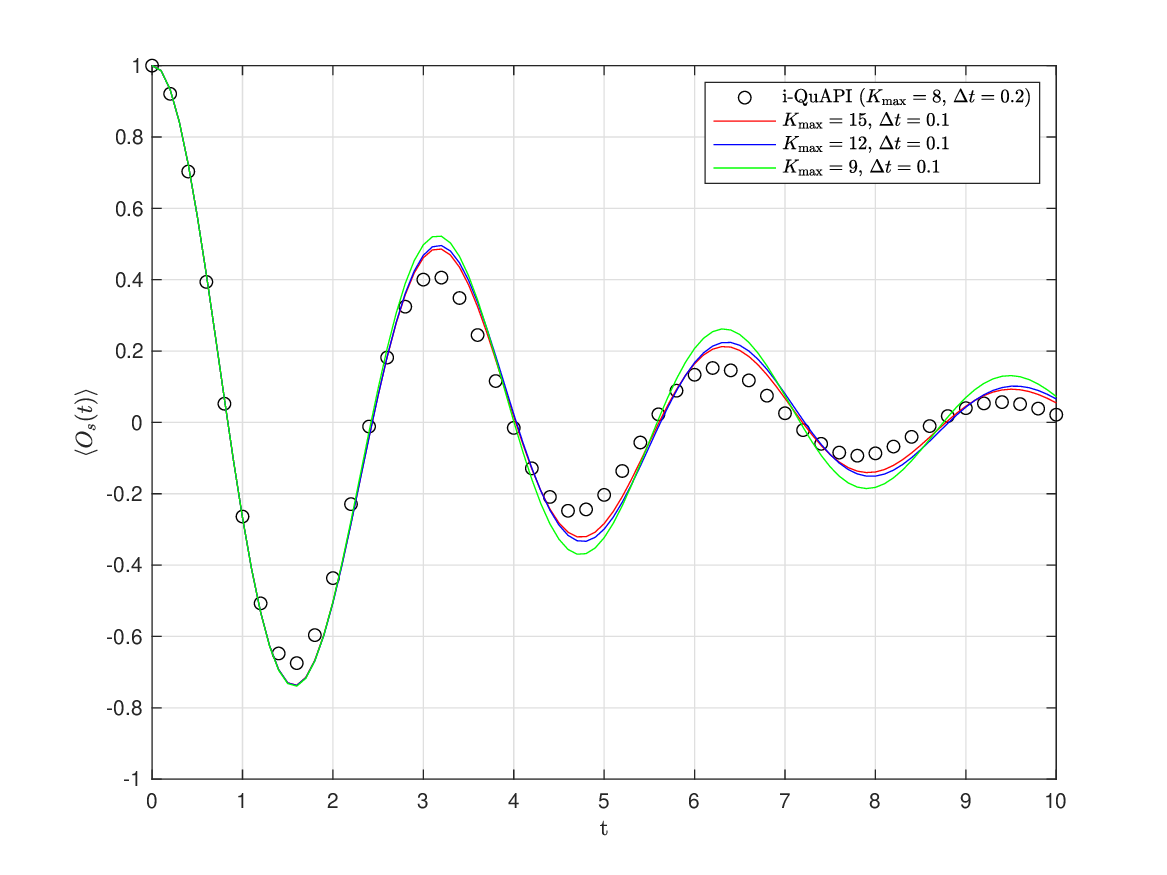}
        \caption{$\xi = 0.2$}
        \label{fig:conv_Kmax_1storder_xi02}
    \end{subfigure}
    \begin{subfigure}[b]{0.45\textwidth} 
        \centering
\includegraphics[width=1.\textwidth]{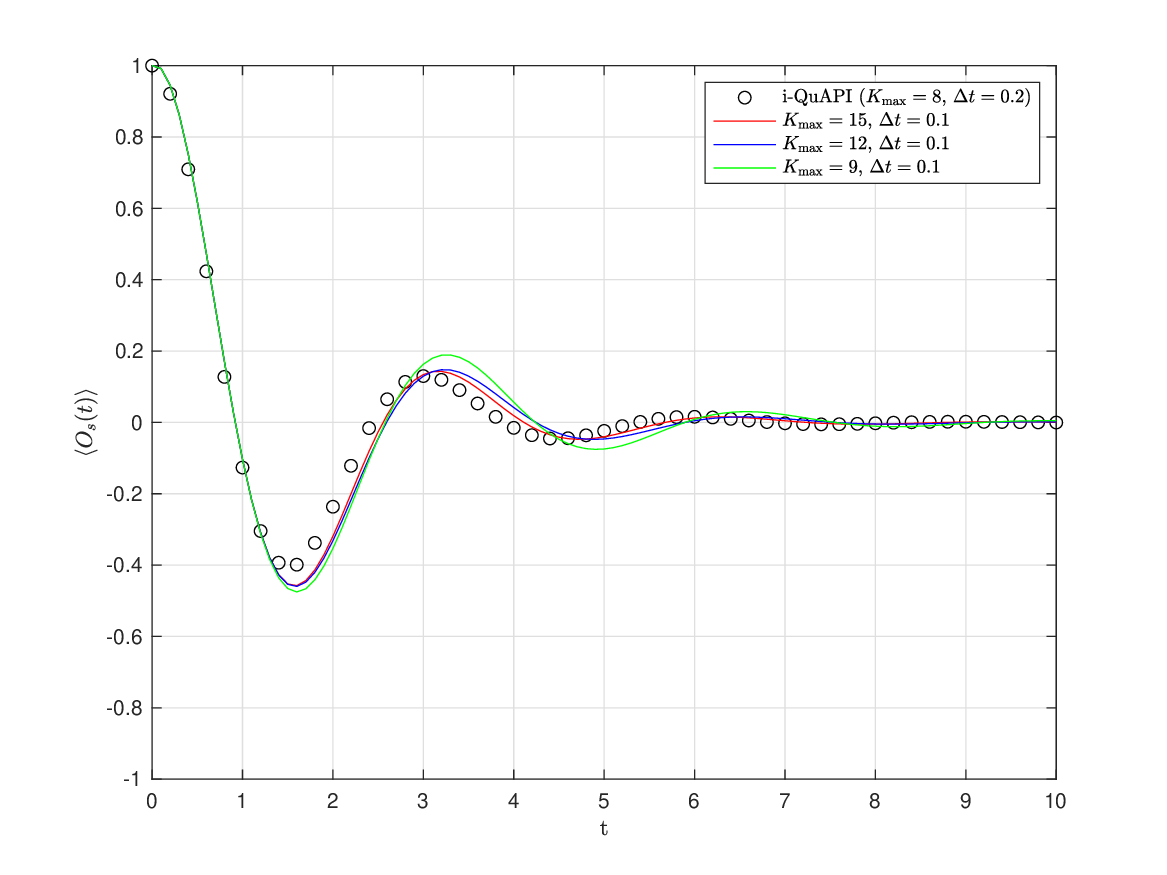}
        \caption{ $\xi = 0.4$}
        \label{fig:conv_Kmax_1storder_xi04}
    \end{subfigure}
    \\
    \begin{subfigure}[b]{0.45\textwidth} 
        \centering
\includegraphics[width=1.\textwidth]{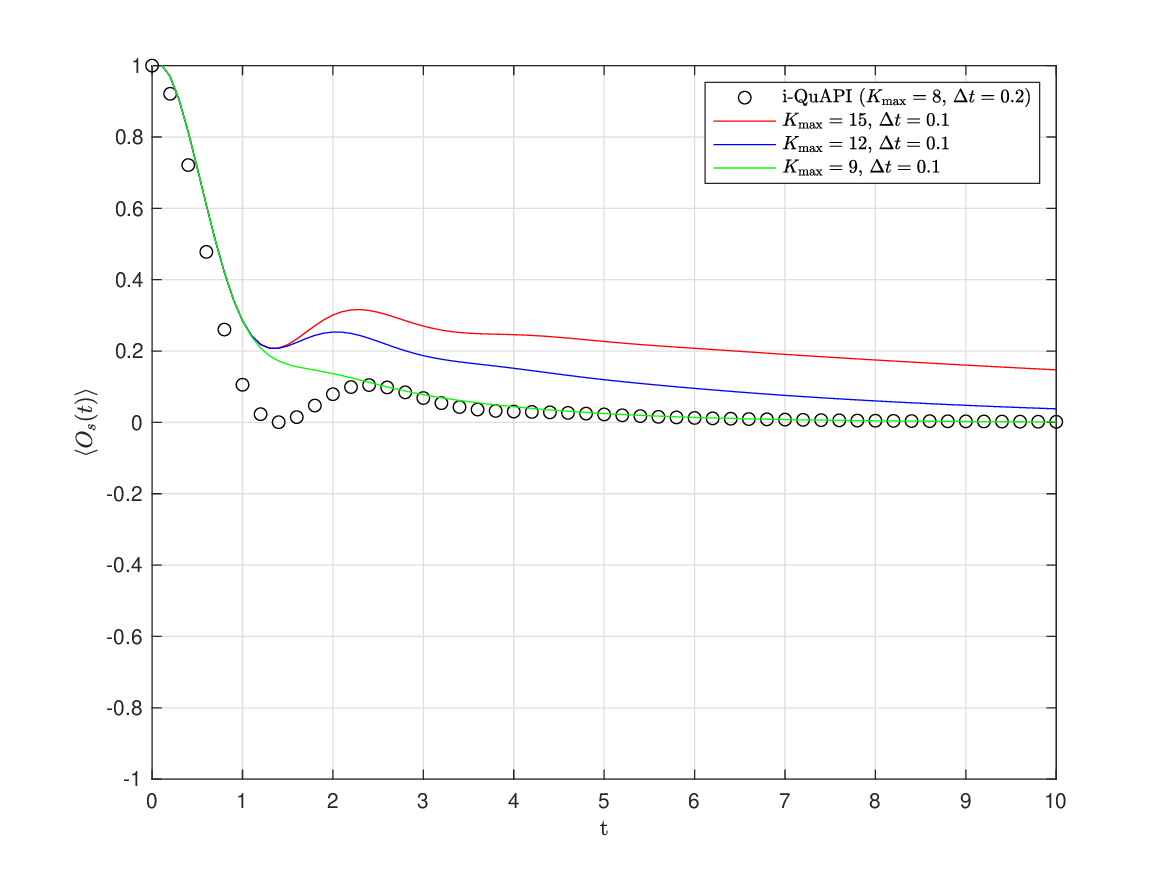}
        \caption{$\xi = 0.8$}
        \label{fig:conv_Kmax_1storder_xi08}
    \end{subfigure}
    \begin{subfigure}[b]{0.45\textwidth} 
        \centering
\includegraphics[width=1.\textwidth]{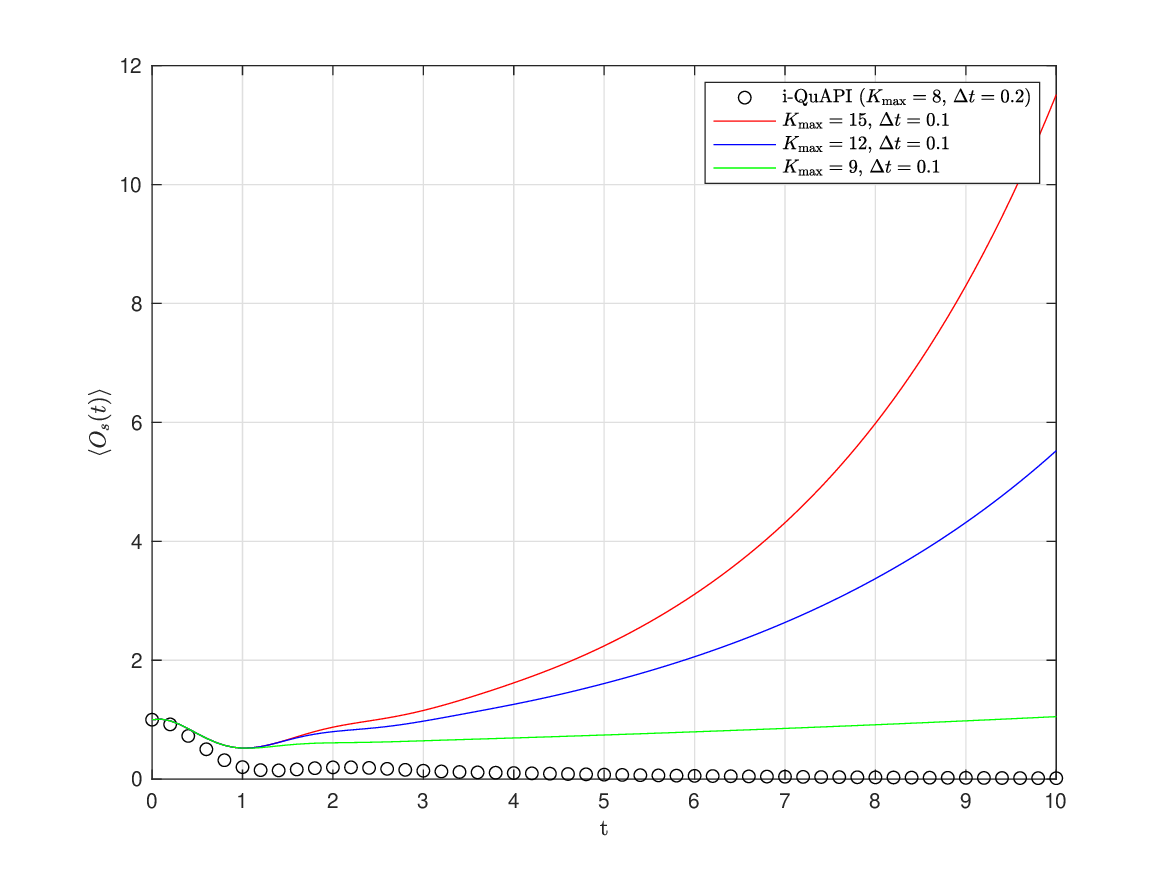}
        \caption{$\xi = 1.0$}
        \label{fig:conv_Kmax_1storder_xi10}
    \end{subfigure}
    \caption{Convergence test of first-order scheme on $K_{\max}$ for $\beta=5$.}
    \label{fig:conv_Kmax_1storder}
\end{figure}

\begin{figure}[!ht]
    \centering
    \begin{subfigure}[b]{0.45\textwidth} 
        \centering
\includegraphics[width=1\textwidth]{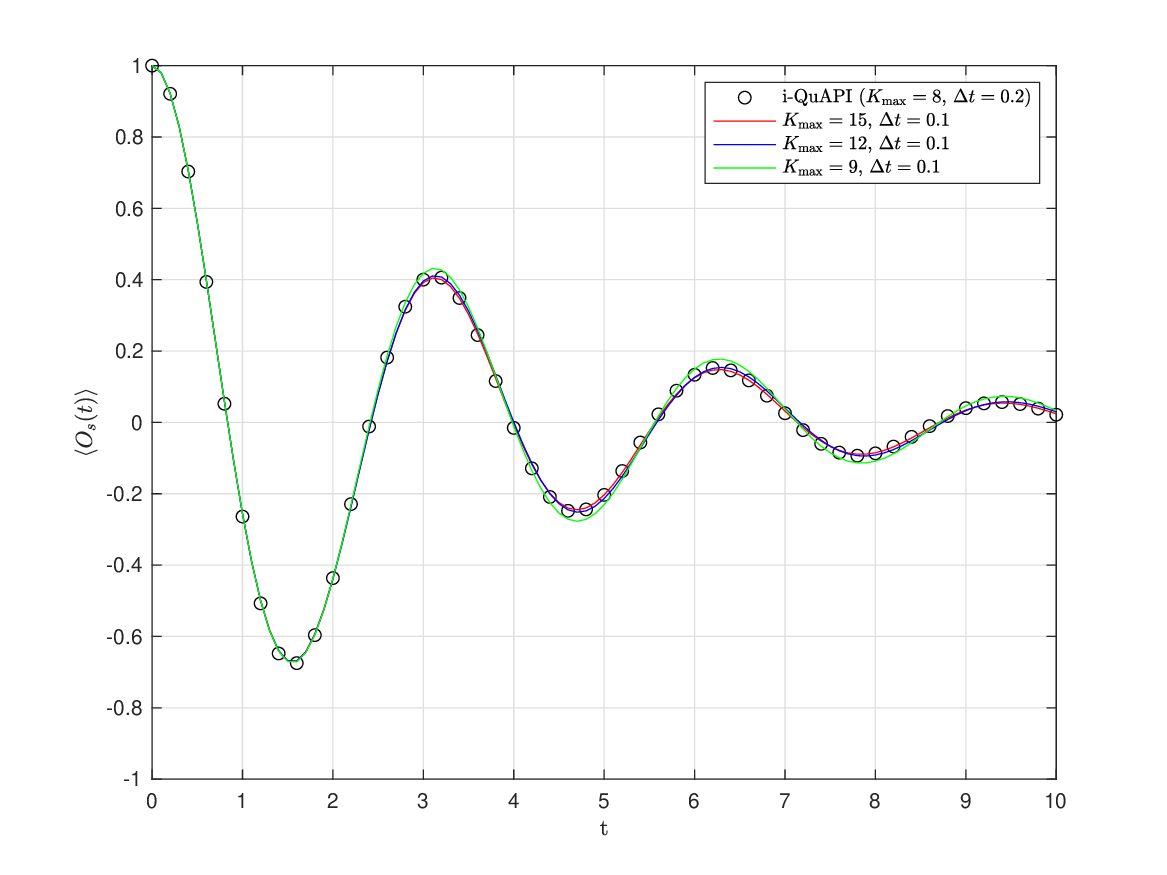}
        \caption{$\xi = 0.2$}
        \label{fig:conv_Kmax_2ndorder_xi02}
    \end{subfigure}
    \begin{subfigure}[b]{0.45\textwidth} 
        \centering
\includegraphics[width=1\textwidth]{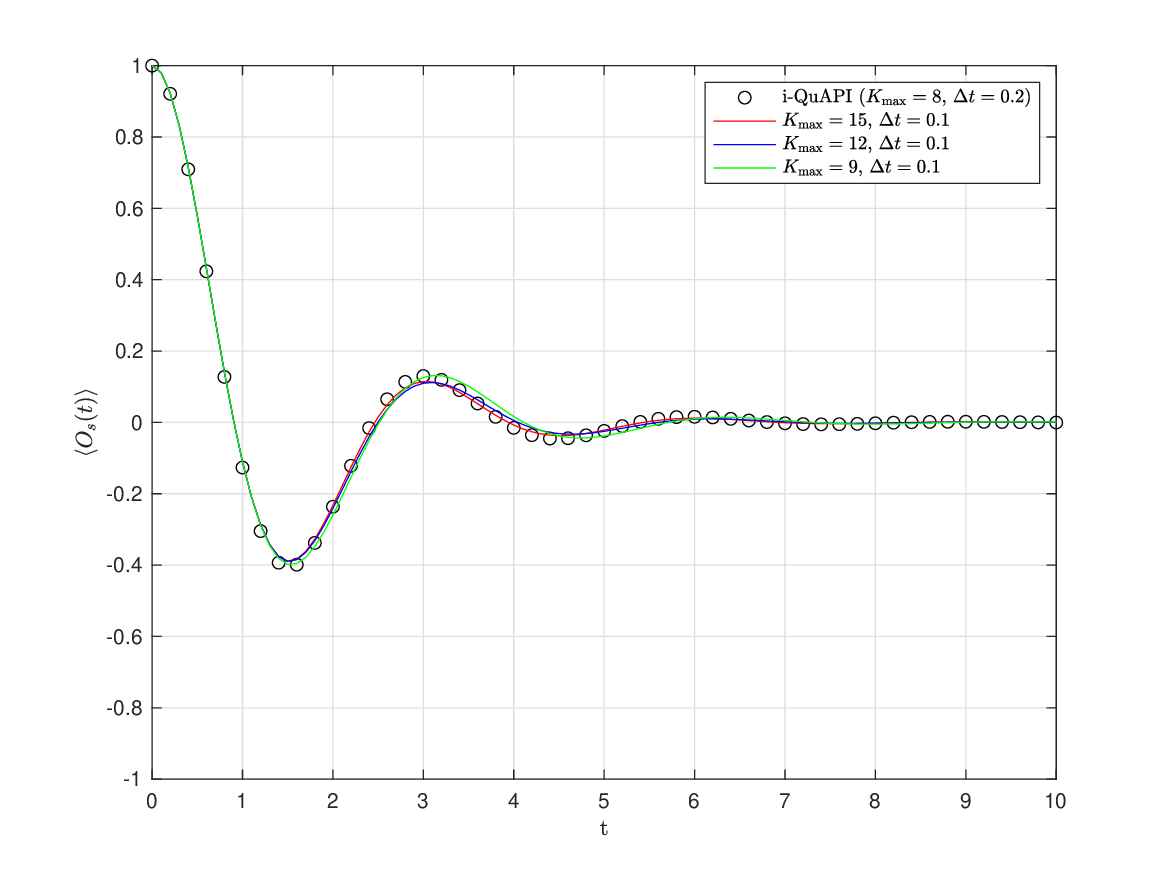}
        \caption{$\xi = 0.4$}
        \label{fig:conv_Kmax_2ndorder_xi04}
    \end{subfigure}
    \\
    \begin{subfigure}[b]{0.45\textwidth} 
        \centering
\includegraphics[width=1\textwidth]{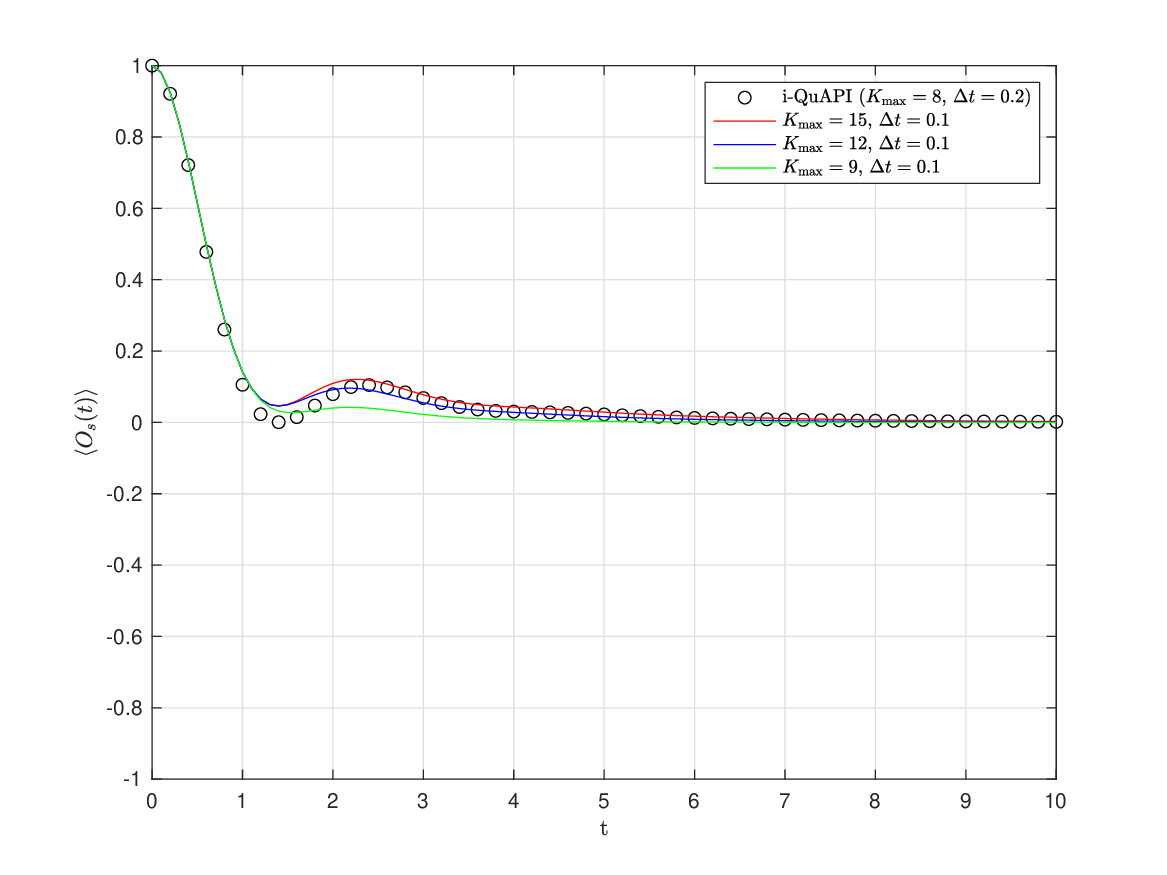}
        \caption{$\xi = 0.8$}
        \label{fig:conv_Kmax_2ndorder_xi08}
    \end{subfigure}
    \begin{subfigure}[b]{0.45\textwidth} 
        \centering
\includegraphics[width=1\textwidth]{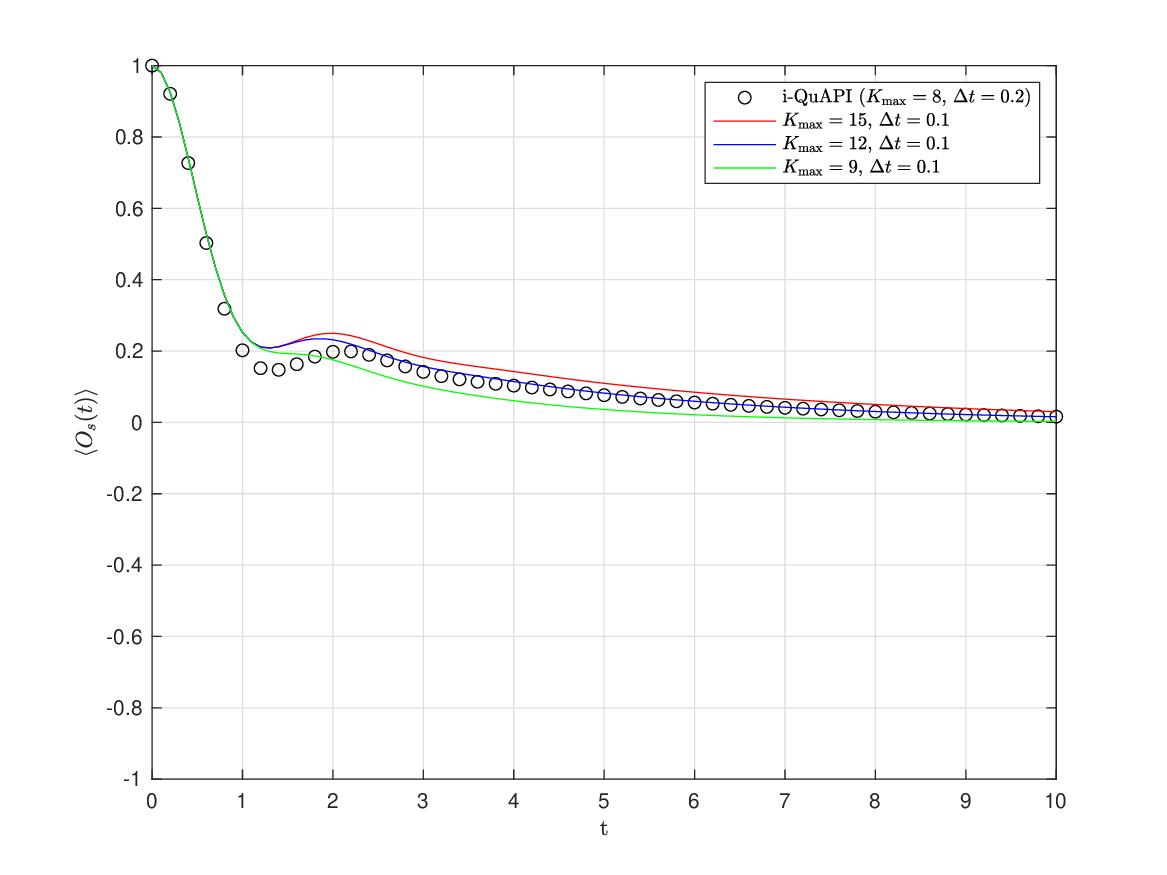}
        \caption{$\xi = 1.0$}
        \label{fig:conv_Kmax_2ndorder_xi10}
    \end{subfigure}

    \caption{Convergence test of second-order scheme on $K_{\max}$ for $\beta=5$.}
    \label{fig:conv_Kmax_2ndorder}
\end{figure}
\begin{figure}[!ht]
    \centering
    \begin{subfigure}[b]{0.45\textwidth} 
        \centering
        \includegraphics[width = 1\textwidth]{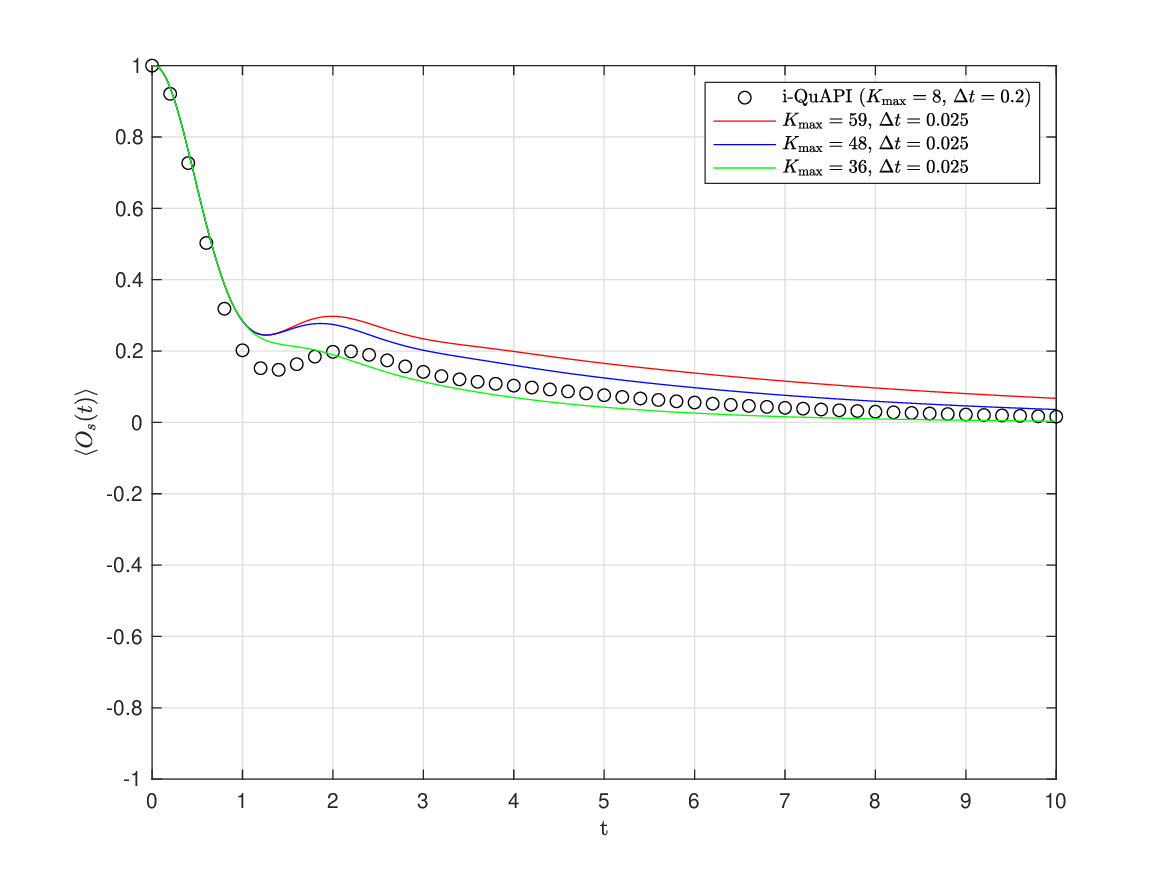}
        \caption{First-order scheme with $\dt = 0.025$}
        \label{fig:conv_kmax_dmax4_beta5_xi10_1storder_dt0025}
    \end{subfigure}
    \begin{subfigure}[b]{0.45\textwidth} 
        \centering
        \includegraphics[width = 1\textwidth]{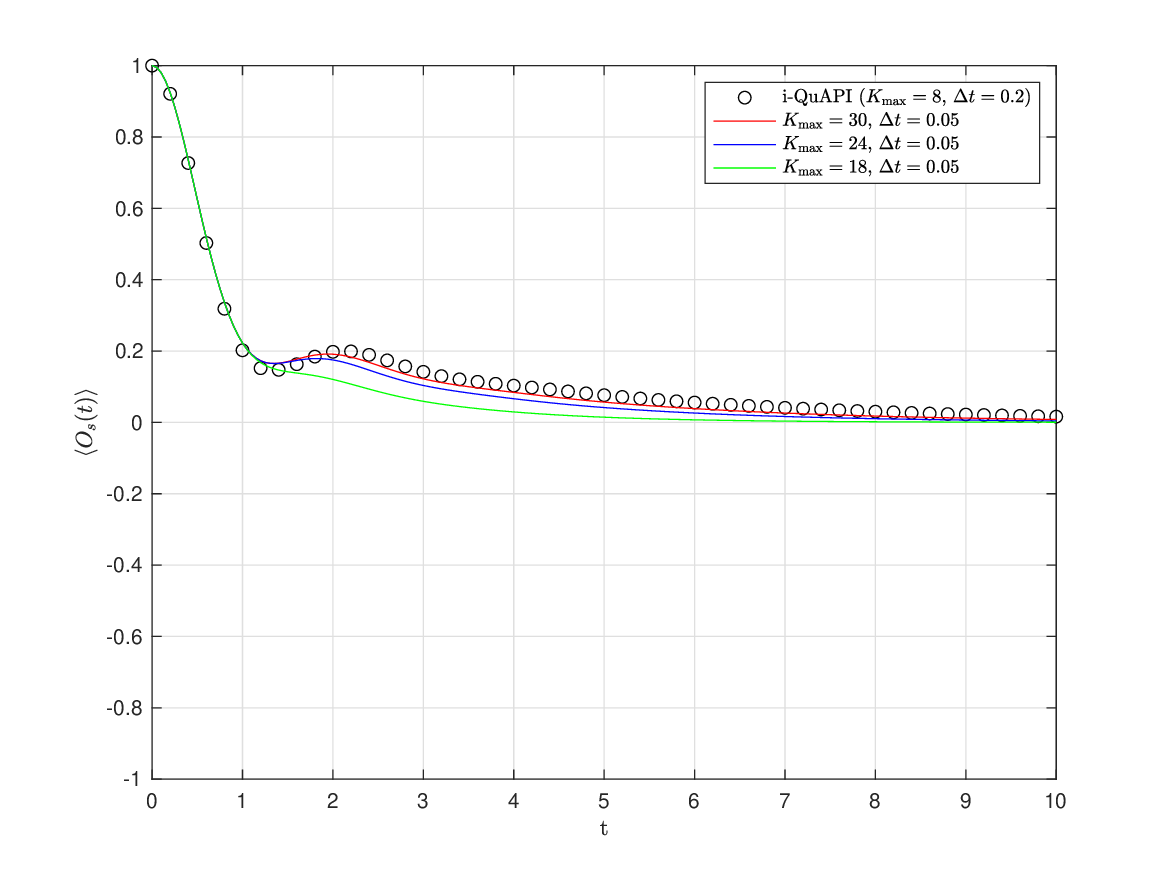}
        \caption{Second-order scheme with $\dt = 0.05$ }
        \label{fig:conv_kmax_dmax4_beta5_xi10_2ndorder_dt005}
    \end{subfigure}
        \caption{Convergence test on $K_{\max}$ for $\beta=5$, $D_{\max}=4$, and smaller time step size $\dt$.}
        \label{fig:conv_kmax_xi10}
\end{figure}

We first examine the convergence of our scheme with respect to the memory length $K_{\max}$. In this test, we fix the time step size $\Delta t = 0.1$ and the diagram truncation parameter $D_{\max} = 4$.
The memory length $K_{\max}$ is set to $9$, $12$, and $15$, corresponding to $T_{\max} = 0.9$, $1.2$, and $1.5$ in \eqref{eq:tilde_B}, respectively 
The results obtained from the first-order and second-order schemes under different Kondo parameter $\xi$ are presented in \cref{fig:conv_Kmax_1storder} and \cref{fig:conv_Kmax_2ndorder}, respectively.
As references, the results computed by i-QuAPI with time step $0.2$ and memory length $1.6$ are also provided.
As shown in \cref{fig:conv_Kmax_1storder_xi02,fig:conv_Kmax_1storder_xi04}, when $\xi$ is relatively small, the two curves for $K_{\max} = 12$ and $15$ are significantly closer compare with the distance between $K_{\max} = 9$ and $12$, implying convergence with respect to $K_{\max}$ in the first-order scheme.
However, the accuracy of temporal discretization is insufficient, reflected by the noticeable difference between our results and i-QuAPI results.
Worse still, for larger $\xi$ (\cref{fig:conv_Kmax_1storder_xi08,fig:conv_Kmax_1storder_xi10}), the first-order error even leads to seemingly divergent results with respect to $K_{\max}$, which is rectified in the second-order results plotted in \cref{fig:conv_Kmax_2ndorder}, indicating correct convergence behavior.
Nevertheless, a gap between the i-QuAPI result and our result still exists for $\xi = 0.8$ and $1.0$.
By our numerical analysis, the error bound grows with $\bar{B} = \max |B(\cdot,\cdot)|$, which is proportional to the Kondo parameter $\xi$.
Hence, a more accurate temporal discretization is required for stronger coupling between the system and the bath.
In particular, for $\xi = 1.0$, we decrease the time step to $\Delta t = 0.025$ for the first-order scheme and $\Delta t = 0.05$ for the second-order scheme, and increase $K_{\max}$ accordingly (see \cref{fig:conv_kmax_xi10}).
The convergence in the first-order results is then recovered, and the second-order results now agree well with the i-QuAPI solution.
This experiment also verifies our claims at the end of \cref{sec:truncation}, stating that the i-QuAPI method is superior in terms of numerical accuracy.

\begin{figure}[!ht]
    \centering
    \begin{subfigure}[b]{0.45\textwidth} 
        \centering
\includegraphics[width=1.\textwidth]{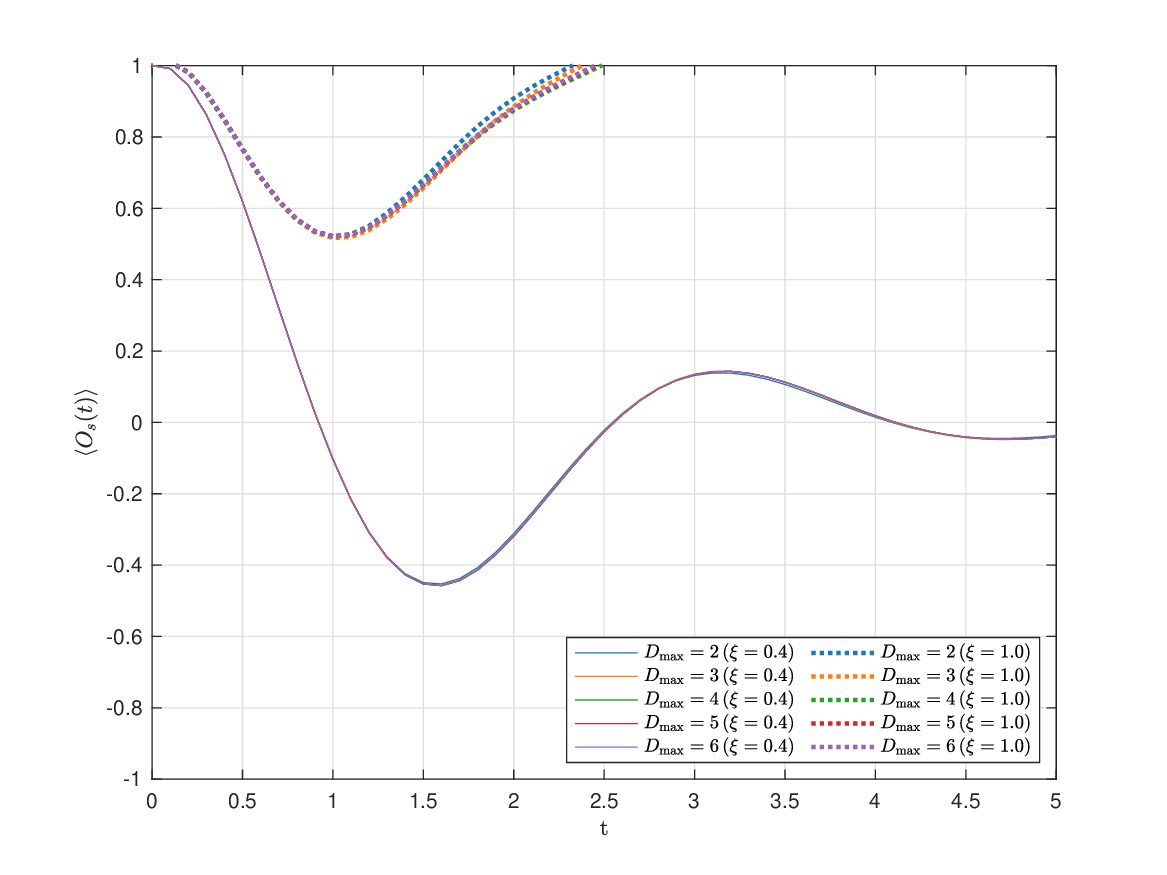}
        \caption{First-order scheme.}
        \label{dif: conv_Dmax_Kmax15_1st}
    \end{subfigure}
    \begin{subfigure}[b]{0.45 \textwidth}
    \centering
    \includegraphics[width=1.\textwidth]{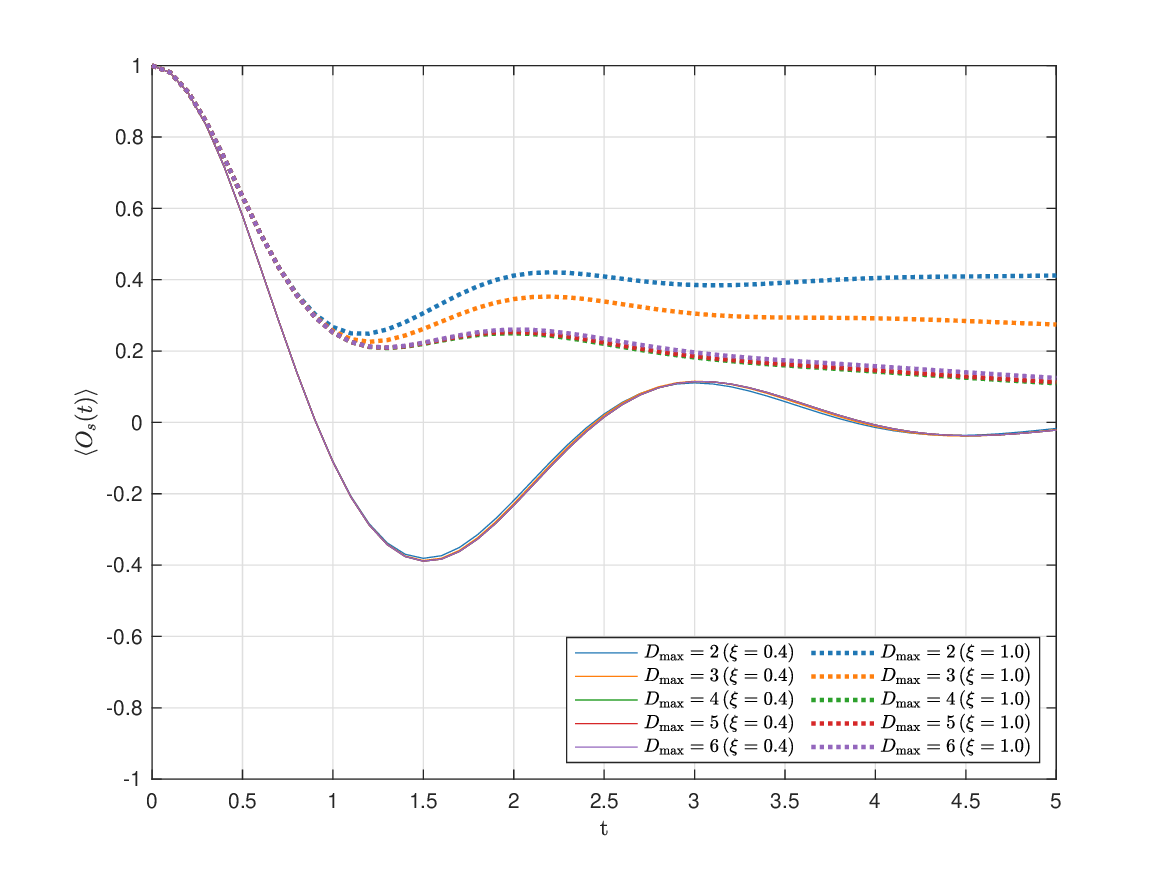}
      \caption{Second-order scheme.}  
      \label{dif: conv_Dmax_Kmax15_2nd}
    \end{subfigure}
    \caption{Convergence test on $D_{\max}$ for $\xi=0.4,1.0$, $K_{\max}=15$, and $\dt = 0.1$}
    \label{dif: conv_Dmax_Kmax15}
\end{figure}

Next, we test the convergence of the method with respect to the parameter $D_{\max}$ by choosing $D_{\max}=2,3,4,5,6$.
We fix $\dt=0.1$ and $K_{\max} = 15$, which means the memory length $T_{\max}$ in \eqref{eq:tilde_B} is set to $1.5$.
In general, it is expected that weaker system-bath coupling requires a smaller value of $D_{\max}$, and this is confirmed by our numerical results in \cref{dif: conv_Dmax_Kmax15}, where the curves corresponding to different values of $D_{\max}$ are almost indistinguishable for $\xi = 0.4$ in plots of both first- and second-order schemes.
When $\xi = 1.0$, the first-order results shown in \cref{dif: conv_Dmax_Kmax15_1st} fail to provide meaning approximations due to insufficient numerical accuracy as in  \cref{fig:conv_Kmax_1storder_xi10}, whereas the second-order results in \cref{fig:conv_Kmax_1storder_xi10} remain reliable and exhibit noticeable differences between curves for $D_{\max} = 2$, $D_{\max} = 3$ and other lines.
The curves for $D_{\max}=4,5,6$ are virtually overlapping, indicating the convergence with respect to $D_{\max}$.
Meanwhile, one can also notice that a larger $D_{\max}$ is needed for a longer simulation time $t$, as the size of the integral domain in the Dyson series \eqref{eq:Ut} is $t^m/m!$, and for a large $m$, this term becomes significant only for large values of $t$.
Recall that $D_{\max}$ can be viewed as an upper bounded for $m$, and hence has to take a greater value for longer simulations.

\begin{figure}[!ht]
    \centering
    \begin{subfigure}[b]{0.45\textwidth}
        \centering
        \includegraphics[width=\textwidth]{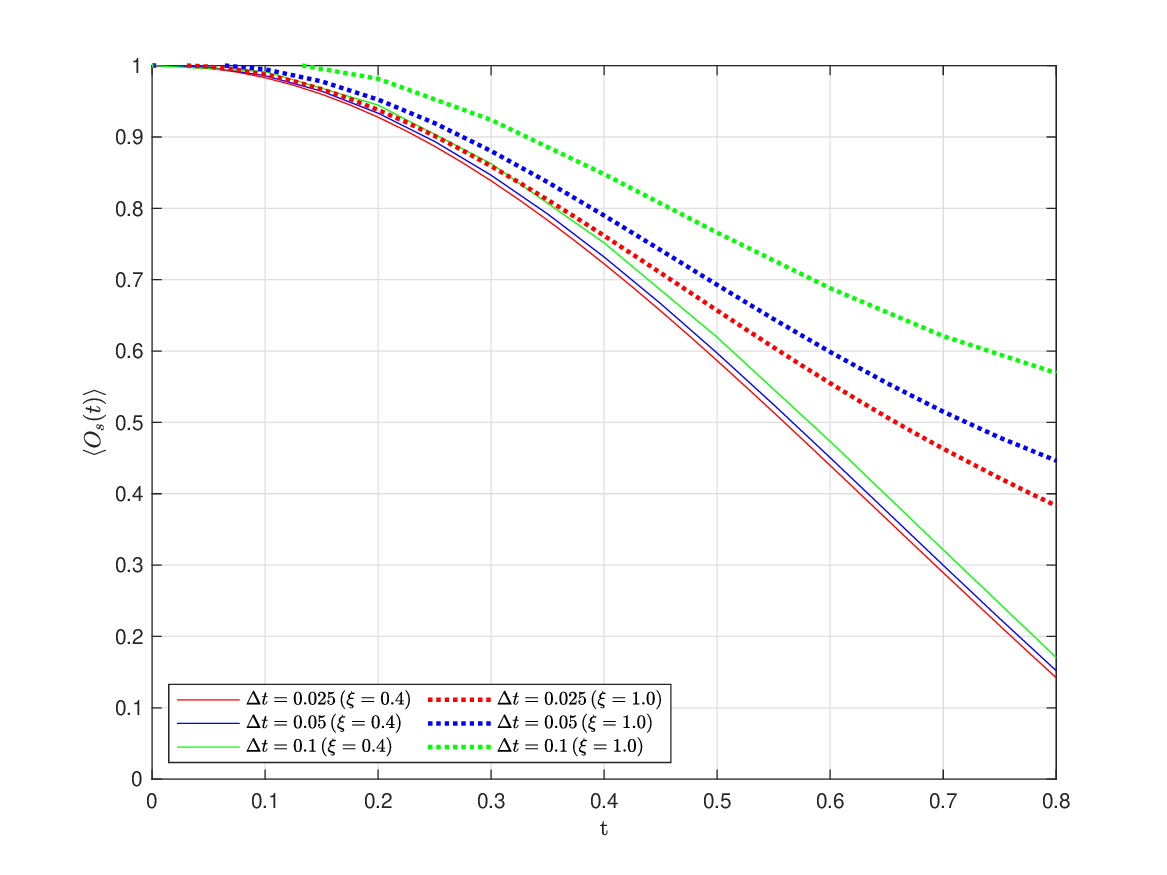}
        \caption{First-order scheme.}
        \label{fig:conv_dt_notrunc_dmax4_1st}
    \end{subfigure}
    \begin{subfigure}[b]{0.45\textwidth}
        \centering
        \includegraphics[width=\textwidth]{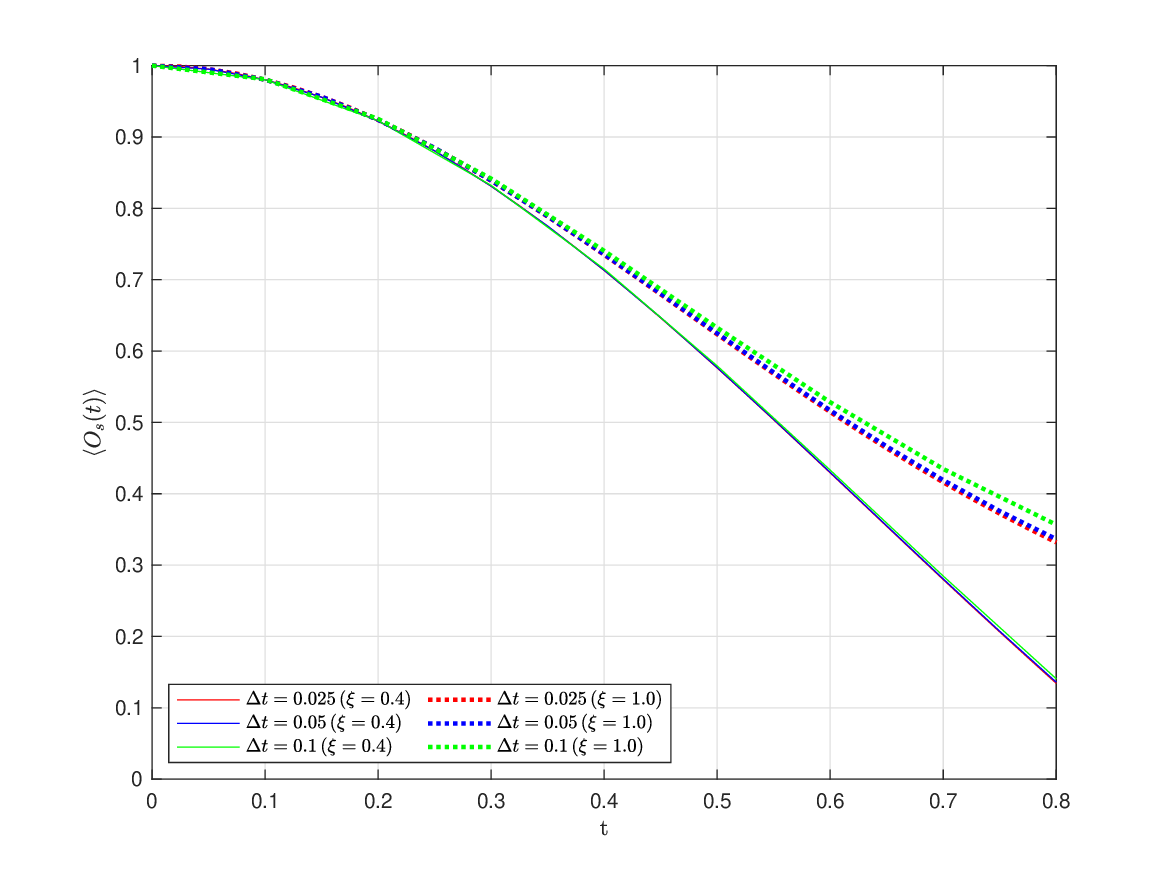}
        \caption{Second-order scheme.}
        \label{fig:conv_dt_notrunc_dmax4_2nd}
    \end{subfigure}
    \caption{Convergence test on $\dt$ for $\xi = 0.4,1.0$ and $D_{\max}=4$.}
    \label{fig:conv_dt_notrunc_dmax4}
\end{figure}

We now perform a convergence test with respect to the time step size $\dt$. 
Our previous results show that $D_{\max} = 4$ suffices for our choices of parameters.
In the experiments below, We will fix $D_{\max}=4$ .
Memory length truncation is not applied in the convergence order test, so that the simulation will be performed only up to $T = 0.8$ to avoid high memory cost.
Let $\expval{O_s(T)}_{\delta t}$, $\expval{O_s(T)}_{2\delta t}$, and $\expval{O_s(T)}_{4\delta t}$ be the expected values of $O_s$ at time $T$ computed using time steps $\delta t$, $2\delta t$, and $4\delta t$, respectively.
Since no exact solutions are available, and it is unaffordable to compute a solution on a very fine grid, we will estimate the convergence order $p$ by
\[
    p=\log_2 \frac{\|\expval{O_s(T)}_{4\delta t}-\expval{O_s(T)}_{2\delta t}\|_2}{ \|\expval{O_s(T)}_{2\delta t}-\expval{O_s(T)}_{\delta t}\|_2}.
\]
The results with $\delta t = 0.025$ are summarized in \cref{table:1st_order_convergence} and \cref{table:2nd_order_convergence}, showing that the desired numerical order can generally be achieved for various choices of $\xi$ and $\beta$.
Some results are also plotted in \cref{fig:conv_dt_notrunc_dmax4}, from which one can observe that the solution at time $T = 0.8$ is monotonic with respect to $\Delta t$, indicating that our estimation of convergence order likely holds.
In \cref{table:2nd_order_convergence}, the figures in the first column drop below $1.77$, possibly due to larger derivatives in the exact solutions for $\beta = 0.2$ (see  \cref{fig:conv_Kmax_2ndorder_xi02}).

\begin{table}[!ht]
    \centering
    \subfloat[First-order scheme]{
    \begin{tabular}{c|cccc}
        &$\xi=0.2$&$\xi=0.4$  &$\xi=0.8$  &$\xi=1.0$  \\
          \hline
       $\beta = 5$&1.0603 & 1.0328 & 1.0071&1.0060\\
       $\beta=2$& 1.0616 &1.0367&1.0230&1.0283\\
       $\beta=1$ & 1.0639&1.0489&1.0696&1.0913
    \end{tabular}
    \label{table:1st_order_convergence}} \qquad
    \subfloat[Second-order scheme]{
    \begin{tabular}{c|cccc}
        &$\xi=0.2$&$\xi=0.4$  &$\xi=0.8$  &$\xi=1.0$  \\
          \hline
       $\beta = 5$&1.7242 & 1.8510 & 2.1225&2.2545\\
       $\beta=2$& 1.7398 &1.8698&2.1665&2.3173\\
       $\beta=1$ & 1.7683&1.9173&2.3094&2.5261
    \end{tabular}
    \label{table:2nd_order_convergence}}

    \vspace{-7pt}
    
    \caption{Convergence orders with respect to $\Delta t$}
    \label{table:convergence}
\end{table}

\begin{figure}
    \centering
    \begin{subfigure}[b]{0.45\textwidth} 
        \centering
\includegraphics[width=1\textwidth]{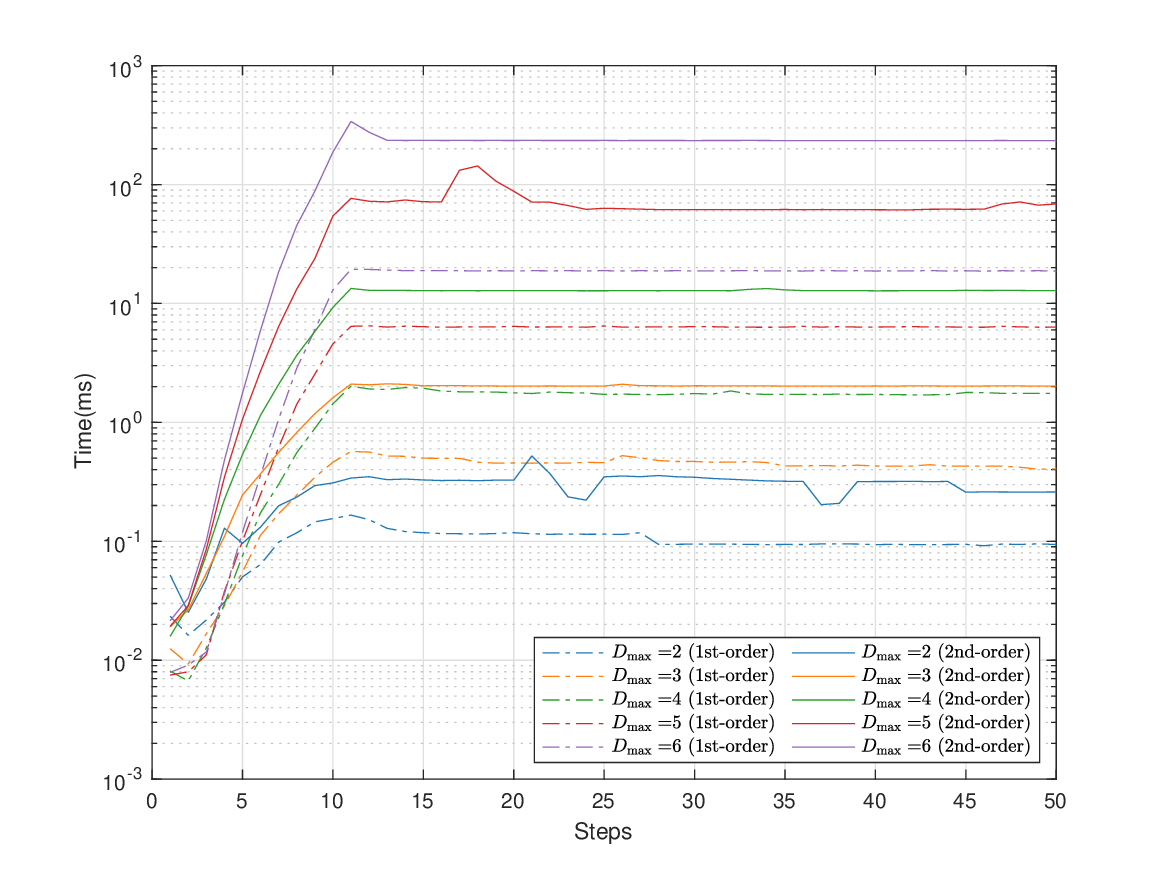}
        \caption{Evaluation time of each step for $K_{\max}=10$.}
        \label{fig:time_kmax10_1st_2nd}
    \end{subfigure}
    \begin{subfigure}[b]{0.45\textwidth} 
        \centering
\includegraphics[width=1\textwidth]{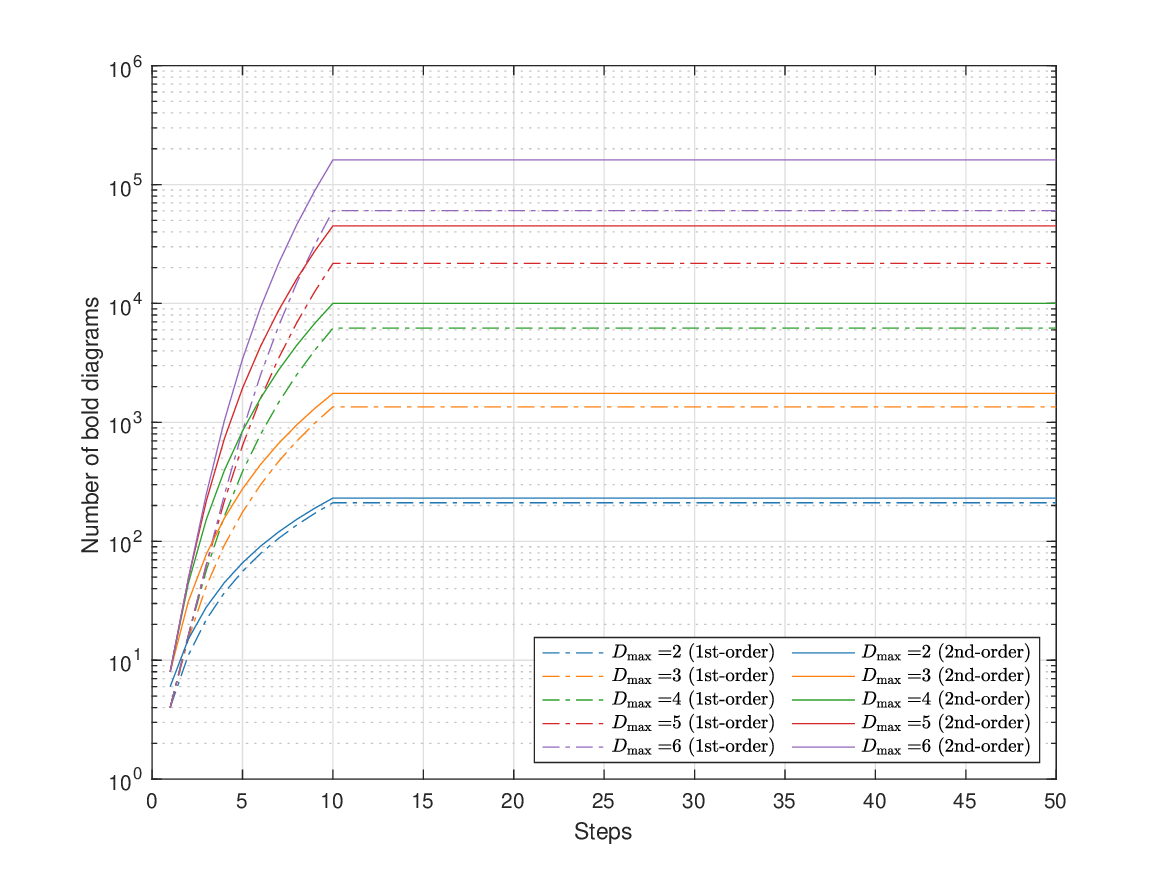}
        \caption{Number of diagrams of each step for $K_{\max}=10$ .}
        \label{fig:ndiag_kmax10_1st_2nd}
    \end{subfigure}

    \caption{Computational time and the number of diagrams in the simulation of the spin-boson model.}
    \label{fig:_cosr}
    \end{figure}

Lastly, we study the computational time and the memory cost in our simulations.
Before the number of time steps reaches $K_{\max}$, the number of diagrams to be computed, as well as the computational time for each time step, increases with the simulation time.
Once the evolution reaches $K_{\max}$ time steps, the computational cost for each time step no longer grows.
The run-time statistics of our experiments, plotted in \cref{fig:_cosr}, confirms this behavior.
\cref{fig:ndiag_kmax10_1st_2nd}
For both first- and second-order schemes, the stable computational time after $K_{\max}$ steps is roughly proportional to the number of diagrams, but the second-order scheme has a larger coefficient of proportionality due to the extra steps such as the two-stage procedure for second-order corrections.

\subsection{Multilevel Systems}
In this section we present the numerical results for multilevel systems obtained using our scheme. We consider an $M$-level open quantum system with nearest-neighbor-coupling Hamiltonian for the system:
\begin{displaymath}
    H_s = \Omega\sum_{i=0}^{M-1} (\dyad{i}{i+1}+\dyad{i+1}{i}),
\end{displaymath}
corresponding to an $M \times M$ matrix whose superdiagonal and subdiagonal elements are all $\Omega$, while all other elements are zero.
The system couples to a bosonic bath $H_b$ through operator $W_s$ defined as follows \cite{makri2020smallMatrixPath}: 
\begin{displaymath}
    W_s = \operatorname{diag}\Big({-1},\, -1+\frac{1}{J},\,\ldots,-1 + \frac{d-1}{J}\Big),\quad J = \frac{d-1}{2}.
\end{displaymath}
We monitor the level/site
populations
\[
P_i(t)\,:=\,\langle i|\rho_s(t)|i\rangle,
\]
Unless stated otherwise, we initialize the system in a localized state
$\rho_s(0)=\dyad{1}{1}$ and set $\rho_b$ as the thermal equilibrium as stated in \cref{eq:thermal_state}.
We will only apply the second-order scheme for all simulations in this section.

\begin{figure}
    \centering
    
    \begin{subfigure}[b]{0.45\textwidth} 
        \centering
\includegraphics[width=1\textwidth]{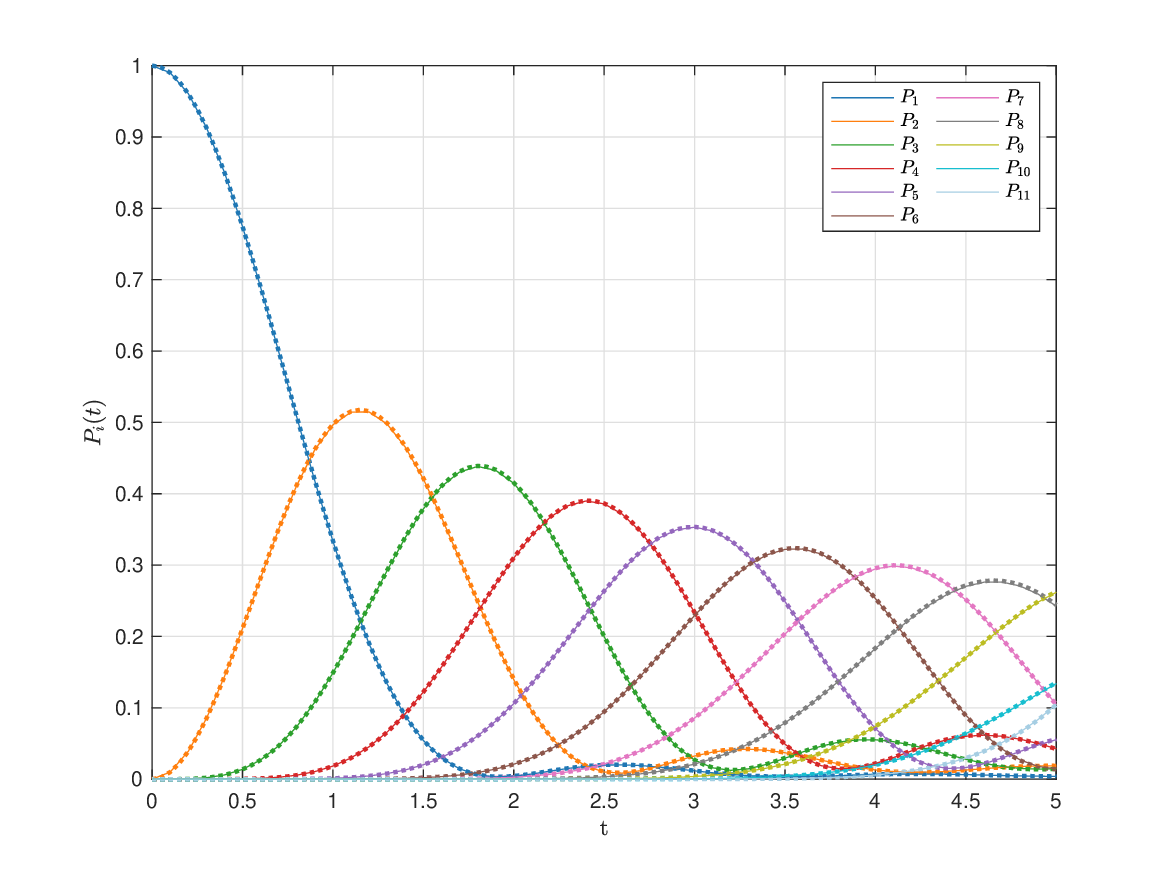}
        \caption{$\xi = 0.4$}
        \label{fig:11_level_vs_inchworm_beta5_xi04_2ndorder_kmax20_dt005_dmax4}
    \end{subfigure}
    \hfill
    \begin{subfigure}[b]{0.45\textwidth} 
        \centering
\includegraphics[width=1\textwidth]{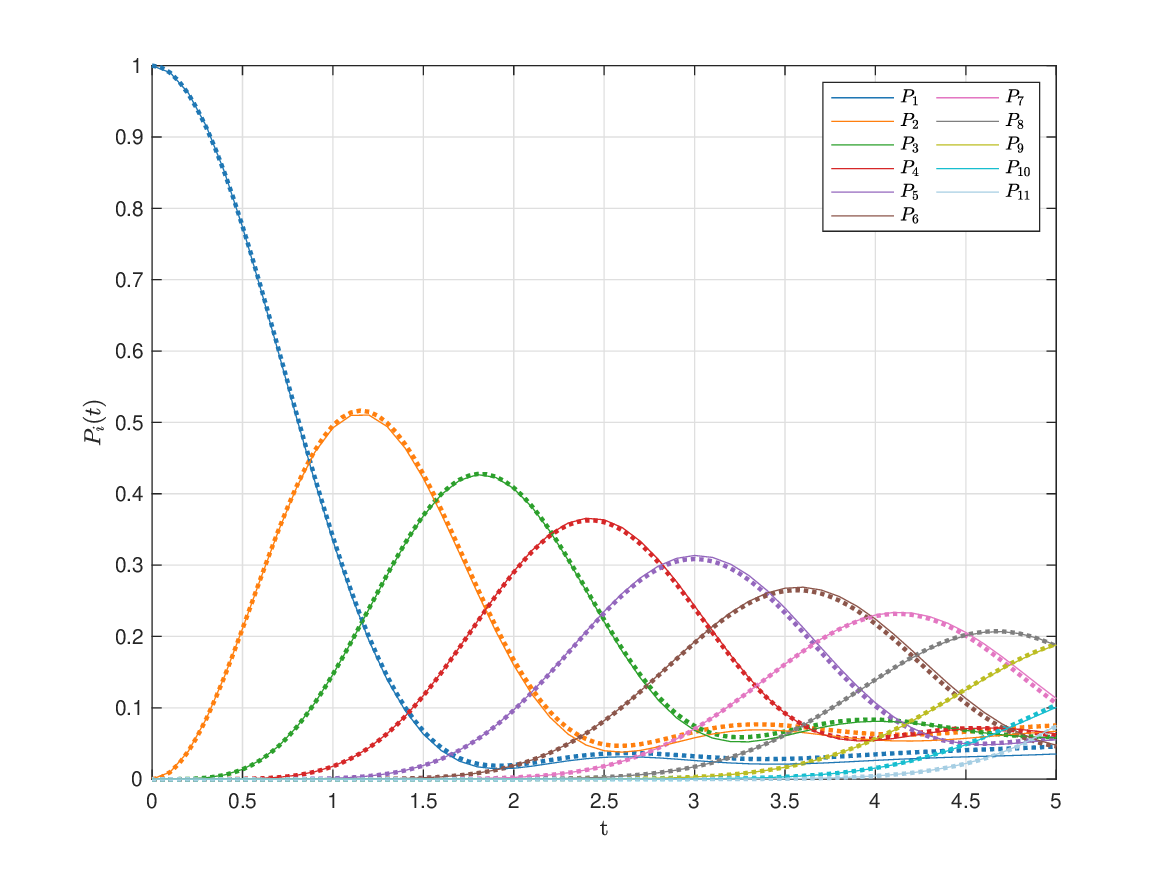}
        \caption{$\xi = 1.0$}
        \label{fig:11_level_vs_inchworm_beta5_xi10_2ndorder_kmax20_dt005_dmax4}
    \end{subfigure}

    \caption{Comparison of second-order scheme and inchworm method for a 11-level system with $\beta=5$, $D_{\max}=4$, $K_{\max}=20$, and $\dt = 0.05$. The solid lines represent the inchworm method; the dashed lines correspond to the second-order scheme.} 
    \label{fig:11_level_vs_inchworm_2ndorder_kmax20_dmax4}
    \end{figure}

We first consider an $11$-level system ($M=11$) with the following parameters: 
\begin{displaymath}
\Omega = 1,\quad \beta = 5,\quad \omega_c = 2.5,\quad \dt = 0.05,\quad K_{\max} = 20,\quad D_{\max} = 4.
\end{displaymath}
The Kondo parameter $\xi$ is chosen as $0.4$ and $1.0$, representing weak and strong couplings, respectively.
For this test case, the i-QuAPI method becomes unaffordable, and we applied the inchworm method \cite{chen2017inchwormITheory} for cross-validation.
The evolution of populations at all levels are plotted in \cref{fig:11_level_vs_inchworm_2ndorder_kmax20_dmax4}, where both methods provide nearly identical results in all cases.
We stopped at $t = 5$ since the inchworm method becomes expensive for longer simulations.
In general, weaker fluctuations of the curves are observed for strong couplings, indicating the effect of quantum dissipation.

\begin{figure}
        \centering
        \begin{subfigure}[b]{0.45\textwidth} 
            \includegraphics[width=\linewidth]{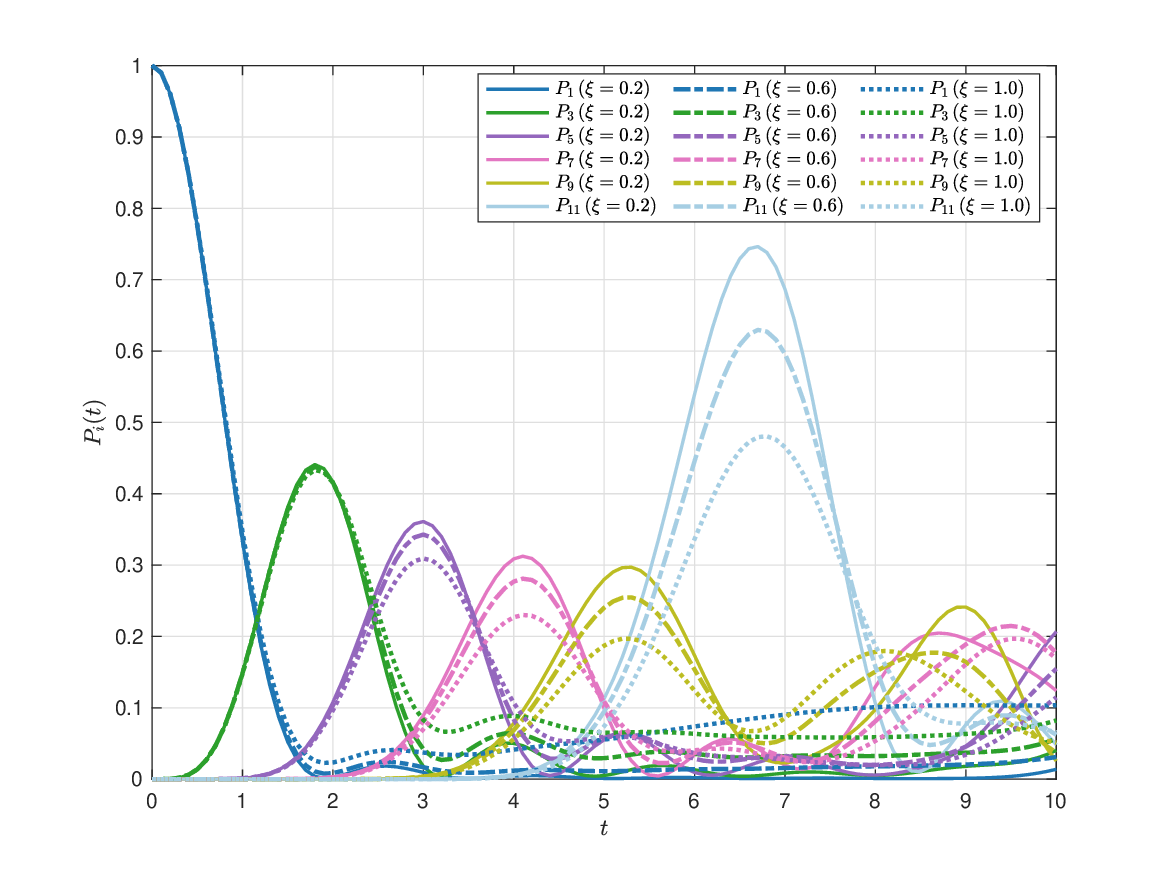}
            \caption{Increasing $\xi$ at fixed $\beta=5,\ \omega_c=2.5,\ \Omega=1.0$.}
            \label{fig:11level_xi}
        \end{subfigure}
        \begin{subfigure}[b]{0.45\textwidth} 
            \includegraphics[width=\linewidth]{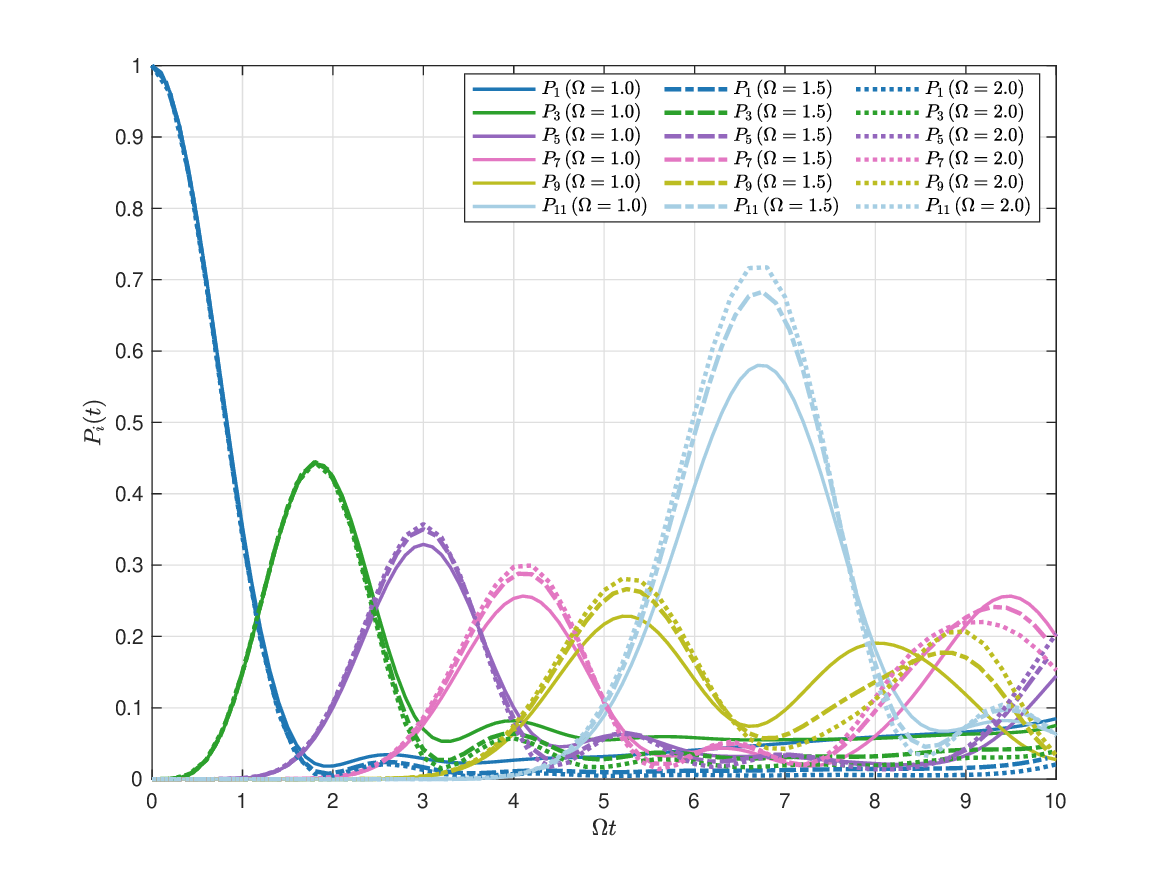}
            \caption{Rescaling time by $\Omega t$ at $\beta=5,\ \omega_c=5.0,\ \xi=0.4$.}
            \label{fig:11level_omega}
        \end{subfigure}
        \\
        \begin{subfigure}[b]{0.45\textwidth} 
            \includegraphics[width = \linewidth]{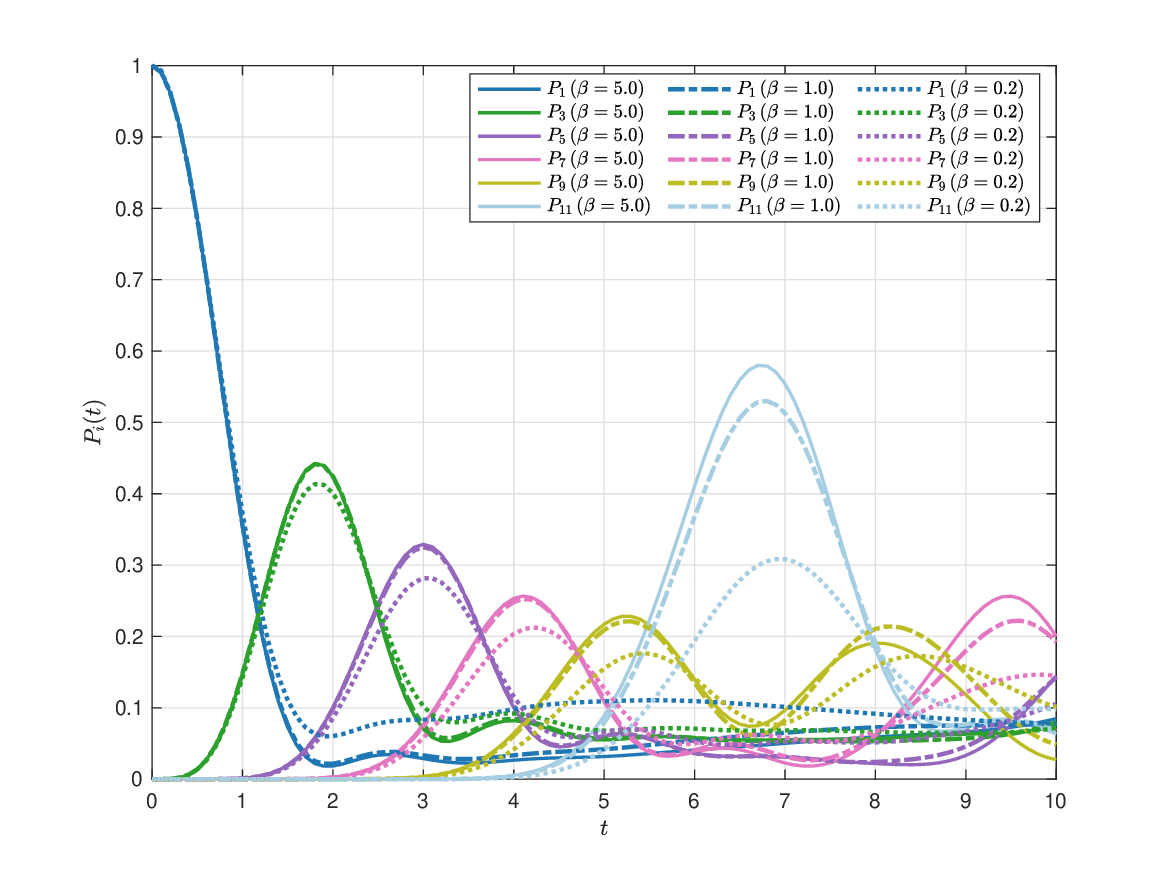}
            \caption{Increasing temperature at $ \omega_c=5.0, \ \xi=0.4,\ \Omega=1.0$.}
            \label{fig:11level_beta}
        \end{subfigure}
        \begin{subfigure}[b]{0.45\textwidth} 
            \includegraphics[width = \linewidth]{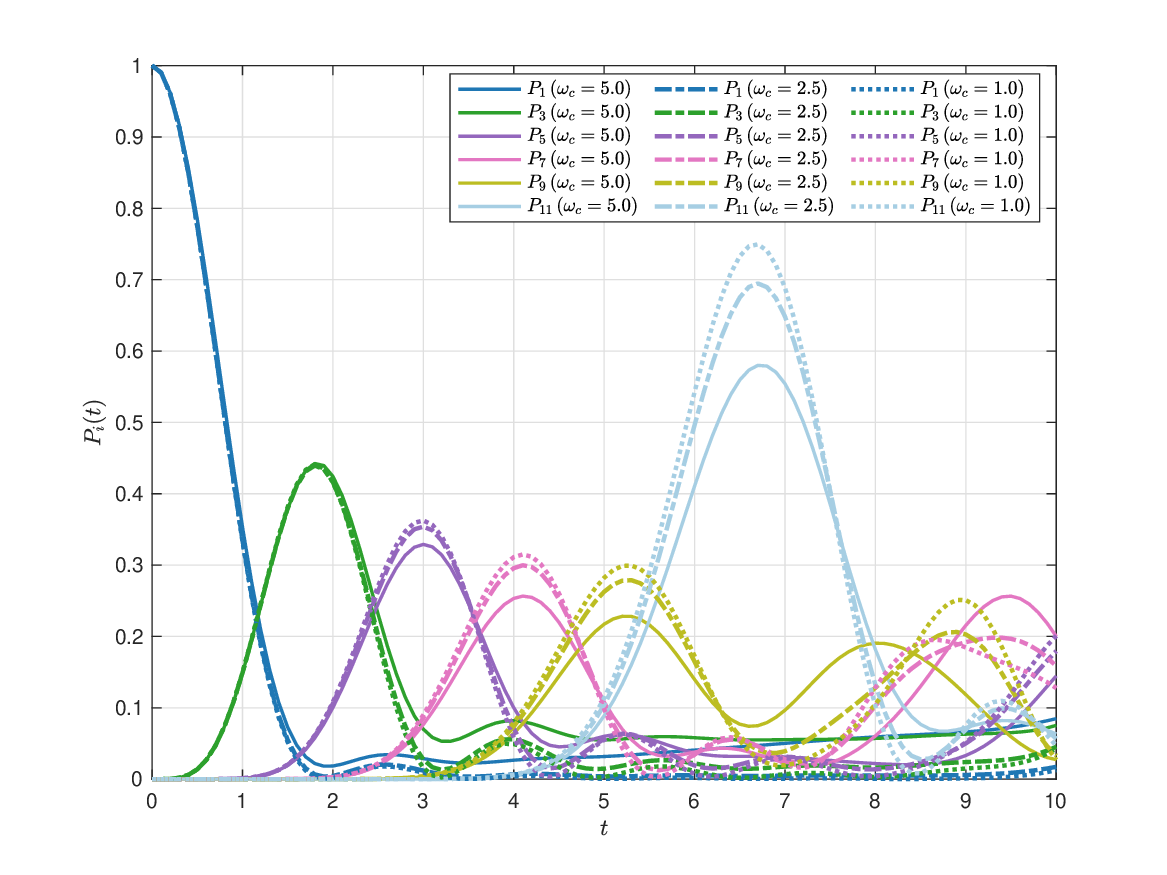}
            \caption{Decreasing $\omega_c$ at $\beta=5,\ \Omega=1,\,\xi=0.4$.}
            \label{fig:11level_wc}
        \end{subfigure}
        \caption{Population dynamics of a $M$-level($M=11$) under varying system-bath parameters.}
        \label{fig:11level_longtime}
    \end{figure}

Next, we present the population dynamics over a longer time span up to $t = 10$, as shown in \cref{fig:11level_longtime}.
In this experiment, we set the time step size to $\Delta t = 0.1$, the memory length to $K_{\max} = 10$, and $D_{\max}=4$.
We investigate how the population evolution depends on the Kondo parameter $\xi$, the inverse temperature $\beta$, the damping rate $\Omega$, and the cutoff frequency $\omega_c$. As shown in \Cref{fig:11level_xi}, increasing $\xi$ leads to faster decay of oscillations, reduced peak heights, and broader peak widths.
The long-time behavior approaches the thermal distribution more quickly at larger $\xi$. In \Cref{fig:11level_omega}, with $\Omega= 1,1.5,2$, the curves nearly collapse when plotted vs $\Omega t$, reflecting the exact scaling of the unitary part $\e^{-\ii H_s t}$. The differences are due to the bath part. The larger $\Omega$ exhibits weaker decay. As $\beta$ decreases (temperature rises), $\coth{(\beta\omega/2)}$ increases for a certain frequency $\omega$ which yields stronger damping of population oscillations and faster relaxation: in \cref{fig:11level_beta} the site populations $P_i(t)$ exhibits lower peaks and a quicker trend to thermal state as $\beta$ varies from 5 to 1. In \Cref{fig:11level_beta}, as $\omega_c$ decreases, the population oscillations are less decayed and persist longer.

\begin{figure}
        \centering
        \begin{subfigure}{0.45\textwidth} 
             \includegraphics[width = \linewidth]{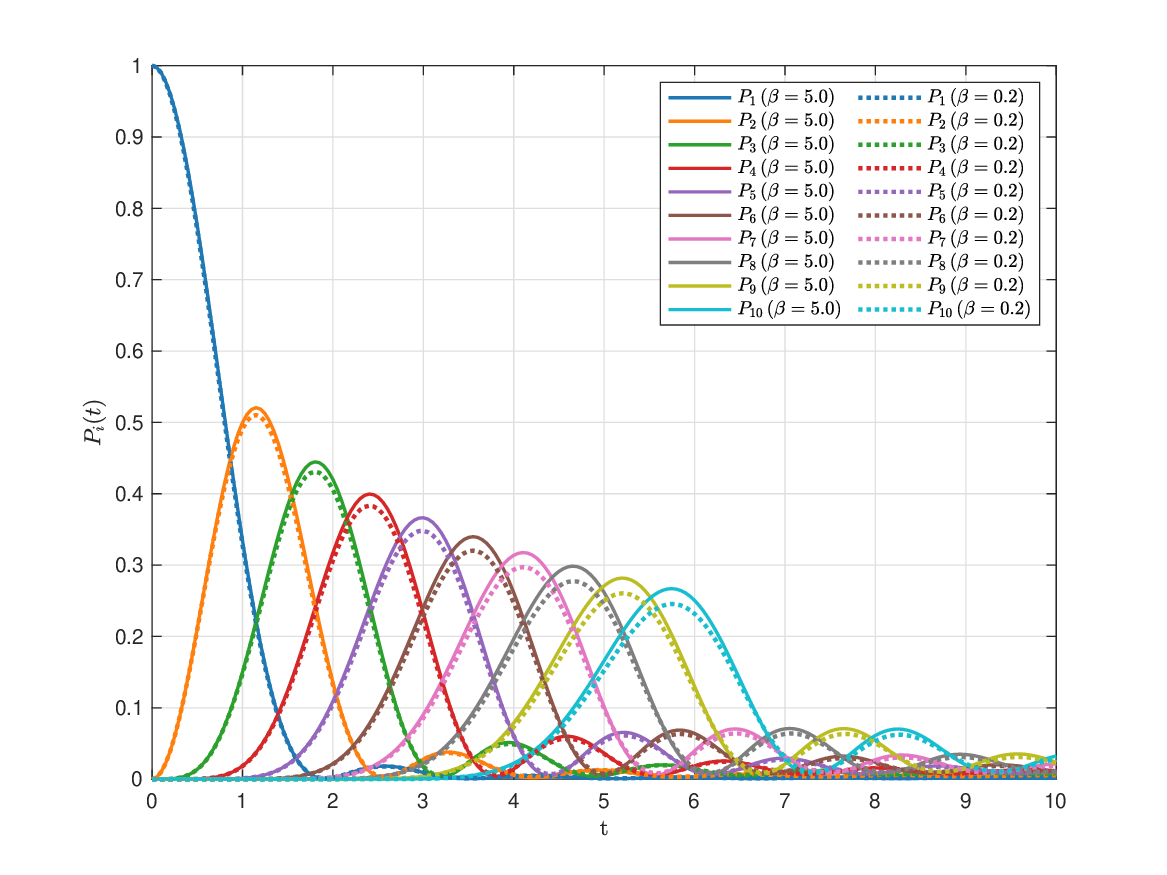}
             \caption{$\rho_s(0) = \dyad{1}{1}$.}
              \label{fig:50level_rho0}
        \end{subfigure}
        \begin{subfigure}{0.45\textwidth} 
             \includegraphics[width = \linewidth]{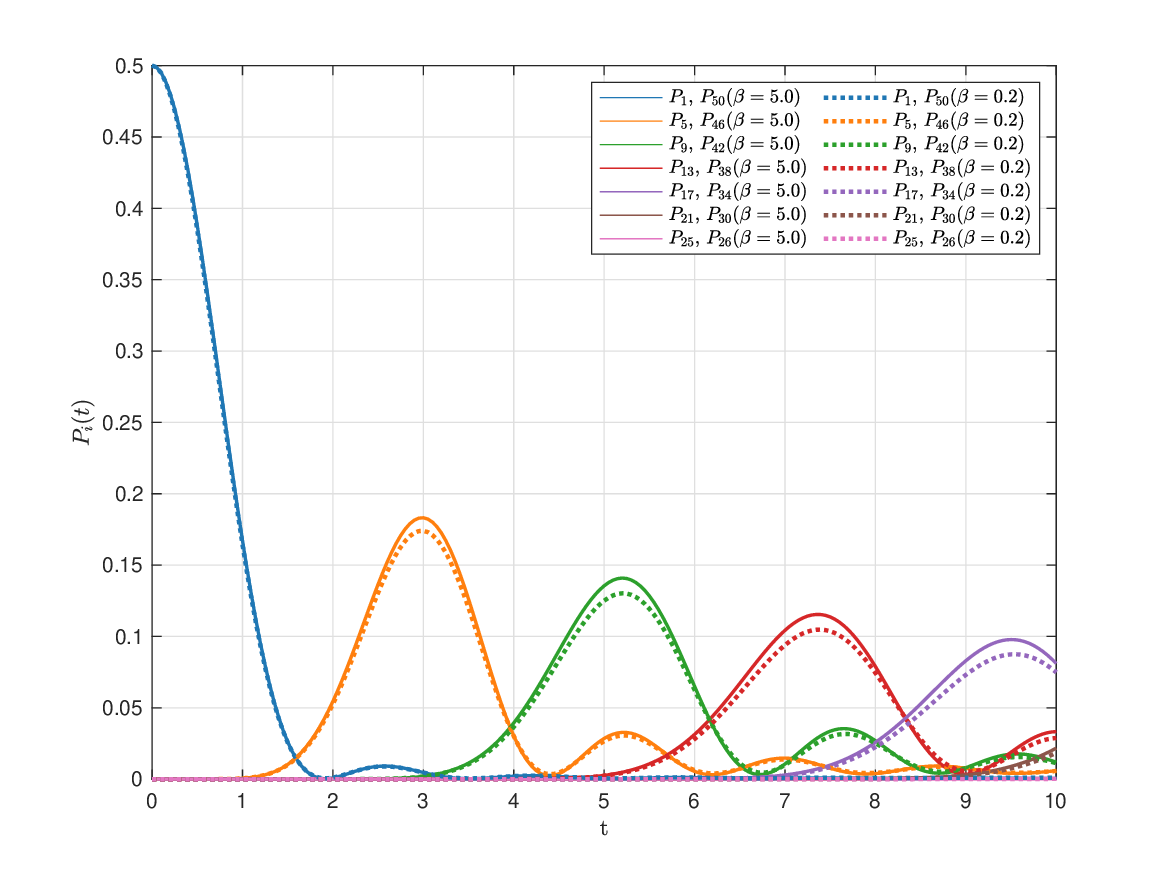}
        \caption{$\rho_s(0)=\tfrac{1}{2}\big(\dyad{1}{1} + \dyad{M}{M}\big)$.}
        \label{fig:50level_symrho0}
        \end{subfigure}
        \caption{Population dynamics of a $M$-level ($M=50$) system under $\xi=0.4$,  $\omega_{c}=5.0$, $\Omega=1.0$, $\dt = 0.05$, $K_{\max} = 20$, $D_{\max} = 4$.}
        \label{fig:50level}
    \end{figure}

Finally, we demonstrate the capability of our scheme for larger multilevel systems. As shown in \Cref{fig:50level}, we simulate a 50-level system for $\xi=0.4$, $\omega_{c}=5.0$, $\Omega=1.0$, $\dt = 0.05$, $K_{\max} = 20$, $D_{\max} = 4$. \Cref{fig:50level_rho0} shows several site populations for the localized initial state
$\rho_s(0) = \dyad{1}{1}$. The peak locations move approximately linearly in $t$ with a slowly decaying envelope due to dissipation. \Cref{fig:50level_symrho0} uses symmetric initial condition $\rho_s(0)=\tfrac{1}{2}\big(\dyad{1}{1} + \dyad{M}{M}\big)$. In this case the populations of symmetric sites coincide, \textit{i.e.}, $P_{i}(t)=P_{M+1-i}(t)$.
In the plot, solid lines denote $\beta=5.0$ and dashed lines denote $\beta=0.2$; for each color we draw the reflection pair $(i,\,M{+}1-i)$ with the same line style, and their overlap certifies the equality.
These results demonstrate that the second-order scheme handles large multilevel system reliably with the modest truncation parameters $K_{\max}$ and $D_{\max}$. Once $K_{\max}=20$ is reached, the computation time per time step is approximately $90\,\mathrm{s}$.

\section{Conclusion and future work}
\label{sec_conclusion_and_future_work}
We have introduced two discrete versions of the Dyson series with first- and second-order accuracy in the time step $\Delta t$. 
Rather than discretizing the high-dimensional integrals in the Dyson series directly, we derived the discrete forms following the derivation of Dyson series from the integral representation of the perturbed Schr\"odinger equation. 
This construction naturally extends to open quantum systems, leading to a novel numerical method for simulating system-bath dynamics. 
The convergence rates of these methods are rigorously proven theoretically and carefully verified numerically.
To make the method computationally feasible, we incorporated several acceleration techniques, including recursive computation, memory-length truncation, and restrictions on the number of coupling operators. 
Together, these strategies yield a scheme with significantly reduced time and space complexities, particularly well-suited for multilevel systems.

In this work, we focused on first- and second-order schemes, but extensions to higher-order methods will be the subject of future research. 
We are also investigating the application of this framework to more complex settings, such as spin-chain models.

\section{Appendix}
\label{sec_appendix}
In this appendix, we provide the complete pseudocode for the second-order numerical scheme.
\Cref{algo:2nd_order} provides the general sketch of the algorithm, and the details of left and right extensions are given in \Cref{algo:left_extension} and \Cref{algo:right_extension}, respectively.

\begin{algorithm}[!ht]
    \begin{algorithmic}[1]
        \State $\Lambda(\,;\,) \gets \rho_{s,0}$
        \For{$k=0,\ldots,N-1$}
        \State $\Lambda\gets\Call{LeftExtension}{\Lambda,k}$
        \State $\Lambda\gets\Call{RightExtension}{\Lambda,k}$
        \State $\Tilde{\rho}_{s,k}\gets\Lambda(0,\ldots,0;0,\ldots,0)$
        \EndFor
\end{algorithmic}
\caption{The second-order scheme for computing $\Tilde{\rho}_{s,N}$}
\label{algo:2nd_order}
\end{algorithm}

\begin{algorithm}[!ht]
\caption{\textsc{LeftExtension}$(\Lambda,k)$}
     \begin{algorithmic}[1]
                 \State $L\gets \{\}$ 
\For{$(j_{1-}, j_{1+}, \ldots, j_{k-}, j_{k+}) \in \{0,1\}^{2k}$}
\State $L(1,j_{k-}, \ldots, j_{1-}; j_{1+}, \ldots, j_{k+})\gets \mathcal{G}_1\Lambda(j_{k-}, \ldots, j_{1-}; j_{1+}, \ldots, j_{k+})$
\State $Op\gets \boldsymbol{0}$
\For{$\ell = 1-,1+,\ldots,k-,k+$ satisfying $j_{\ell}=0$}
\State $j_{\ell}\gets 1$
\State $Op \gets Op + \mathcal{G}_1 \Bcal{(k+1)-}{\ell}\Lambda(j_{k-}, \ldots, j_{1-}; j_{1+}, \ldots, j_{k+})$
\State $j_{\ell}\gets 0$
\EndFor
\State $L(0,j_{k-}, \ldots, j_{1-}; j_{1+}, \ldots, j_{k+})\gets Op$
\EndFor
\For{$(j_{1-}, j_{1+}, \ldots, j_{k-}, j_{k+}) \in \{0,1,2\}^{2k}$}

\State $Op_0 \gets \mathcal{P}_{s,0}\Lambda(j_{k-}, \ldots, j_{1-}; j_{1+}, \ldots, j_{k+})$
\For{$\ell = 1-, 1+, \ldots, k-, k+$ satisfying $j_{\ell} = 0$ or $j_{\ell} = 1$}
\State $j_{\ell} \gets j_{\ell}+1$
\State $Op_0 \gets Op_0 + \mathcal{P}_{s,1} \Bcal{(k+1)-}{\ell} \Lambda(j_{k-}, \ldots, j_{1-}; j_{1+}, \ldots, j_{k+})$
\State $j_{\ell} \gets j_{\ell}-1$
\EndFor
\State $Op_0 \gets Op_0 + \mathcal{G}_2\Bcal{(k+1)-}{(k+1)-} L(1,j_{k-}, \ldots, j_{1-}; j_{1+}, \ldots, j_{k+})$
\State $Op_1 \gets \mathcal{P}_{s,1}\Lambda(j_{k-}, \ldots, j_{1-}; j_{1+}, \ldots, j_{k+})$
\State $Op_1 \gets Op_1 + \mathcal{G}_2 L(0,j_{k-}, \ldots, j_{1-}; j_{1+}, \ldots, j_{k+})$
\For{$\ell = 1-,1+,\ldots,k-,k+$ satisfying $j_{\ell}=0$}
\State $j_{\ell}\gets1$
\State $Op_0 \gets Op_0 + \mathcal{G}_2 \Bcal{(k+1)-}{\ell} L(0,j_{k-}, \ldots, j_{1-}; j_{1+}, \ldots, j_{k+})$
\State $Op_1 \gets Op_1 + \mathcal{G}_2 \Bcal{(k+1)-}{\ell}L(1,j_{k-}, \ldots, j_{1-}; j_{1+}, \ldots, j_{k+})$
\State $j_{\ell}\gets 0$
\EndFor
\State $\Lambda(0,j_{k-}, \ldots, j_{1-}; j_{1+}, \ldots, j_{k+})\gets Op_0$
\State $\Lambda(1,j_{k-}, \ldots, j_{1-}; j_{1+}, \ldots, j_{k+})\gets Op_1$
\State $\Lambda(2,j_{k-}, \ldots, j_{1-}; j_{1+}, \ldots, j_{k+})\gets 2\mathcal{G}_2L(1,j_{k-}, \ldots, j_{1-}; j_{1+}, \ldots, j_{k+})$
\EndFor
     \end{algorithmic}
     \label{algo:left_extension}
\end{algorithm}

\begin{algorithm}[!ht]
\caption{\textsc{RightExtension}$(\Lambda,k)$}
     \begin{algorithmic}[1]
\State $R\gets\{\}$
\For{$j_{1-},j_{1+},\ldots,j_{k-},j_{k+},j_{(k+1)-}\in\{0,1\}^{2k+1}$}
\State $R(j_{(k+1)-},j_{k-},\ldots,j_{1-};j_{1+},\ldots,j_{k+},1)\gets \Lambda(j_{(k+1)-},\ldots,j_{1-};j_{1+},\ldots,j_{k+})\mathcal{G}_1^\dagger$
\State $Op\gets 0$
\For{$\ell =1-,1+,\ldots,k-,k+,(k+1)-$ satisfying $j_\ell=0$}
\State $j_\ell\gets1$
\State $Op\gets Op+\Lambda(j_{(k+1)-},\ldots,j_{1-};j_{1+},\ldots,j_{k+})\mathcal{G}_1^\dagger\Bcal{\ell}{(k+1)+}$
\State $j_\ell \gets0$
\EndFor
\State $R(j_{k-},\ldots,j_{1-};j_{1+},\ldots,j_{k+},0)\gets Op)$
\EndFor

\For{$(j_{1-}, j_{1+}, \ldots, j_{k-}, j_{k+},j_{(k+1)-}) \in \{0,1,2\}^{2k+1}$}

\State $Op_0 \gets \mathcal{P}_{s,0}\Lambda(j_{(k+1)-}, \ldots, j_{1-}; j_{1+}, \ldots, j_{k+})$
\For{$\ell = 1-, 1+, \ldots, k-, k+,(k+1)-$ satisfying $j_{\ell} = 0$ or $j_{\ell} = 1$}
\State $j_{\ell} \gets j_{\ell}+1$
\State $Op_0 \gets Op_0 + \Lambda(j_{(k+1)-}, \ldots, j_{1-} j_{1+}, \ldots, j_{k+})\mathcal{P}_{s,1}^\dagger \Bcal{\ell}{(k+1)+}$
\State $j_{\ell} \gets j_{\ell}-1$
\EndFor
\State $Op_0 \gets Op_0 +  R(j_{(k+1)-}, \ldots, j_{1-}; j_{1+}, \ldots, j_{k+},1)\mathcal{G}_2^\dagger\Bcal{(k+1)+}{(k+1)+}$
\State $Op_1 \gets \Lambda(j_{(k+1)-}, \ldots, j_{1-}; j_{1+}, \ldots, j_{k+})\mathcal{P}_{s,1}^\dagger$
\State $Op_1 \gets Op_1 + R(j_{(k+1)-}, \ldots, j_{1-}; j_{1+}, \ldots, j_{k+},0)\mathcal{G}_2^\dagger $
\For{$\ell = 1-,1+,\ldots,k-,k+,(k+1)-$ satisfying $j_{\ell}=0$}
\State $j_{\ell}\gets1$
\State $Op_0 \gets Op_0 +  R(j_{(k+1)-}, \ldots, j_{1-}; j_{1+}, \ldots, j_{k+},0)\mathcal{G}_2^\dagger \Bcal{\ell}{(k+1)+}$
\State $Op_1 \gets Op_1 + R(j_{(k+1)-}, \ldots, j_{1-}; j_{1+}, \ldots, j_{k+},1)\mathcal{G}_2^\dagger \Bcal{\ell}{(k+1)+}$
\State $j_{\ell}\gets 0$
\EndFor
\State $\Lambda(j_{(k+1)-}, \ldots, j_{1-}; j_{1+}, \ldots, j_{k+},0)\gets Op_0$
\State $\Lambda(j_{(k+1)-}, \ldots, j_{1-}; j_{1+}, \ldots, j_{k+},1)\gets Op_1$
\State $\Lambda(j_{(k+1)-}, \ldots, j_{1-}; j_{1+}, \ldots, j_{k+},2)\gets 2R(j_{(k+1)-}, \ldots, j_{1-}; j_{1+}, \ldots, j_{k+},1)\mathcal{G}_2^\dagger$
\EndFor

\end{algorithmic}
     \label{algo:right_extension}
\end{algorithm}

\bibliographystyle{amsplain}
\bibliography{myBib_abbr}

\end{document}